\documentclass[twocolumn]{aastex631}

\usepackage{rotating}
\usepackage{amsmath}
\usepackage{amssymb}

\shorttitle{The massive cluster IRS 13}
\shortauthors{Pei{\ss}ker et al.}

\begin{document}

\title{The Evaporating Massive Embedded Stellar Cluster IRS 13 Close to Sgr~A*. II. Kinematic structure}

\correspondingauthor{Florian Pei{\ss}ker}
\email{peissker@ph1.uni-koeln.de}

\author[0000-0002-9850-2708]{Florian Pei$\beta$ker}
\affil{I.Physikalisches Institut der Universit\"at zu K\"oln, Z\"ulpicher Str. 77, 50937 K\"oln, Germany}

\author[0000-0001-6450-1187]{Michal Zaja\v{c}ek}
\affil{Department of Theoretical Physics and Astrophysics, Faculty of Science, Masaryk University, Kotl\'a\v{r}sk\'a 2, 611 37 Brno, Czech Republic}

\author[0009-0000-2832-8989]{Mat\'u\v{s} Labaj}
\affil{Department of Theoretical Physics and Astrophysics, Faculty of Science, Masaryk University, Kotl\'a\v{r}sk\'a 2, 611 37 Brno, Czech Republic}

\author{Lauritz Thomkins}
\affil{I.Physikalisches Institut der Universit\"at zu K\"oln, Z\"ulpicher Str. 77, 50937 K\"oln, Germany}

\author{Andreas Elbe}
\affil{I.Physikalisches Institut der Universit\"at zu K\"oln, Z\"ulpicher Str. 77, 50937 K\"oln, Germany}

\author[0000-0001-6049-3132]{Andreas Eckart}
\affil{I.Physikalisches Institut der Universit\"at zu K\"oln, Z\"ulpicher Str. 77, 50937 K\"oln, Germany}
\affil{Max-Plank-Institut f\"ur Radioastronomie, Auf dem H\"ugel 69, 53121 Bonn, Germany}

\author[0000-0001-5342-5713]{Lucas Labadie}
\affil{I.Physikalisches Institut der Universit\"at zu K\"oln, Z\"ulpicher Str. 77, 50937 K\"oln, Germany}

\author[0000-0002-5760-0459]{Vladim\'ir Karas}
\affil{Astronomical Institute, Czech Academy of Sciences, Bo\v{c}n\'{i} II 1401, CZ-14100 Prague, Czech Republic}

\author[0000-0001-7134-9005]{Nadeen B. Sabha}
\affil{Institut f\"ur Astro- und Teilchenphysik, Universit\"at Innsbruck, Technikerstr. 25, 6020 Innsbruck, Austria}

\author[0000-0002-5859-3932]{Lukas Steiniger}
\affil{I.Physikalisches Institut der Universit\"at zu K\"oln, Z\"ulpicher Str. 77, 50937 K\"oln, Germany}

\author[0000-0002-4902-3794]{Maria Melamed}
\affil{I.Physikalisches Institut der Universit\"at zu K\"oln, Z\"ulpicher Str. 77, 50937 K\"oln, Germany}


\begin{abstract}

The existence of two distinct and apparently unrelated populations of dusty stellar objects in the {Nuclear Stellar Cluster (NSC) of the Milky Way}, namely IRS~13 and the S-cluster, are potentially prone to a general process describing the star formation history in the Galactic Center (GC). The former cluster is thought to be entangled in the clockwise and counterclockwise disks, a large-scale stellar distribution revealed by the analysis of stars at different distances from Sgr~A*, the supermassive black hole in the GC. Recently, this large-scale distribution was reported to exhibit a multi-disk structure with at least four components. Motivated by this finding, we revisit the anisotropic IRS~13 cluster and find strong evidence for a disk-like structure. An examination of about 50 individual stellar orbits reveals a new structure that does not follow any trend known in the literature. Furthermore, we investigate the possibility of an inspiral cluster undergoing star formation processes, as proposed by several authors. Using a simplified N-body simulation to reproduce our observational results, we conclude that, under certain conditions, a massive cluster can migrate from the Circum Nuclear Disk toward the inner parsec. Based on this classification, we revisit the large-scale NACO (VLT) observations of IRS~13 and find evidence for a separation of the cluster into a gravitationally stable core remnant and a dissipating part. With the velocity-resolved H30$\alpha$ line and the broadband spectral energy distribution of IRS~13 E3, we provide tentative support for the existence of an intermediate-mass black hole of $\sim 3\times 10^4\,M_{\odot}$ surrounded by a hot gaseous stream. 
\end{abstract}


\section{Introduction} \label{sec:1}

The Galactic center hosts the closest supermassive black hole Sgr~A* which underlines the unique character of this gravitational laboratory. In this environment, the interaction between stellar objects and Sgr~A* can be studied in detail to reveal the imprints of Sgr~A* resulting in breakthrough discoveries such as the Schwarzschild precession of the S-cluster star S2 \citep{Parsa2017, gravity2018, Do2019S2} or an upper limit on the dark mass inside its orbit \citep{Heissel2022, Peissker2022}. One of the most intriguing discoveries was the unexpected detection of young O/B stars in the direct vicinity ($\sim$ 40 mpc) of Sgr~A* \citep{Ghez2003, Habibi2017}. This is in strong contrast to the cluster dynamics on the scales of several parsecs as stated by \cite{Morris1993}, who predicts the migration timescales of stars from further away ($\sim$ 2-4 pc) towards Sgr~A* of about $10^{10}$yrs. The majority of the {brigthest} members of the S-cluster, the stars closest to Sgr~A*, have lifetimes of the order of $10^6$yrs \citep{Habibi2017}, raising the question about possible star formation channels in the dominant tidal field of the supermassive black hole. To date, the question of a possible {\it in situ} star formation scenario has not been solved but it is required to explain the timescale inconsistency. However, several detections of candidate Young Stellar Objects (YSOs) with notable bipolar outflows at a distance of 0.5 pc towards Sgr~A* have been reported by \cite{Yusef-Zadeh2017-ALMAVLA}, increasing the need for a suitable explanation. Recently, an even closer candidate YSO has been reported by \cite{peissker2023b}, following up on the proposed observation of massive protostars in the inner parsec \citep{Yusef-Zadeh2013}. This candidate YSO is located at a distance of about 0.1 pc towards Sgr~A* which is comparable to the projected distance of {the embedded cluster} IRS 13.\newline 
{At a distance of a few arcseconds further away from the S-cluster, the} authors of \cite{Genzel1996}, \cite{Maillard2004}, and especially \cite{Paumard2006} focused on a large collection of stars revealing a non-randomized stellar distribution that is arranged in the clockwise and counterclockwise disks (CWD and CCWD, respectively), which has been updated by \cite{Fellenberg2022}, who suggested a multi-component setup for the inner parsec. These authors investigated almost 3000 stars and found in total four disks including the CWD and CCWD. However, they did not symmetrically center their stellar sample around Sgr~A* due to observational constraints, which is similar to \cite{Paumard2006}, who contradicts generalizing statements about {\it all} stars in the inner parsec. Given the size of the influence radius of Sgr A*, which can be estimated as follows
\begin{align}
    r_{\rm inf}\,&\sim\,\frac{G{M_{\rm SgrA*}}}{\sigma_{\star}^2}\,\notag\\
    &\sim 1.7\,\left(\frac{M_{\rm SgrA*}}{4\times 10^6\,M_{\odot}} \right)\left(\frac{\sigma_{\star}}{100\,{\rm km\,s^{-1}}} \right)^{-2}\,{\rm pc}\,,
\end{align}
where $G$ defines the gravitational constant, $M_{\rm SgrA*}$ is the mass of Sgr~A*, and $\sigma_{\star}$ is the velocity dispersion of stars, sample sizes that describe the composition of the inner parsec should encompass a fraction of $r_{\rm inf}$, i.e $\sim 0.1$ pc \citep{Peissker2020c}. If not, the sample size could also be associated with a distinct population such as the S-cluster, for which \cite{Ali2020}, for example, suggested a non-randomized stellar distribution.
While subclusters, {such as the S-cluster,} with a length scale of a few mpc do not hinder the formation of large-scale distributions such as the CWD/CCWD, it is important to note that these different-sized components can be considered as the building blocks of the stellar members of the inner parsec that might have partially formed in dense clumps at a distance of a few parsec \citep{Dinh2021}. Hence, precise and continuous observations are needed to disentangle the different large- and small-scale components.\newline

{Taking a closer look at one of the largest coherent structures of the inner parsec, the} IRS 13 cluster is particularly interesting due to its embedded dust nature, although historically it was identified as a single source denoted IRS 2 \citep[][]{Rieke1978, Becklin1978, Smith1990}. With increasing resolving power, the science community realized a complex structure and called the region IRS 13E and IRS 13W to distinguish between the different components \citep{Simon1990}. Due to technological evolution and the introduction of active optics that allowed larger telescope dishes \citep{Wilson1987}, high spatial resolution observations with the SHARP camera showed a dense collection of stars that already implied a cluster arrangement \citep{Krabbe1995, Menten1997}. Taking into account a rather poor coverage of the interesting cluster IRS 13, we concentrated on the analysis of single stellar objects in \cite{peissker2023c}, hereafter Paper I. In this paper, we identified over 30 new stellar objects that we classified to a large extent as the YSO candidates that coexist with main-sequence stars, such as E1-E4 \citep{Maillard2004, muzic2008}. {In addition, we implemented YSO candidates that are know from the literature \citep{Eckart2004a}.} This suggested that IRS 13 underwent two distinct star-formation epochs which were triggered by individual processes. We furthermore found indications that the size of IRS 13 is much larger than it is anticipated by the literature. Consequently, we will treat IRS 13 as a dissolving cluster that is gravitationally captured by Sgr~A*. {Furthermore, a classification between IRS 13N and IRS 13E becomes unnecessary because all these two regions are part of IRS 13. Confusingly, IRS 13W is a single star located to the west of the cluster.}
In order to present a comprehensive analysis of IRS 13, we will estimate in this work the 3d distance of the cluster to Sgr~A*. In addition, we focus on the trajectory of the {dusty sources (DS, see Paper I)}, the greek labeled YSO candidates \citep{Eckart2004a}, and the E-stars\footnote{{The core of the IRS 13 cluster harbors the massive stars IRS 13E1-E7 \citep{Fritz2010}. Based on the nomenclature for the S-star in the S-cluster, we will use the term E-star to reflect on the membership to IRS 13.}} to present for the first time comprehensive Keplerian solutions for {about 50} orbits using a data baseline that covers almost two decades. Based on these orbits, we identify a substructure of the deeply embedded cluster that categorizes IRS 13 as a disk-like system following the work of \citet{Genzel1996}, \citet{Paumard2006}, \citet{Ali2020}, and \cite{Fellenberg2022}. Furthermore, we present Doppler-shifted ALMA data that show an ionized arrangement of H30$\alpha$ enveloping the projected position of IRS 13E3 \citep{Murchikova2019, tsuboi2019}. For the region corresponding to this source, we collect radio/mm, infrared, and X-ray data from the literature \citep{Tsuboi2017b,tsuboi2019,Zhu2020,peissker2023c} and find that the broad-band spectral properties can be described by the hot accretion flow model, which provides tentative support for the hypothetical presence of the intermediate-mass black hole (IMBH).

The paper is structured as follows. In Section~\ref{sec:analysis}, we describe used data sets and methods. The results concerning the kinematical structure are presented in Section~\ref{sec:results}. We discuss the overall length-scale of IRS 13, its origin and dynamics, its relation to other nuclear structures, as well as the putative association of IRS 13 E3 source with the IMBH in Section~\ref{sec:discuss}. We conclude with Section~\ref{sec:conclusion}.

\section{Data and Tools} 
\label{sec:analysis}
The analysis presented in this work is solely based on the data listed in Paper I. Hence, we refer the reader to this paper for the description of the data reduction and analysis in detail. All the data discussed in this work are equivalent to the archival observations listed in the Appendix of Paper I. {However, we indicate a short overview of the used data in Table \ref{tab:data} and refer the interested reader to Paper I for detailed information.}
\begin{table}
\setlength{\tabcolsep}{0.9pt}
\centering
\begin{tabular}{|cccc|}
\hline
Year & Telescope/Instrument & Band& Purpose\\
\hline
2002-2018 & VLT/NACO    & J       & Q, C\\
2002-2018 & VLT/NACO    & H       & Si, C, SED\\
2002-2018 & VLT/NACO    & K       & Si, C, SED, K\\
2002-2018 & VLT/NACO    & L       & Si, C, SED, K\\
2010      & VLT/NACO    & M       & Si, SED\\
2014      & VLT/SINFONI & H+K     & D   \\
2016      & ALMA        & 343 GHz & Si, SED\\
2017      & ALMA        & 232 GHz & Si, SED\\
\hline
\end{tabular}
\caption{{Summary of the data used in Paper I and this work. We list the related data baseline of the instruments and telescopes used, including their designated band. The individual {purposes} are indicated with the following abbreviations: (Q) Q-factor, (C) Color analysis, (Si) Source identification, (SED) Spectral Energy distribution, (K) Keplerian fit, (D) 3D distance.}}
\label{tab:data}
\end{table}

\subsection{Keplerian fit}

The orbital solutions for the sources presented in this work are derived based on the Keplerian approximation. We use the Sequential Least Squares Programming (SLSQP) method to estimate a Keplerian solution for the data points derived from a PSF-sized Gaussian fit. The SLSQP method calculates the reduced $\chi ^2$ and aims to minimize/optimize the distance between the calculated fit and the input positions \citep[see][]{Kraft1988}. Our reference is the position of Sgr~A* calculated from the orbit of the S-cluster star S2 \citep[][]{Do2019S2}, which initially was identified using a MASER triangulation by \cite{Plewa2015} and \cite{Parsa2017}. For the central gravitational potential of the fit, we use the compact mass of Sgr~A*, which is in the order of $4.02\times 10^6 M_{\odot}$ \citep{Peissker2022, eht2022}. Whenever suitable, we use both the K- and L-band NACO data to fit the trajectory of the dusty sources. We refer to Paper I for a detailed discussion about crowding problems \citep[also, see][]{peissker2020a, Peissker2020d}. {Furthermore, we adopt the Hill radius of 22 mpc estimated from the velocity dispersion of the IRS 13 cluster members. The enclosed mass is estimated to be 3.9$\times$10$^4\,M_{\odot}$ which is used as a justification for the use of the Keplerian approximation of the investigated sources. In Sec. \ref{sec:results}, we additionally estimate the 3d distance of the cluster to show, that the orbit of all sources can be solved with a Keplerian approximation.} 

\subsection{Linear transformation}

Due to the stellar density of the cluster \citep{Paumard2006} and the proper motion of the dusty sources of IRS 13 between 2002 and 2018, the data points suffer from deviations {imposed in every epoch that could} affect the derived trajectory. Despite the high cluster density, which results in increased confusion, a considerable contributor to deviations are distortions problems of the NACO detector \citep{Plewa2015, Plewa2018}. {While a constant deviation should be canceled over the entire time span, \cite{Plewa2018} showed a non-linear effect of the NACO instrument.} 
The authors of Plewa et al. estimate distortion effects for the S27 camera of NACO {(L-band)} in the range of $\sim 30\%$ of one pixel (27 mas). For bright stars inside the S-cluster, \cite{Gillessen2009}, \cite{Plewa2015}, and \cite{Parsa2017} estimate an uncertainty of about 1 mas for the positions. Due to the non-linearity distortion effect of the NACO imager, this effect can be as high as 10 mas depending of the location IRS 13 on the detector chip. {Considering the typical size of a NACO PSF with an FWHM of 4 pixels = 108 mas}, crowding effects might be of the same order or even higher \citep{Sabha2012, peissker2020a}. {Taking into account a putative source confusion, this scenario demands an individual detection of the objects in the first place at a distance that is at least half of the FWHM observed with NACO (54 mas).}
\begin{figure}
	\centering
	\includegraphics[width=.5\textwidth]{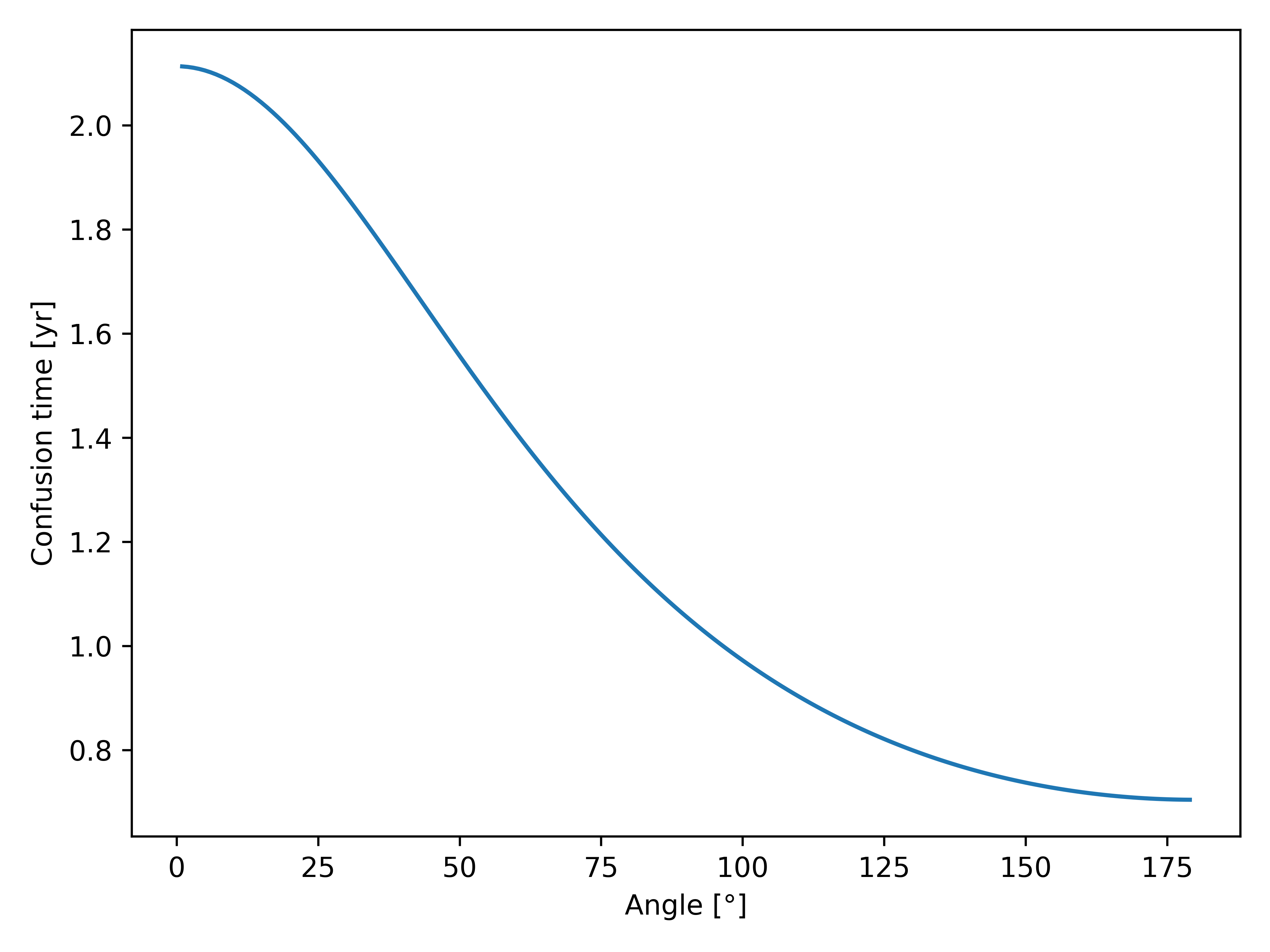}
	\caption{{Confusion time for two stars as a function of the angle between their trajectories. The angles $0^{\circ}$ and $180^{\circ}$ are excluded because they represent a parallel trajectory without blending. The initial distance of both stars is 54 mas.}}
\label{fig:source_confusion}
\end{figure}
If we now assume two stars with $\rm v_{\rm star\,1}=100\,km/s$ and $\rm v_{\rm star\,2}=200\,km/s$, we can calculate the confusion time of both objects as a function of the angle between their individual trajectories. {Considering the range of observed proper motions (Fig. 4, Paper I) of a few km/s up to 500 km/s, the proposed setting is justified.}
Excluding an angle of $0^{\circ}$ and $180^{\circ}$ which means that both stars move parallel to each other, we find a confusion time of about 1-2 years (Fig. \ref{fig:source_confusion}) assuming an initial distance of 54 mas. Although different distances and velocities do impact the confusion time, i.e. where both stars cannot be detected individually, the core information is a timescale of only 1-2 years. Therefore, the confusion is the main uncertainty contributor on short timescales ($\rm \sim\,2\,yr$) whereas the distortion impacts the analysis over the complete data baseline.\newline
{Because this distortion mainly impacts the S27 camera of NACO,} we apply a linear transformation to the investigated sources observed in the L-band.For that, we determine the standard deviation $\sigma_{pos}$ for all positional aberrations from a linear trajectory. Due to the distance of the dusty sources to Sgr~A* and the therefore expected {long} orbits compared to the S-stars {(a few thousand years versus a few ten years)}, the assumption of a local linear trajectory for the analyzed data baseline is justified. 
In the final step, we apply $\sigma_{pos}$ to the Gaussian estimated positions of the dusty sources. In Table \ref{tab:LinTrans}, we list the applied averaged deviations from all the analyzed objects to the derived L-band data points.
\begin{table}
\centering
\begin{tabular}{|ccccc|}
\hline
Date & R.A. & $\Delta$ R.A. & Dec & $\Delta$ Dec \\
 & (mas) & (mas) & (mas) & (mas) \\
\hline
2002.660 & 4.13  & 2.66 & -13.56 & 2.47 \\
2003.353 & 9.06  & 1.03 & -9.52  & 0.96 \\
2004.314 & 2.03  & 0.95 & 14.12  & 0.68 \\
2005.364 & -8.69 & 1.67 & -14.14 & 0.66 \\
2006.405 & -8.64 & 1.07 & -3.99  & 0.89 \\
2007.247 & 2.26  & 2.09 & -3.45  & 1.18 \\
2007.367 & 0.63  & 0.73 & 4.51   & 0.98 \\
2007.370 & 0.04  & 2.25 & -6.94  & 2.42 \\
2007.373 & 3.42  & 1.45 & -6.96  & 3.27 \\
2007.375 & 1.02  & 1.61 & -7.75  & 2.21 \\
2007.386 & -0.25 & 1.48 & -6.27  & 1.14 \\
2007.389 & 5.10  & 2.06 & -10.65 & 1.48 \\
2008.399 & 0.18  & 1.09 & 17.47  & 1.04 \\
2008.410 & -3.16 & 0.76 & 12.65  & 0.50 \\
2008.413 & -2.67 & 2.98 & 12.94  & 2.28 \\
2008.418 & -0.79 & 1.60 & 10.82  & 1.70 \\
2008.421 & 0.75  & 1.35 & 19.25  & 1.07 \\
2011.395 & -6.91 & 3.13 & 14.68  & 1.34 \\
2012.372 & -8.12 & 0.94 & 14.69  & 1.15 \\
2013.351 & -5.79 & 1.81 & 25.39  & 1.49 \\
2016.224 & 2.02  & 1.62 & -8.62  & 1.52 \\
2018.304 & 2.52  & 1.67 & -24.08 & 1.33 \\
2018.310 & 8.50  & 1.20 & -7.41  & 1.36 \\ 
\hline
\end{tabular}
\caption{Averaged deviations for positional data points that are related to the (dusty) sources of IRS 13. {As mentioned in the text, the uncertainty is less than one pixel and in agreement with the analysis of \cite{Plewa2015} and \cite{Parsa2017}.}}
\label{tab:LinTrans}
\end{table}
Since the spatial pixel scale of the NACO L-band data is $~$27 mas, our estimated deviations listed in Table \ref{tab:LinTrans} are in the subpixel domain and, therefore, reasonable. The derived aberrations are further in agreement with the results of \cite{Plewa2015}, although the authors used SiO Maser high-precision measurements to estimate the location of Sgr~A*. However, for deviation values larger than one pixel, we apply a quality control and discard the data due to poor quality. We emphasize that the linear transformation is also valid for the Keplerian solution presented in this work since the curvature of the orbit is diminished.


\subsection{Modelling cluster dynamics}

To model the dynamics of the cluster, we apply the Astrophysical Multipurpose Software Environment \citep[AMUSE, see][]{Portegies-Wart2009, Portegies-Wart2013, Pelupessy2013, Portegies-Wart2018}, {which} is a PYTHON-based project that can be used, among other applications, for N-body simulations. We use the AMUSE-PhiGRAPE \citep{Harfst2007} module to simulate a cluster of a certain mass $M_{\rm cluster}$ in the influence sphere of Sgr~A* with $M_{\rm SgrA^*}$. This module uses the King model \citep{King1966} that requires three parameters to describe the simulated cluster, namely the core radius $r_{\rm c}$, the related brightness $L_{\rm c}$, and the distribution $W_0$ of the stars inside the cluster. Whereas the luminosity and the radius are directly estimated from the observations, the King parameter describes the density of the cluster and may be challenging to derive. However, the overall shape of the investigated cluster may indicate a specific range of possible values for $W_0$. For the simulation, a numerical value of $W_0\,=\,0$ results in a dissolved cluster because the stars are not bound to the system. Consequently, higher numerical values represent an increasing stellar density \citep[see, e.g.,][]{Chernoff1987, Joshi2001} and hence a gravitationally bound system. We refer to \cite{Portegies-Wart2018} for a detailed description of the code.

\section{Results}
\label{sec:results}
In this section, we present the findings of the analysis of IRS 13. We present a NIR spectrum observed with SINFONI and fit a Keplerian orbit to the positions of the DS sources listed in Paper I. Furthermore, we investigate the resulting orbital elements to check for systematic trends.

\subsection{3d distance of IRS 13 to Sgr~A*}
\label{sec:3d_distance}

\begin{figure}
	\centering
	\includegraphics[width=.5\textwidth]{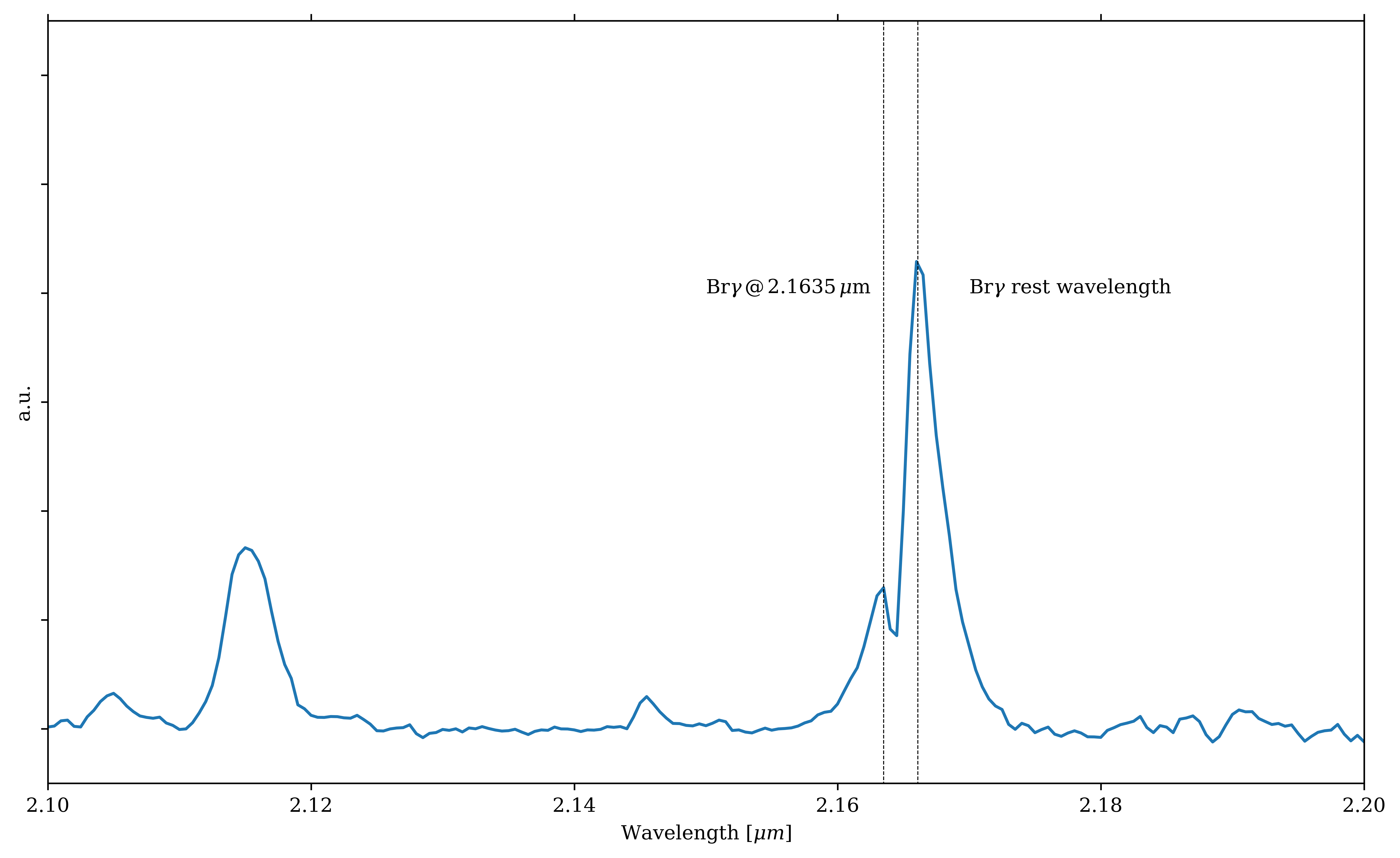}
	\caption{Near-infrared spectrum of the IRS 13 cluster observed with SINFONI/VLT in 2014. The spectrum covers the wavelength range between 2.1-2.2$\mu$m and shows the prominent Br$\gamma$ rest wavelength and a related Doppler-shifted emission line at $\sim$2.1635$\mu$m. The aperture used to create the indicated spectrum has a diameter of 0.5" which covers the complete core region and partially the tip of IRS 13. The total width of the Br$\gamma$ line is almost 2000 km/s between 2.159-2.173$\mu$m.}
\label{fig:spec_irs13}
\end{figure}

{To perform a Keplerian analysis, we have to ensure that the investigated cluster members are gravitational bound to Sgr~A*.} According to \cite{Tsuboi2020b}, the 3d distance of IRS 13 is $\geq$ 0.4 pc away from Sgr~A* and therefore outside the Bondi sphere with the radius of $r_{\rm Bondi}$,
\begin{align}
    r_{\rm Bondi} &\sim \frac{2GM_{\rm SgrA*}}{c_{\rm s}^2}\,\notag\\
    &\sim 0.21 \left(\frac{M_{\rm SgrA*}}{4\times 10^6\,M_{\odot}} \right)\left(\frac{T}{10^7\,{\rm K}} \right)^{-1} \,{\rm pc}\,,
\end{align}
where $c_{\rm s}$ is the sound speed corresponding to the temperature $T$ of the hot, X-ray emitting plasma.
With the estimate of Tsuboi et al., the stellar cluster members should still, however, be gravitationally bound to Sgr~A* since it is inside the influence radius, and hence a possible interaction with the SMBH is not excluded.
Due to the large projected distance of about 0.14 pc to Sgr~A*, we assume a close-to-circular bound trajectory of IRS 13 around the central black hole. We furthermore place a circular-annular aperture (radius1: 0.25", radius2: 0.26")\footnote{The aperture consists of a circular area with a radius of 0.25". Around this circular area, a ring with a radius of 0.01" is concatenated. The total radius of the circular annular aperture is 0.26".} on the core region of IRS 13 and extract the spectrum shown in Fig. \ref{fig:spec_irs13} where we indicate the Br$\gamma$ rest wavelength at 2.1661$\mu$m, the related Doppler-shifted emission line at 2.1635$\mu$m. {Since the electron temperature of Br$\gamma$ is in the range of 10$^4$K \citep{gravitycollaboration2023}, we assume a connection between the radial velocity of the members of the cluster and the gas.}
With the resulting LOS velocity of {v$_{\rm LOS}\,=\-,346.24\pm27.5\,$km/s} and a proper motion of v$_{\rm prop}\,=\,128.86\pm\,0.14$km/s\footnote{{This proper motion is based on the velocity dispersion of all cluster members. For the conversion from mas/year to km/s, we assumed a distance of 8 kpc. Please see Paper I for further details.}}, we use
\begin{equation}
    r_{\rm IRS13}\,=\,GM_{\rm SgrA*}/(\sqrt{v_{\rm LOS}^2+v_{\rm prop}^2})^2
    \label{eq:distance}
\end{equation}
and get {r$_{\rm IRS13}\,=\,0.12\,$pc for a bound trajectory of IRS 13 and r$_{\rm IRS13}\,=\,0.25\,$pc for a parabolic orbit \citep{Peissker2020c}. For the gravitational bound case, IRS 13 with its stellar members and its gas and dust depot is within both the influence radius and the Bondi sphere, and thus orbits Sgr~A*. If IRS 13 exhibits a parabolic orbit, the cluster is still in the influence sphere of Sgr~A* but outside its Bondi sphere. We would like to add that the assumed link between the gas and the cluster members may be premature. However, the listed proper motion and radial velocity in \cite{Paumard2006}, \cite{2009ApJ...697.1741B}, and \cite{Jia2023} reveal with Eq. \ref{eq:distance} a similar 3d distance for the E-stars\footnote{IRS 13E1: 0.35-0.39 pc, IRS 13E2: 0.17 pc, IRS 13E3: 0.37-0.43 pc, IRS 13E4: 0.15 pc} as derived in this section. Using the velocity dispersion of the cluster members of v$_{\rm prop}\,=\,128.86\pm\,0.14$km/s, we derive an upper limit for the 3d distance of about 1 pc with Eq. \ref{eq:distance} by setting the radial velocity to 0 km/s.}\newline
{Since we estimate the enclosed mass in Paper I to be 22 mpc, we find that the cluster is inside the influence sphere of Sgr~A* independent of the exact LOS velocity. Therefore, we can safely assume a Keplerian motion of all cluster members}, which will be presented in the next section. Furthermore, we exclude a possible interference of the IRS 13 region and the mini-cavity due to the difference of the LOS velocity of about $\Delta v_{LOS}\,=\,200-300\,$km/s \citep{Lutz1993, Ciurlo2016}. However, there may be a more complex relation between IRS 13 and the mini-cavity involving the IRS 16 cluster (please consider the Appendix for further discussion), which will be explored in a forthcoming article.

\subsection{Orbits}

\begin{figure}
	\centering
	\includegraphics[width=.5\textwidth]{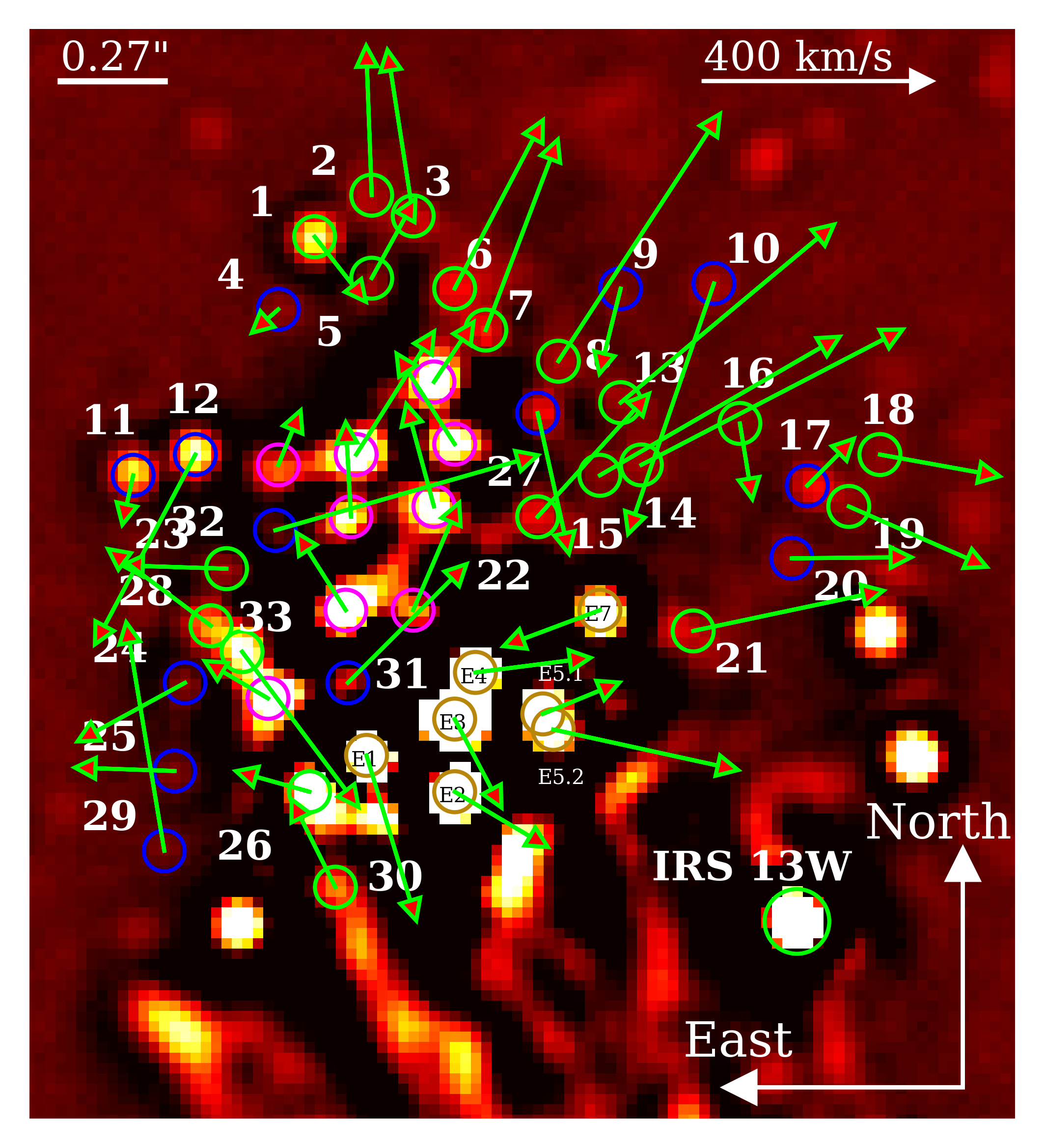}
	\caption{Finding chart of the investigated cluster members of IRS 13. This finding chart is observed in the L-band with NACO in 2004 and shows all sources including their proper motion vector (see Paper I). {Every velocity is multiplied by a factor of 0.05 to improve the representation of the arrows for the reader. The length of velocity arrow at the top of the image scales with the size of the proper motion arrows for each source.} The circles at the position of the individual sources indicate their nature and age. While all lime (DS objects) and pink (greek-named objects) colored sources are young objects with an estimated age of $\rm <\,1\,Myr$, the brown (E-stars) and blue (DS objects) circles show main-sequence stars with an age of $\rm >\,1\,Myr$.}
\label{fig:irs13_finding_chart}
\end{figure}

\begin{figure*}%
	\centering
	\includegraphics[width=1.\textwidth]{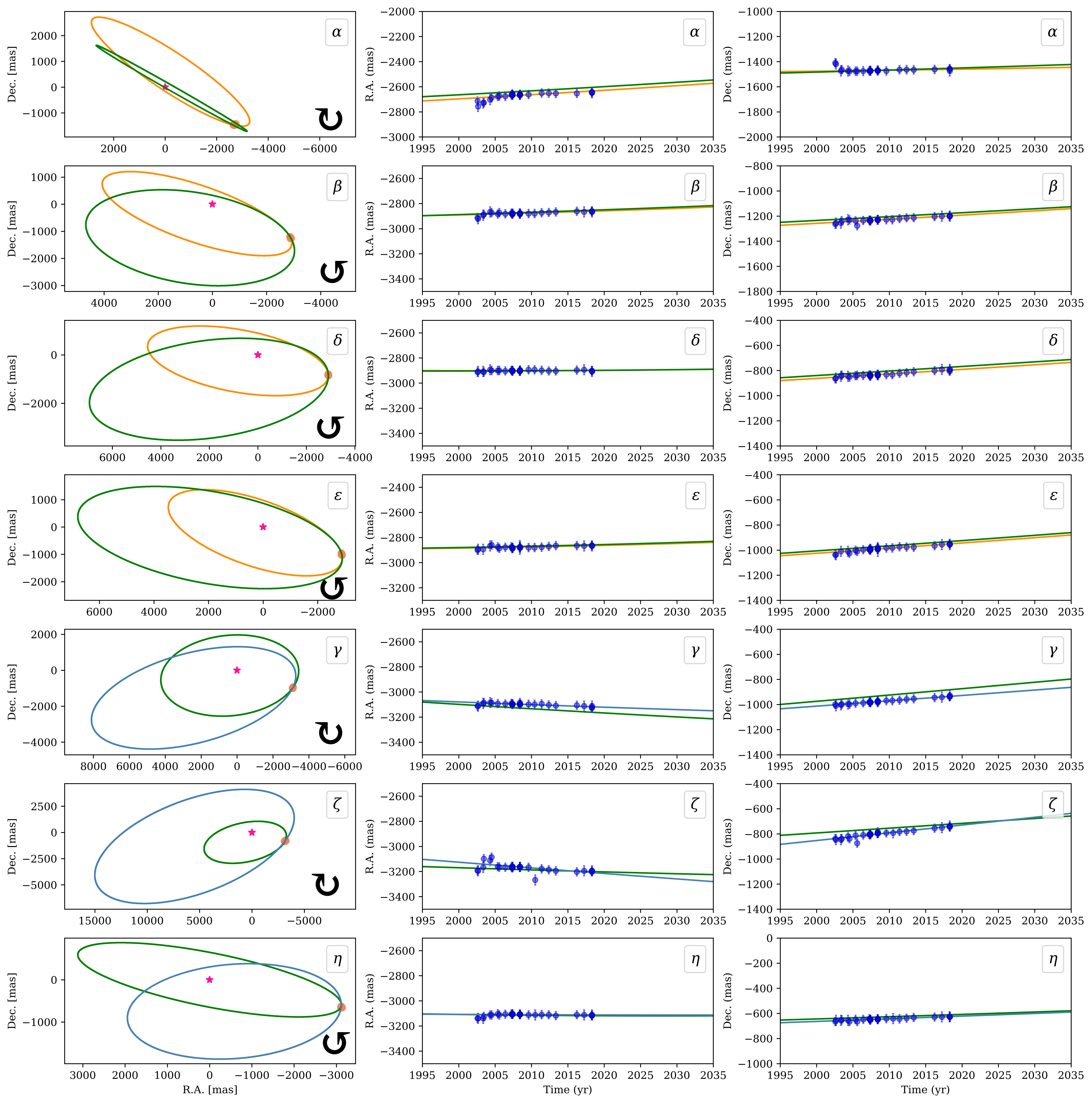}
	\caption{Best-fit Keplerian orbits for the dusty sources analyzed in this work. We include $K$- and $L$-band data observed with NACO between 2002 and 2018 to estimate the best-fit solution to describe the trajectory of the dusty sources. The left column shows the projected orbit on the sky where we mark the position of Sgr~A* with a pink-colored asterisk. The middle and right columns exhibit the R.A. and DEC plots as a function of time, respectively. In addition, we include blue dots representing the data with their respective astrometric uncertainty. The black arrows indicate the direction of the orbital trajectory of the related dusty object. To acknowledge the challenging orbital coverage of the data points, we implement two different solutions, both describing the trajectory of the sources. The resulting orbital elements of these fits are listed in Table \ref{tab:orbit_table}. These Keplerian parameters suggest a grouped arrangement noticeable in the LOAN. We acknowledge the membership of the related source by using orange (N2) and blue (N1). Both solutions are the best-fit result, where the green-colored fit always indicates a solution for the Keplerian approximation with decreased significance. The statistically poor green orbit is randomly picked from a set that represents a reduced significance for the Keplerian approximation. The R.A. and DEC plots for all the sources investigated in Paper I and this work, including the complete orbital positions, are publicly available and released along with this manuscript. Please see Appendix \ref{sec:sup_mat_app} for a description of the supplementary material.}
\label{fig:orbits_1}
\end{figure*}

For the Keplerian orbits, we select all E-stars {(seven sources)} and the DS sources {(33 sources)} including the Greek-named objects $\alpha$, $\beta$, $\gamma$, $\delta$, $\epsilon$, $\gamma$, $\zeta$, and $\eta$ {(seven sources)} of IRS 13 that are analyzed in Paper I Fig. \ref{fig:irs13_finding_chart}. Initially, these sources were selected due to their colors {(L-band $<$ 16.5 mag)}, proper motions {($>$ 25 km/s)}, and projected position {(2.5 as $<$ distance $<$ 4.5 as)}. 
We chose the K- and L-band data for the orbital analysis of these sources because the data covers almost two decades of continuous observations between 2002 and 2018. For the fits, we assume that all the sources are gravitationally bound to Sgr~A* (see Section \ref{sec:3d_distance}) and adopt the mass of 4.02$\,\times\,10^{6}$M$_{\odot}$ for the supermassive black hole \citep[][]{eht2022, Peissker2022}. {Due to the resolution of the SINFONI IFU data with a spatial pixel scale of 0.1" and a distance of the individual sources less than 0.1", we are limited to the NACO K- and L-band data.\newline}
Due to the low proper motion of the sources (Paper I), the orbital data coverage is less than 2$\%$ for some of the cluster members. In addition to crowding effects, $\iota$ and $\vartheta$ are blended with nearby sources that hinder a confusion-free analysis. Hence, we exclude both sources from the Keplerian analysis. In contrast, we find orbital Keplerian solutions for all {other} dusty objects, which are listed in Table \ref{tab:orbit_table} and Table \ref{tab:orbit_table_all_ds}. In Fig. \ref{fig:orbits_1}, we highlight the results of the orbital analysis, where we reflect the data coverage mentioned above by incorporating two different Keplerian solutions that represent the same data sets. As indicated in Table \ref{tab:orbit_table}, we use the reduced $\chi^2$ parameter as a qualitative parameter to identify the best-fit Keplerian solution to the data set. In comparison, we selected another Keplerian solution with an increased $\chi^2$ indicating a statistically poor solution for the given data set. As shown in Fig. \ref{fig:orbits_1}, both solutions simultaneously demonstrate the limitations of this analysis which can be resolved by incorporating the so far missing LOS velocities for the individual DS sources using, for example, ERIS (VLT) or MIRI (JWST).
In Table \ref{tab:orbit_table}, we list all related orbital parameters for the dusty sources shown in Fig. \ref{fig:orbits_1}.
\begin{table*}[]
\setlength{\tabcolsep}{0.9pt}
\centering
\begin{tabular}{|cccccccc|}
\toprule
Source     & $a$ (mpc)&	$e$  &	$i$($^o$)&$\omega$($^o$)&	$\Omega$($^o$)&	$t_{\rm closest}$(yr) & $\chi^2$\\
\hline \hline
$\alpha$ (orange) & 255.31$\pm$6.20 & 0.8201$\pm$0.0064 & 96.78$\pm$3.13 & 88.30$\pm$1.96 & 236.84$\pm$1.60 & 2312.33$\pm$4.16 & 2.65\\
$\alpha$ (green)   & 242.91$\pm$6.20 & 0.8330$\pm$0.0064 & 90.52$\pm$3.13 & 92.24$\pm$1.96 & 240.06$\pm$1.60 & 2304.00$\pm$4.16 & 4.01\\
\hline
$\beta$ (orange)  & 169.42$\pm$47.40 & 0.5032$\pm$0.1550 & 75.05$\pm$0.28 & 78.49$\pm$2.86 & 249.80$\pm$3.72 & 2333.59$\pm$2.20 &  1.15\\
$\beta$ (green)    & 264.23$\pm$47.40 & 0.8130$\pm$0.1550 & 74.48$\pm$0.28 & 84.22$\pm$2.86 & 257.25$\pm$3.72 & 2338.00$\pm$2.20 & 13.92\\
\hline
$\delta$ (orange)  & 159.04$\pm$49.17 & 0.3648$\pm$0.1755 & 70.47$\pm$0.57 & 58.55$\pm$2.52 & 257.83$\pm$5.73 & 2302.77$\pm$27.00 & 0.51 \\
$\delta$ (green)    & 257.39$\pm$49.17 & 0.7160$\pm$0.1755 & 69.32$\pm$0.57 & 63.59$\pm$2.52 & 269.29$\pm$5.73 & 2356.79$\pm$27.00 & 10.55\\
\hline
$\epsilon$ (orange) & 137.64$\pm$42.40 & 0.2463$\pm$0.1626 & 69.24$\pm$0.61 & 75.96$\pm$10.19 & 250.77$\pm$2.37 & 2337.51$\pm$37.62 & 0.80\\
$\epsilon$ (green)   & 222.45$\pm$42.40 & 0.5716$\pm$0.1626 & 70.47$\pm$0.61 & 55.57$\pm$10.19 & 255.53$\pm$2.37 & 2262.26$\pm$37.62 & 3.68\\
\hline
$\gamma$ (green)  & 155.03$\pm$50.07 & 0.1582$\pm$0.2108 & 53.85$\pm$4.96 & 224.59$\pm$0.10 & 94.53$\pm$2.06 & 2501.31$\pm$2.56 & 1.11 \\
$\gamma$ (blue)    & 255.18$\pm$50.07 & 0.5798$\pm$0.2108 & 63.77$\pm$4.96 & 224.80$\pm$0.10 & 98.66$\pm$2.06 & 2496.18$\pm$2.56 & 0.43 \\
\hline
$\zeta$ (green)   & 177.16$\pm$115.30 & 0.4688$\pm$0.0551 & 64.17$\pm$3.06 & 247.51$\pm$43.74 &  97.97$\pm$8.28 & 2549.58$\pm$128.71 & 9.04\\
$\zeta$ (blue)     & 407.77$\pm$115.30 & 0.5791$\pm$0.0551 & 58.05$\pm$3.06 & 160.02$\pm$43.74 & 114.53$\pm$8.28 & 2292.16$\pm$128.71 & 6.06 \\
\hline
$\eta$ (green) & 127.10$\pm$10.10 &0.00001$\pm$0.3686 & 77.92$\pm$2.95 & 258.40$\pm$10.72 & 78.49$\pm$10.07 & 2480.69$\pm$27.26 & 6.44\\
$\eta$ (blue)   & 147.30$\pm$10.10 & 0.7374$\pm$0.3686 & 72.01$\pm$2.95 & 279.85$\pm$10.72 & 98.63$\pm$10.07 & 2535.22$\pm$27.26 & 0.68 \\

\hline
\end{tabular}
\caption{Best-fit orbital elements for the Keplerian solutions shown in Fig. \ref{fig:orbits_1}. We distinguish between two independent orbital solutions expressed by orange and blue (Fig. \ref{fig:orbits_1}) to emphasize the orbital coverage of the K- and L-band data observed between 2002 and 2018. The standard deviation of the derived orbits denotes the indicated uncertainties. We register a pattern for the numerical values of the argument of periapsis $\omega$ and longitude of the ascending node $\Omega$ describing the projected on-sky distribution of the dusty sources $\gamma$, $\zeta$, and $\eta$ compared to $\alpha$, $\beta$, $\delta$, and $\epsilon$. In Sec. \ref{sec:discuss}, we will explore the possibility for a pattern that would suggest a non-randomized distribution of the analyzed sources. In the last column, we indicate the minimized $\chi^2$ estimate for the related Keplerian solution, where lower values represents a better fit.
{Please note that the minimized $\chi^2$ is not a global parameter since the distribution of the astrometric measurements for the individual sources is different. Since the uncertainties of the standard deviation and the minimized $\chi^2$ analysis may over- or underestimate the error range, we will employ MCMC simulations to cover possible fluctuations of the Keplerian approximation (see Appendix \ref{sec:mcmc-app}). Here, the minimized $\chi^2$ value of the related source of interest is solely used as a qualitative measure to pick the best-fit orbit. Worse minimized $\chi^2$ values do not necessarily have to represent a bad solution (see $\eta$ in Fig. \ref{fig:orbits_1}).}}
\label{tab:orbit_table}
\end{table*}
Based on the derived orbital solutions, we find that some sources, such as $\alpha$, $\beta$, and $\eta$, are close to the projected apoapsis (Fig. \ref{fig:orbits_1}). For all the sources, we derive an almost constant acceleration between 2002 and 2018 that results in the proper motion listed in Paper I. We emphasize that a linear fit applied to the data points represents the trajectory of the DS sources as well. Due to the arguments indicated in Sec. \ref{sec:3d_distance}, we prefer Keplerian solutions for the cluster members, which is consistent with other large-scale analyses, such as the one presented in \cite{Fellenberg2022}.
In addition, we provide the minimized reduced $\chi^2$ that is defined as the weighted sum of the squared deviations related to the two orbital solutions shown in Fig. \ref{fig:orbits_1}. These deviations are calculated by comparing the theoretical and observed data points. Therefore, a minimized value reflects a higher probability such that the estimated Kepler fit reflects the observed positions of the object. {We discuss the limitations of this approach in Sec. \ref{sec:discuss}.}
For the orbital elements listed in Table \ref{tab:orbit_table_all_ds}, we find a wide spread of suitable parameters representing the Keplerian motion of the DS sources.
\begin{table*}[]
\centering
\begin{tabular}{|ccccccccc|}
\toprule
Source     & $a$ (mpc)&	$e$  &	$i$($^o$)&$\omega$($^o$)&	$\Omega$($^o$)&	$t_{\rm closest}$(yr) & $\chi^2$ & Group\\
\hline \hline
DS1  & 480.98 $\pm$  3.68 & 0.7602 $\pm$ 0.0110 & 100.47 $\pm$  8.59 & 173.13 $\pm$ 14.89 &  81.75 $\pm$  4.52 & 2031.06 $\pm$  3.89 & 0.20 & N1 \\
DS2  & 447.15 $\pm$  2.87 & 0.7360 $\pm$ 0.0100 &  61.35 $\pm$  5.21 & 186.10 $\pm$  5.32 &  87.21 $\pm$  2.11 & 2032.11 $\pm$  5.23 & 1.98 & N1 \\
DS3  & 486.14 $\pm$  3.30 & 0.7495 $\pm$ 0.0230 &  57.29 $\pm$  4.06 & 195.26 $\pm$  5.44 &  84.54 $\pm$  5.50 & 2060.74 $\pm$ 21.89 & 1.47 & N1 \\
DS4  & 433.65 $\pm$  2.40 & 0.7289 $\pm$ 0.0290 &  90.52 $\pm$  7.90 & 194.23 $\pm$ 10.77 &  80.01 $\pm$  6.35 & 1976.00 $\pm$  7.07 & 3.25 & N1 \\
DS5  & 456.68 $\pm$  1.95 & 0.7397 $\pm$ 0.0220 &  74.08 $\pm$  6.93 & 182.29 $\pm$  6.70 &  83.80 $\pm$  6.64 & 2038.10 $\pm$  7.99 & 1.23 & N1 \\
DS6  & 490.00 $\pm$  2.13 & 0.7074 $\pm$ 0.0230 &  51.79 $\pm$ 10.08 & 155.40 $\pm$  2.80 & 106.36 $\pm$  2.92 & 2062.76 $\pm$  6.45 & 2.00 & N1 \\
DS7  & 490.01 $\pm$  1.77 & 0.7125 $\pm$ 0.0130 &  45.83 $\pm$  7.16 & 156.01 $\pm$  4.29 & 104.95 $\pm$  2.57 & 2050.12 $\pm$ 12.67 & 2.07 & N1 \\
DS8  & 490.83 $\pm$  2.76 & 0.6624 $\pm$ 0.0350 &  50.97 $\pm$ 22.28 & 143.23 $\pm$  6.76 & 111.15 $\pm$  3.38 & 2051.29 $\pm$ 17.53 & 2.21 & N1 \\
DS9  & 341.15 $\pm$  2.04 & 0.6434 $\pm$ 0.0120 & 120.32 $\pm$  3.72 & 143.23 $\pm$ 10.37 & 104.83 $\pm$  3.78 & 1590.00 $\pm$ 14.16 & 1.51 & N1 \\
DS10 & 539.91 $\pm$  2.23 & 0.7094 $\pm$ 0.0290 & 157.13 $\pm$  5.21 & 233.56 $\pm$  6.41 & 120.32 $\pm$  6.35 & 2097.13 $\pm$  6.18 & 1.53 & N1 \\
DS11 & 438.65 $\pm$  3.96 & 0.8502 $\pm$ 0.0370 &  97.48 $\pm$  7.96 & 126.05 $\pm$  8.93 & 75.19  $\pm$  9.68 & 1750.41 $\pm$ 11.14 & 0.23 & N1 \\
DS12 & 394.04 $\pm$  2.26 & 0.7009 $\pm$ 0.0350 & 127.10 $\pm$  5.55 & 183.34 $\pm$ 12.03 & 117.57 $\pm$  6.18 & 1739.27 $\pm$ 16.51 & 0.85 & N1 \\
DS13 & 597.12 $\pm$  1.72 & 0.7426 $\pm$ 0.0250 &  57.29 $\pm$  6.70 &  90.78 $\pm$  8.02 & 105.00 $\pm$  6.76 & 1693.43 $\pm$  7.20 & 4.19 & N1 \\
DS14 & 450.00 $\pm$  2.89 & 0.8248 $\pm$ 0.0300 &  34.37 $\pm$  7.90 &  46.72 $\pm$  4.69 & 108.23 $\pm$  7.10 & 1647.31 $\pm$ 12.20 & 3.25 & N1 \\
DS15 & 443.32 $\pm$  2.76 & 0.7900 $\pm$ 0.0130 &  40.10 $\pm$ 15.64 &  58.02 $\pm$ 15.98 & 104.67 $\pm$  3.03 & 1650.11 $\pm$  8.38 & 3.61 & N1 \\
DS16 & 999.27 $\pm$  3.76 & 0.8436 $\pm$ 0.0110 & 103.13 $\pm$  4.23 & 195.07 $\pm$  4.69 &  72.68 $\pm$  2.29 & 2263.13 $\pm$ 11.45 & 1.46 & N1 \\
DS17 & 600.61 $\pm$  3.10 & 0.8175 $\pm$ 0.0150 &  78.00 $\pm$  7.79 & 121.47 $\pm$  7.10 &  71.04 $\pm$ 10.42 & 1550.00 $\pm$  6.84 & 0.89 & N1 \\
DS18 & 617.14 $\pm$  3.62 & 0.8058 $\pm$ 0.0380 &  88.96 $\pm$ 13.29 & 110.13 $\pm$ 13.80 &  78.56 $\pm$ 11.00 & 1620.61 $\pm$ 18.42 & 1.96 & N1 \\
DS19 & 619.11 $\pm$  3.29 & 0.8449 $\pm$ 0.0190 &  98.83 $\pm$ 15.58 &  94.97 $\pm$ 43.20 &  77.55 $\pm$ 10.31 & 1614.57 $\pm$  6.24 & 1.68 & N1 \\
DS20 & 596.84 $\pm$  4.33 & 0.8107 $\pm$ 0.0480 &  81.38 $\pm$ 15.81 & 111.81 $\pm$ 22.11 &  73.41 $\pm$  7.84 & 1638.66 $\pm$ 19.44 & 1.53 & N1 \\
DS21 & 617.91 $\pm$  7.42 & 0.7957 $\pm$ 0.0380 &  63.58 $\pm$  9.22 &  96.49 $\pm$ 11.74 &  79.75 $\pm$  3.66 & 1661.65 $\pm$ 11.75 & 1.73 & N1 \\
DS22 & 507.61 $\pm$  2.12 & 0.7932 $\pm$ 0.0280 &  51.74 $\pm$  9.11 & 132.55 $\pm$ 13.63 &  58.66 $\pm$  6.07 & 1705.20 $\pm$  8.31 & 0.89 & N1 \\
DS23 & 539.57 $\pm$  7.54 & 0.7874 $\pm$ 0.0370 & 100.72 $\pm$ 16.61 & 160.42 $\pm$ 73.91 &  75.56 $\pm$  9.11 & 1715.05 $\pm$ 22.25 & 1.75 & N1 \\
DS24 & 517.57 $\pm$  8.17 & 0.7690 $\pm$ 0.0690 & 108.46 $\pm$  9.91 & 162.36 $\pm$ 24.46 &  78.86 $\pm$  8.53 & 1677.17 $\pm$ 10.62 & 1.90 & N1 \\
DS25 & 509.04 $\pm$  5.15 & 0.7463 $\pm$ 0.0330 &  99.01 $\pm$ 19.53 & 163.70 $\pm$ 19.25 &  64.23 $\pm$  9.45 & 1660.47 $\pm$ 17.10 & 1.31 & N1 \\
DS26 & 530.78 $\pm$ 12.41 & 0.7453 $\pm$ 0.0720 &  93.18 $\pm$ 14.66 & 162.71 $\pm$ 42.85 &  60.30 $\pm$  9.62 & 1705.01 $\pm$ 10.89 & 0.62 & N1 \\
DS27 & 214.93 $\pm$  0.87 & 0.6078 $\pm$ 0.0080 & 133.56 $\pm$  3.89 & 116.70 $\pm$  4.92 & 107.27 $\pm$  1.94 & 1587.54 $\pm$  4.20 & 14.53& N1 \\
DS28 & 372.41 $\pm$  1.72 & 0.8391 $\pm$ 0.0170 &  77.96 $\pm$  5.50 &  74.48 $\pm$  8.99 & 247.35 $\pm$  1.60 & 2242.89 $\pm$ 10.20 & 2.34 & N2 \\
DS29 & 616.23 $\pm$  1.89 & 0.6557 $\pm$ 0.0190 &  68.18 $\pm$  3.03 & 177.77 $\pm$ 10.14 & 120.32 $\pm$  6.35 & 3076.97 $\pm$  7.89 & 4.19 & N1 \\
DS30 & 489.02 $\pm$  1.48 & 0.7063 $\pm$ 0.0190 &  99.75 $\pm$  7.56 & 179.47 $\pm$  6.07 &  51.73 $\pm$  2.29 & 2104.85 $\pm$ 10.01 & 0.18 & N1 \\
DS31 & 146.66 $\pm$  1.93 & 0.9900 $\pm$ 0.0220 & 107.80 $\pm$  4.98 & 120.67 $\pm$  4.46 & 220.60 $\pm$  2.97 & 2430.91 $\pm$  8.13 & 0.15 & N2 \\
DS32 & 437.06 $\pm$  2.04 & 0.7597 $\pm$ 0.0080 & 110.71 $\pm$  3.38 &  27.07 $\pm$  5.04 & 244.57 $\pm$  1.37 & 2108.44 $\pm$ 14.84 & 0.16 & N2 \\
DS33 & 433.16 $\pm$  2.48 & 0.7411 $\pm$ 0.0160 &  95.96 $\pm$  5.32 &  24.43 $\pm$  8.19 & 242.04 $\pm$  0.97 & 2107.78 $\pm$ 11.92 & 0.08 & N2 \\
\hline
\end{tabular}
\caption{Orbital elements of the DS sources. The uncertainties are adapted from MCMC simulations. In the last column of the table, we indicate the associated group of the related DS source. Please consult the text for details.}
\label{tab:orbit_table_all_ds}
\end{table*}
In addition, we analyze the trajectory of the E-stars and determine the Keplerian solutions listed in Table \ref{tab:orbit_table_estars}.
\begin{table*}
\centering
\begin{tabular}{|ccccccccc|}
\toprule
Source     & $a$ (mpc)&	$e$  &	$i$($^o$)&$\omega$($^o$)&	$\Omega$($^o$)&	$t_{\rm closest}$(yr) & $\chi^2$ & Group\\
\hline \hline
E1   & 510.61 $\pm$ 0.97 & 0.5018 $\pm$ 0.0090 & 108.94 $\pm$  1.94 &  89.38 $\pm$  3.72 &  25.46 $\pm$ 4.35 & 1652.25 $\pm$  4.98 & 0.02 & N1\\
E2   & 659.80 $\pm$ 2.63 & 0.0592 $\pm$ 0.0110 &  90.54 $\pm$ 18.79 &  97.11 $\pm$ 23.66 &  59.03 $\pm$ 7.21 & 1612.85 $\pm$ 16.35 & 0.36 & N1\\
E3   & 612.54 $\pm$ 1.71 & 0.0010 $\pm$ 0.0150 &  97.75 $\pm$  3.43 &  93.89 $\pm$  1.37 &  29.15 $\pm$ 5.50 & 1571.66 $\pm$  3.53 & 0.49 & N1\\
E4   & 910.39 $\pm$ 1.02 & 0.5299 $\pm$ 0.0040 &  72.56 $\pm$  1.03 & 104.65 $\pm$  0.63 & 130.77 $\pm$ 1.20 & 2199.90 $\pm$  2.34 & 4.29 & N1\\
E5.1 & 649.68 $\pm$ 0.66 & 0.0010 $\pm$ 0.0070 &  84.13 $\pm$  0.85 &  97.92 $\pm$  0.85 &  91.67 $\pm$ 1.48 & 1731.26 $\pm$  4.41 & 1.66 & N1\\
E5.2 & 620.53 $\pm$ 1.51 & 0.7862 $\pm$ 0.0120 &  90.94 $\pm$  1.60 & 101.42 $\pm$  6.18 &  65.00 $\pm$ 1.14 & 1700.99 $\pm$  8.16 & 1.13 & N1\\
E7   & 395.61 $\pm$ 1.34 & 0.0001 $\pm$ 0.0290 & 105.42 $\pm$  3.38 &  96.74 $\pm$  7.84 & 292.58 $\pm$ 3.66 & 2753.28 $\pm$ 11.98 & 0.95 & N2\\
\hline
\end{tabular}
\caption{Best fit orbital elements of the E-stars. The derived Keplerian parameters for the E-stars 1-5 are in agreement with the analysis of \citep{muzic2008}, except for E7. As for the DS sources, we adapt the uncertainties from the MCMC simulations.}
\label{tab:orbit_table_estars}
\end{table*}
{Due to the main-sequence nature of the E-stars, we will group them by inspecting the orbital elements responsible for their 3D orientation. Hence, by averaging} the inclination and the {longitude of the ascending node (LOAN)} of the E-stars, we find $i\,=\,(92.89\,\pm\,11.57)^{\circ}$ and $\Omega\,=\,(99.09\,\pm\,85.83)^{\circ}$ where the uncertainty is based on the standard deviation. Although the LOAN is suffering from an increased uncertainty due to the orbital solution for E7, we find a satisfying agreement with the analysis of \cite{muzic2008} where the authors derive $\Omega\,=\,100^{\circ}$ and $i\,=\,83^{\circ}$ for the E-stars (Paper I). We will investigate the set of orbital elements for systematic trends in the next section. 

\subsection{Cluster analysis}

\begin{figure*}%
	\centering
	\includegraphics[width=1.\textwidth]{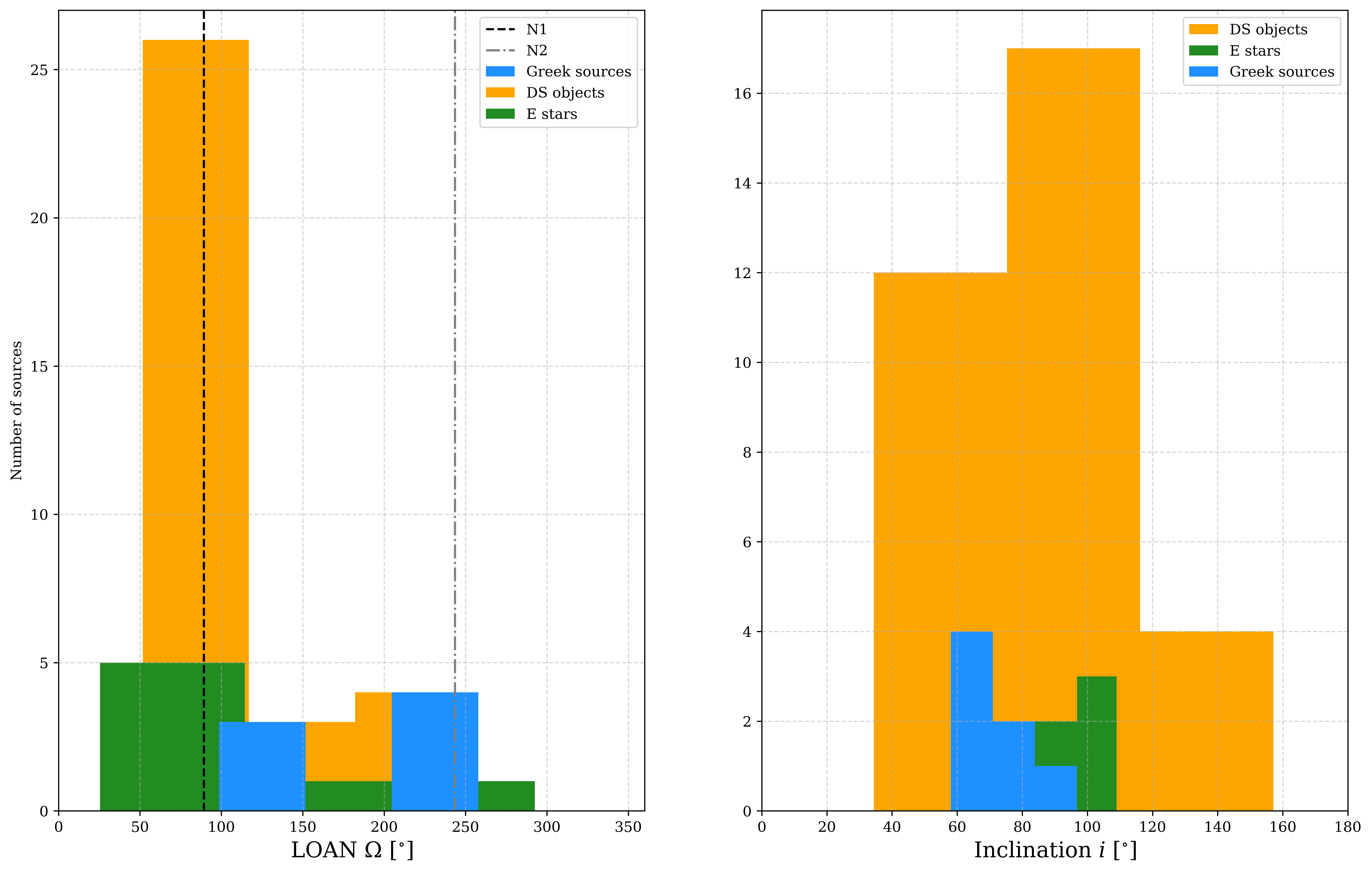}
	\caption{Orbital elements that represent the location of the dusty sources of IRS 13 in 3D space. The related orbital elements are listed in Table \ref{tab:orbit_table_all_ds}-\ref{tab:final_orbit_table}. {Motivated by the distribution of the longitude of the ascending node of the Greek sources (blue bars in the shown histogram)}, we divide the cluster members into two groups, namely, N1 and N2. {Furthermore, we include the green-colored orbital parameter of the E-stars}. In the presented histogram, the bin width is half of the square root of the number of cluster members. {The range of the x-axis is related to the allowed values of the corresponding orbital parameter.}} 
\label{fig:cluster_structure}
\end{figure*}

\begin{figure}%
	\centering
	\includegraphics[width=.5\textwidth]{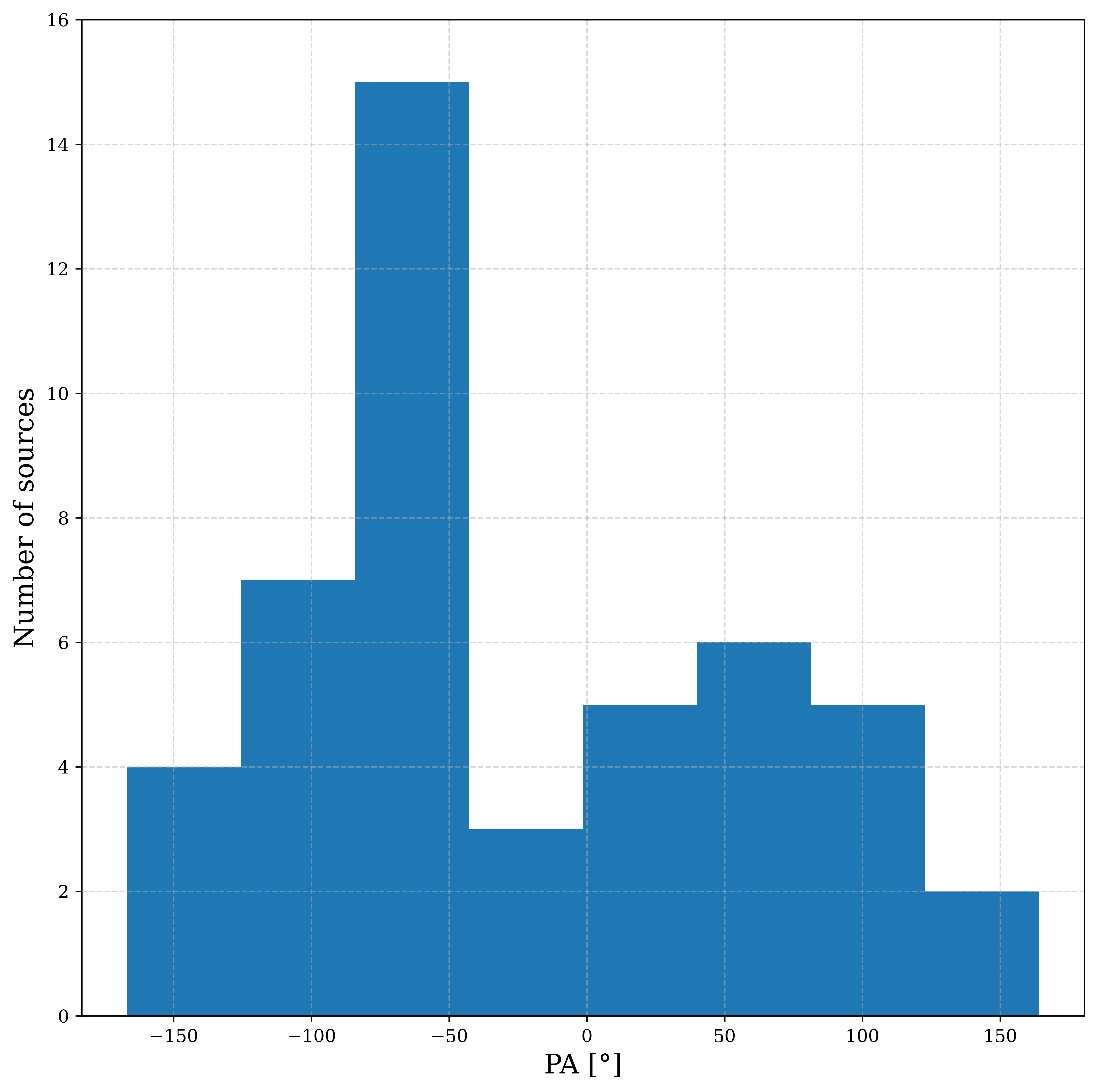}
	\caption{Position angle of all the cluster members of IRS 13. In analogy to the inclination and the LOAN distributions displayed in Fig. \ref{fig:cluster_structure}, we find two distinct peaks implying a non-randomized distribution of the stellar objects in the IRS 13 cluster. However, the difference in significance between the two peaks is pronounced. Here, we set the bin width to 40$^{\circ}$.} 
\label{fig:cluster_structure_pa}
\end{figure}

For the orbital parameters of the Greek-named dusty sources, we plot in Fig. \ref{fig:orbits_1} two different Keplerian solutions colored in {\it orange} and {\it blue}, reflecting on the number of possible individual trajectories \citep{Portegies-Zwart2023}. In addition, it is plausible that the orbital solutions derived for the dusty sources will potentially evolve with time depending on the absence or presence of an IMBH \citep{Schoedel2005, Tsuboi2017}. The possible ongoing elongation leading to cluster disruption could have an impact on the orbits of the dusty sources of IRS 13 \citep{Portegies-Zwart2003}. {For example, \cite{Tsuboi2017} shows the H30$\alpha$ distribution observed with ALMA. The authors find different velocities for the components of the cluster. While this implies a putative global direction of the cluster}, there are too many unknown parameters to provide a qualitative estimate about the evolution of the orbits. It is further possible, that the cluster itself undergoes an evolution due to the interaction with other components of the NSC, such as Sgr~A* \citep{Ali2020} or the CWD/CCWD \citep{Paumard2006, Fellenberg2022}.\newline
However, we select the orbits with the highest probability given by the minimized reduced $\chi ^2$ (Table \ref{tab:orbit_table}) and summarize them in Table \ref{tab:final_orbit_table}. Undoubtedly, the orbital solution for the dusty sources would have a different shape if the local dominant gravitational potential, represented by an IMBH, were located inside the IRS 13 cluster. Although we are not excluding the idea of the presence of an IMBH and inspected in Sec. \ref{sec:IMBH}, we assume that IRS 13 and its members are gravitationally bound to Sgr~A*. In agreement with \cite{Portegies-Zwart2023} and shown in Fig. \ref{fig:orbits_1}, multiple orbital solutions with comparable probabilities are acceptable. Despite the difficulty of the Keplerian analysis due to crowding effects {and the missing LOS velocities}, a closer look at the related orbital elements listed in Table \ref{tab:orbit_table} draws attention to the limitation of the quantity of possible results. With
\begin{equation}
 r\,=\,\frac{a(1-e^2)}{e\,cos(\theta)+1}
 \label{eq:kepler}
\end{equation}
where $r$ defines the distance of a celestial object to the central gravitational potential, it is evident that for a given {true anomaly} $\theta$\footnote{{The true anomaly $\theta$ indicates the position of a celestial body on its orbit around a gravitational potential.}}, various solutions for the size parameters $a$ and $e$ are allowed {which are explored in Appendix \ref{sec:parameter_space_e_a_app}. 
Despite the obvious influence of the semi-major axis a on the eccentricity (and vice versa) for a fixed distance r and $\theta$, we will use in the following the reduced $\chi ^2$ as a qualitative quantity to measure the goodness of the fit. For the analyzed data set, the reduced $\chi ^2$ poses a local minimum of the Keplerian solutions.} This can directly be transferred to the uncertainty range based on the MCMC simulations in Table \ref{tab:final_orbit_table}. The listed error spectrum represents a local solution and may be updated with the potential access to Doppler-shifted emission lines that should have an impact on the significance of the Keplerian orbits.
\begin{table*}
\centering
\begin{tabular}{|cccccccc|}
\toprule
ID     & $a$ (mpc)&	$e$  &	$i$($^o$)&$\omega$($^o$)&	$\Omega$($^o$)&	$t_{\rm peri}$(yr) & Group\\
\hline \hline
$\alpha$   & 255.31$\pm$14.12 & 0.8201$\pm$0.1180 & 96.78$\pm$31.97 & 88.30$\pm$38.27 & 236.84$\pm$16.78 & 2312.33$\pm$16.52& N2 \\
\hline
$\beta$    & 169.42$\pm$2.94 & 0.5032$\pm$0.0210 & 75.05$\pm$42.97 & 78.49$\pm$68.75 & 249.80$\pm$22.34 & 2333.59$\pm$0.90 & N2 \\
\hline
$\delta$   & 159.04$\pm$17.04 & 0.3648$\pm$0.0570 & 70.47$\pm$8.59 & 58.55$\pm$6.41 & 257.83$\pm$12.03 & 2302.77$\pm$1.85 & N2 \\
\hline
$\epsilon$ & 137.64$\pm$15.65 & 0.2463$\pm$0.0860 & 69.24$\pm$39.07 & 75.96$\pm$46.12 & 250.77$\pm$35.00 & 2337.51$\pm$14.37 & N2 \\
\hline
\hline
$\gamma$   & 255.18$\pm$18.67 & 0.5798$\pm$0.0730 & 63.77$\pm$30.13 & 224.80$\pm$32.83 & 98.66$\pm$11.57 & 2496.18$\pm$1.52 & N1 \\
\hline
$\zeta$    & 407.77$\pm$8.67  & 0.5791$\pm$0.0240 & 58.05$\pm$15.75 & 160.02$\pm$15.87 & 114.53$\pm$11.97 & 2292.16$\pm$3.11 & N1 \\
\hline
$\eta$     & 147.30$\pm$8.98  & 0.7374$\pm$0.0890 & 72.01$\pm$21.77 & 279.85$\pm$33.91 & 98.63$\pm$7.27   & 2535.22$\pm$1.76 & N1 \\
\hline
\end{tabular} 
\caption{Final orbital elements of the dusty sources of IRS 13. We group $\alpha$, $\beta$, $\delta$, and $\epsilon$ to N2, $\gamma$, $\zeta$, and $\eta$ to N1 (see the text for details). The given uncertainties are based on the {MCMC simulations in which we used the indicated orbital elements as the prior (see Appendix \ref{sec:mcmc-app})}. Due to the inevitable confusion, the pericenter passage t$_{peri}$ can differ significantly for increased data baselines. Without the line of sight velocity, t$_{\rm peri}$ should be classified as an initial parameter. We will overcome this limitation through the upcoming JWST/MIRI observations.}
\label{tab:final_orbit_table}
\end{table*}
Following the analysis of \cite{Witzel2017}, \cite{Ali2020}, and \cite{Fellenberg2022}, we use the inclination and the LOAN as reliable quantities for estimating the distribution of the investigated sources in the 3-dimensional space. 
From Table \ref{tab:final_orbit_table}, Table \ref{tab:orbit_table_all_ds}, and Table \ref{tab:orbit_table_estars}, it is implied that the inclination $i$ and the longitude of the ascending node $\Omega$ seem to be grouped, which is displayed in Fig. \ref{fig:cluster_structure} representing a non-randomized distribution of the key parameters that describe the 3d orientation in space. 
We follow up on the naive assumption of a grouped distribution of the sources by assigning $\alpha$, $\beta$, $\delta$, and $\epsilon$ to group N2 and $\gamma$, $\zeta$, and $\eta$ to group N1 by using the LOAN as a distinguishing parameter. In addition, we apply this classification on the DS sources (Table \ref{tab:orbit_table_all_ds}) as well as the E-stars (Table \ref{tab:orbit_table_estars}) and find that the majority of these sources belong to N1. 
Although we are aware of the uncertainties of the orbits (Fig. \ref{fig:orbits_1}) and, therefore, the high risk of a biased analysis, the clustering of the objects is in agreement with the non-randomized distribution of the dusty sources that follow the trend given by the anisotropic parameter $\gamma_{TR}$ discussed in Paper I. 
Considering the orbital parameters for the individual groups, we estimate for {N1} an average LOAN value of $(89.29\pm 19.10)^{\circ}$ whereas {N2} is characterized by $(243.72\pm 10.53)^{\circ}$. The related uncertainties are based on the standard deviation. Therefore, we estimate a difference of the LOAN between the two groups to $\Delta\Omega_{N1/N2}\,=\,(154.43\pm 14.81)^{\circ}$.\newline
{In contrast, the inclination of the dusty sources shown in Fig. \ref{fig:cluster_structure} suggests a classification that questions the grouping introduced by the distribution of the LOAN. In detail, we find numerical values of the inclination in Table \ref{tab:orbit_table_all_ds}-\ref{tab:final_orbit_table} that exhibit a trend that peaks at about $100^{\circ}$.
If we assume the presence of two distinct groups, we find an average inclination of $i=(73.88\pm 19.45)^{\circ}$ and $i=(121.02\pm 16.79)^{\circ}$ for the two groups. Consequently, we estimate a difference between the two groups of $\Delta i\,=\,(47.14\pm 18.11)^{\circ}$ where the combined average inclination is $(97.44\pm 23.57)^{\circ}$ for the Greek sources and the DS sources.
The same clustered arrangement is imprinted on the Keplerian parameters $i$ and the LOAN for the E-stars (Figure \ref{fig:cluster_structure} and Table \ref{tab:orbit_table_estars}). We note that the here derived orbital elements are in agreement with the averaged parameters for the E-stars presented in \cite{muzic2008}. However, we find that most of these evolved stars belong to the N1 group, given their rather low LOAN (Table \ref{tab:orbit_table_estars}).
Given the anisotropic distribution of the cluster members found in Paper I, the clear non-randomized distribution of the LOAN and inclination displayed in Fig. \ref{fig:cluster_structure} is expected. However, the significance for a two-disk structure, and hence two groups of cluster members, is low considering only the LOAN and the inclination (except for E7). To explore the presence of substructures, the position angle (PA) imposes an additional tool for the analysis of a cluster.   
With the definition of the PA using atan2, which is the 2nd argument of the arctangent \citep{Ali2020}, we estimate the orientation of all the dusty sources including the E-stars. As displayed in Fig. \ref{fig:cluster_structure_pa}, we find in agreement with the indicated distribution in Fig. \ref{fig:cluster_structure} a non-randomized orientation of the PA for all the investigated cluster members in favor of a disk-like distribution.}
Despite a rather shallow maximum around 60$^{\circ}$, the presence of a more prominent peak at -60$^{\circ}$ suggests the proposed non-randomized distribution.
Based on this orbital analysis, we are allowed to explore the classification of the members of the cluster presented in Paper I and incorporate the proposed clustering, which results in the properties derived listed in Table \ref{tab:generation}.
\begin{table}
        \centering
        \begin{tabular}{|c|cc|}
        \hline\hline
              & 1. Generation  & 2. Generation   \\  
         \hline\hline     
        Approx. age [Myr] & 4 &  $<$ 1  \\
        \hline
        Birthplace      & CND & Bow-shock shell \\
        \hline
        Current location  &  &  \\     
        inside IRS 13  & Core & Tip \\        
        \hline 
        Mean $K$-$L$ index & 2.49 & 5.72 \\
        \hline 
        Group  & N1 $\&$ N2  & N1 $\&$ N2 \\
        Group-Quantity & 85$\%$ $\&$ 15$\%$ & 82$\%$ $\&$ 18$\%$\\
        \hline
        \end{tabular}   
\caption{Two distinct populations in the IRS 13 cluster. We adapt this table from Paper I and extend it with the proposed grouping of the two populations. As suggested by the data, the majority of sources are located in N1. The distribution of the groups implies either two warped disks that are inclined by about 50$^{\circ}$ to each other or one disk for which we observe ascending and descending part in projection.}
\label{tab:generation}
\end{table}
As indicated in Table \ref{tab:generation}, both the E-stars and the dusty sources show a comparable sample distribution into N1 and N2 groups. We will explore the significance of both groups in Sec. \ref{sec:discuss}. 

\subsection{N-body simulations}
\label{sec:simulations}

The Keplerian solution of the orbits of the dusty sources is obtained by assuming that Sgr~A* dominates the gravitational potential. This, however, confronts the analysis with the question of the characterization of IRS 13 as a gravitationally stable cluster since one would expect that the dusty sources are not bound to Sgr~A* but rather to the cluster itself or a hypothetical IMBH. However, we can treat IRS 13 as a disturber of a purely Keplerian orbit that describes the dusty sources' trajectory. Hence, we disentangle the system by defining
\begin{equation}
    \rm True\,Orbit\,=\,Keplerian\,Orbit\,\otimes\,IRS\,13\,trajectory
\end{equation}
where the Keplerian orbit corresponds to the solution presented in Table \ref{tab:final_orbit_table}, Table \ref{tab:orbit_table_all_ds}, and Table \ref{tab:orbit_table_estars}. Furthermore, the IRS 13 trajectory represents the in-spiral path of the cluster. 
As shown by \cite{Zajacek2017}, the orbital energy of an in-spiraling (stellar) system decreases towards Sgr~A*. The authors of Zaja\v{c}ek et al. use G2/DSO \citep[][]{Peissker2020b,peissker2021c,peissker2023a} as a probe to trace the decrease in orbital energy due to dynamical friction \citep[][]{Morris1993, Jalali2014}. Using the same argument, it is plausible that the cluster evaporates during its in-spiral as argued in addition by others \citep[][]{Portegies-Zwart2003, Maillard2004}. Applying simple Keplerian mechanics, it is obvious that the ratio $m_2/m_1$ of both involved masses, i.e., Sgr~A* ($m_1$) and IRS 13 ($m_2$), converges to zero if the ongoing evaporation continues, hence,
\begin{equation}
a^3/T^2(m_1,m_2)\,\neq\,{\rm const}
\end{equation}
where the above equation represents the non-preserved third Keplerian law. In other words, the evaporation or mass-loss results in a continuous change of the in-spiral trajectory of the cluster and its orbital energy. In the following, we will explore a time efficient N-body simulation that could represent the evolution of IRS 13. {Our main objective is a quantitative estimate about the migration timescales of a potential cluster into the NSC.}
To ensure {computational efficient} setup, stellar winds or local density fluctuations are not included which could impact the evolution of the model cluster at distances smaller than the estimated and current 3D distance for IRS 13. 
{As mentioned,} we want to investigate the sink timescales of a cluster entering the inner parsec. {Therefore,} we treat the global trajectory of the IRS 13 as a free parameter. Due to the location of the cluster at (0,3) pc with respect to Sgr~A* which is located at (0,0) pc, the initial trajectory of the cluster is directed from west to east. Table \ref{tab:parameter_space} lists all the used input parameters for the N-body simulations.\newline
Since the massive E-stars in IRS 13 must have formed at large distances \citep{Morris1993}, the presence of candidate YSOs suggests a second triggered star formation epoch in the cluster. 
\begin{table}[hbt!]
    \centering
    \begin{tabular}{|cc|}
         \hline 
         \hline
           Parameter & Numerical values  \\
         \hline
         Cluster Mass     &  4.5$\times$10$^4 M_{\odot}$ \\
         King parameter   &  16 \\
         Cluster size     &  0.1 pc \\
         Initial location &  (0,3) pc \\
         Particle number  & 64, 248, 512\\
         Gravitational potential    &  4.5$\times$10$^6 M_{\odot}$ \\
       \hline
    \end{tabular}
    \caption{{Input parameter for the simulations. The particle number includes the 50 analyzed sources and includes the possibility of faint stars that are below our detection limit. The initial location of the cluster is with respect to Sgr~A* which is located at (0,0) pc.}}
    \label{tab:parameter_space}
\end{table}
Due to their proposed classification of the second generation cluster members \citep[Paper I and][]{Eckart2004a}, the age of most dusty sources is less than $<\,1$Myr. Since we assumed a Class I classification for the dusty sources, we limit the simulations to a time span of about 0.0-0.3 Myr, assuming that star formation processes started already before or during the cluster entered the {\it inner parsec}. We note that the overall process of an inspiralling cluster or molecular cloud is well documented in the literature \citep{Portegies-Zwart2002, Maillard2004, Bonnell2008, Hobbs2009, Jalali2014}.\newline
Furthermore, we set the number of particles used for the model to 64, 248, and 512. This set of particles represents the approximate amount of sources investigated in this work and includes the possibility that some cluster members could have been expelled due to cluster dynamics \citep[][]{Maillard2004, Paumard2006, Zwart2010}. Not only do we acknowledge former and expelled cluster members with the chosen particle number, we also tribute faint stars below the detection limit. By using different particle numbers, we can furthermore investigate if the migration timescales depend on the number of stars in the cluster.\newline
For the model, we use the modules AMUSE-PhiGRAPE \citep{Harfst2007}, AMUSE-BHtree \citep{Barnes1986}, and AMUSE-ph4 \citep{Portegies-Wart2018} to perform the simulations. The cluster with a total mass of $M_{\rm IRS 13}\,=\,4.5\times 10^4\,M_{\odot}$ is located at a distance of 3 pc from Sgr~A* with an initial size of 0.1 pc \citep[comparable conditions are used by][]{Petts2017}. {Considering that the enclosed mass is measured to be $M_{\rm IRS 13}\,\sim\,4\times 10^4\,M_{\odot}$ (Paper I) and assuming that cluster members have be expelled during the migration, the assumed initial cluster mass is justified. Furthermore, we} set the King parameter to 16, equal to a highly dense and compact cluster consistent with the exceptional high core mass density of $\rm 3\,\times\,10^8\,M_{\odot}pc^{-3}$ estimated by \cite{Paumard2006}. We adapt the Ansatz of \cite{Hobbs2009} using a stellar cusp mass of about $5\,\times\,10^5\,M_{\odot}$. We set the mass of Sgr~A* to $4\times10^6\,M_{\odot}$ and neglect the low mass of the CND of about $2.5\,\times\,10^4\,M_{\odot}$ estimated by \cite{Hsieh2021}. With the combined mass M$_{{\rm comb}}$ of Sgr~A* and the enclosed mass representing the stellar cusp, we use a static potential as a function of distance with
\begin{equation}
    M(r)\,=\,M_{{\rm comb}}\left(\frac{r}{R_{\rm enc}}\right)^{\alpha}
\end{equation}
where $R_{\rm enc}$ denotes the size of the enclosed region that we set to 3 pc whereas the mass distribution exponent is fixed to $\alpha\,=\,1.2$ that is valid for $R\,\leq\,500$pc \citep{Mezger1996}.
\begin{figure*}%
	\centering
	\includegraphics[width=1.\textwidth]{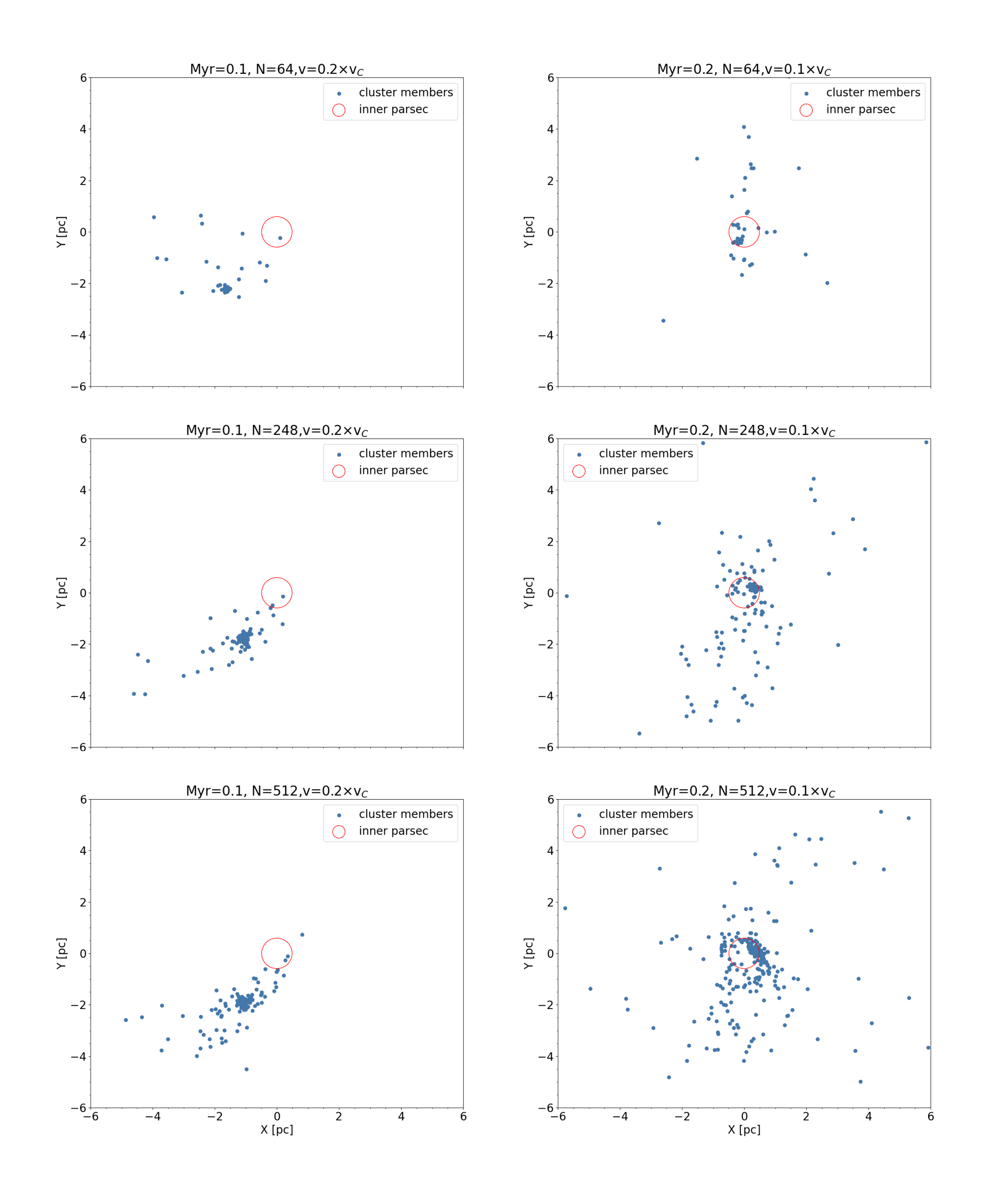}
	\caption{N-body simulations of a dense cluster with the properties derived in Paper I and this work. For every setup of the simulations, we find that the modeled cluster quickly ($<$0.3 Myr) migrates from the CND region ($>$2 pc) to the inner parsec that we indicate with a red circle. Since the cluster mass is equally scattered among the stellar members, the simulations suggested that especially high-mass stars tend to migrate in the NSC. Sgr~A* is located at the center of each red circle at (0,0) pc. {We refer to Appendix \ref{sec:nbody_mock_app} where we present a mock image of these simulations.}}
\label{fig:irs13_evo}
\end{figure*}
We demand that the in-spiraling cluster must have undergone a violent event that resulted in the loss of its orbital angular momentum, e.g. due to the initial cloud-cloud collision within the CND or the Central Molecular Zone \citep{2000ApJ...536..173T,Jalali2014}. Furthermore, we assume that high-mass stars are not prone to this violent interaction which is motivated by the analysis of hard and soft binary systems \citep[][]{Heggie1975, Hills1975a, Hills1975b, Zwart2010}\footnote{{Based on the presence of the E-stars in the cluster core of IRS 13, the massive stars were cluster members before, during, and after the violent interaction. Therefore, they are part of the cluster with the start of our simulations.}}. Hence, the modeled cluster encloses particles with the same mass based on $M_{\rm IRS 13}$.\newline
For a stable orbit, stars have a virial velocity v$_{\rm virial}$ after the cluster relaxes at $t_{\rm R}$. We mimic the in-spiral motion caused by the loss of angular momentum of the cluster by setting $v_{\rm virial}$ to 20$\%$ of v$_{\rm relax}$ defined as the root mean square velocity $\langle v^2 \rangle ^{1/2}$ of all cluster members. The relation between the relaxation time scale $t_{\rm R}$ and the velocity dispersion $\langle v^2 \rangle$ is given by
\begin{equation}
    t_{\rm R}\,\simeq\,\frac{0.065\langle v^2 \rangle ^{3/2}}{nm^2G^2{\rm ln}\Lambda}
\end{equation}
where we again assume identical particles with the same mass $m$, the number density $n$, and $\Lambda$ describing the impact parameter \citep{Spitzer1987, Portegies-Wart2018}. However, we note that these parameters undergo a multi-dimensional transformation because not only will the velocity dispersion $\langle v^2 \rangle$ differ for the single objects inside of the cluster depending on their global habitat (CND and {\it inner parsec}), but also the impact parameter $\Lambda$ will change due to the ongoing dissolution of IRS 13. In addition, the number density between the CND and {\it inner parsec} increases by almost 100$\%$ \citep{Vollmer2022} which should result in a decreased orbital energy \citep{Zajacek2017}, which we mimic by setting $v_{\rm virial}$ to 10$\%$ as soon as the cluster enters the NSC. Based on these settings, we show in Fig. \ref{fig:irs13_evo} the snapshots for a cluster with different numbers of particles moving towards the inner parsec at 0.1 Myrs and 0.2 Myrs. {Hence, we find a migration timescale for a cluster of $\leq$ 0.3 Myrs in agreement with the numerical simulations presented in \cite{Bonnell2008}, \cite{Hobbs2009}, and \cite{Jalali2014}.}

\section{Discussion} 
\label{sec:discuss}
In this section, we discuss the findings presented in this work. By additionally incorporating the results from Paper I, we propose a comprehensive view of the arrangement of the IRS 13 cluster. Furthermore, we will explore the presence of a putative IMBH using the photometric results of Paper I. 

\subsection{Dimensions of IRS 13}
\label{sec:dimensions_discussion}

\begin{figure*}%
	\centering
	\includegraphics[width=1.\textwidth]{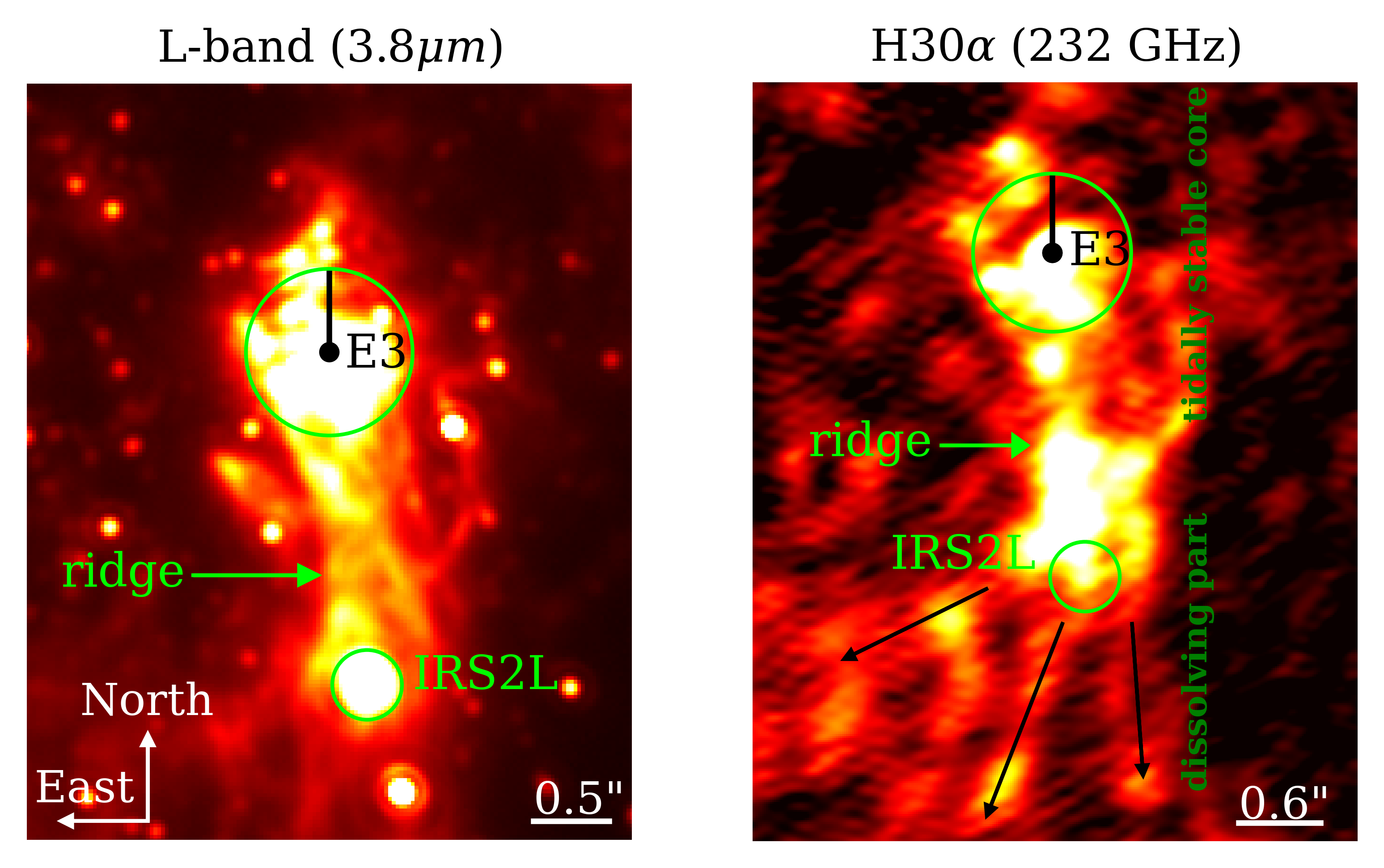}
	\caption{Observation of IRS 13 with NACO and ALMA. The left image shows the $L$-band observation of the region associated with the IRS 13 cluster in 2004. The right panel exhibits a collapsed data cube observed in the radio/submm regime of the same region at 232 GHz ($H30\alpha$). In both plots, we indicate the position of E3 with a black dot and mark the ridge that connects the IRS 13 region with IRS 2L \citep{Buchholz2013} with a lime-colored arrow. Furthermore, we illustrate the tidal (Hill) radius by the lime-colored circle around E3 which has been associated with a candidate IMBH in the literature (left image). It shows that the southern section is the potentially dissolved and detached part of the northern, tidally stable core (right image). The black arrows in the same image indicate the direction of the evaporation.}
\label{fig:irs13_dimensions}
\end{figure*}

\begin{figure}%
	\centering
	\includegraphics[width=.5\textwidth]{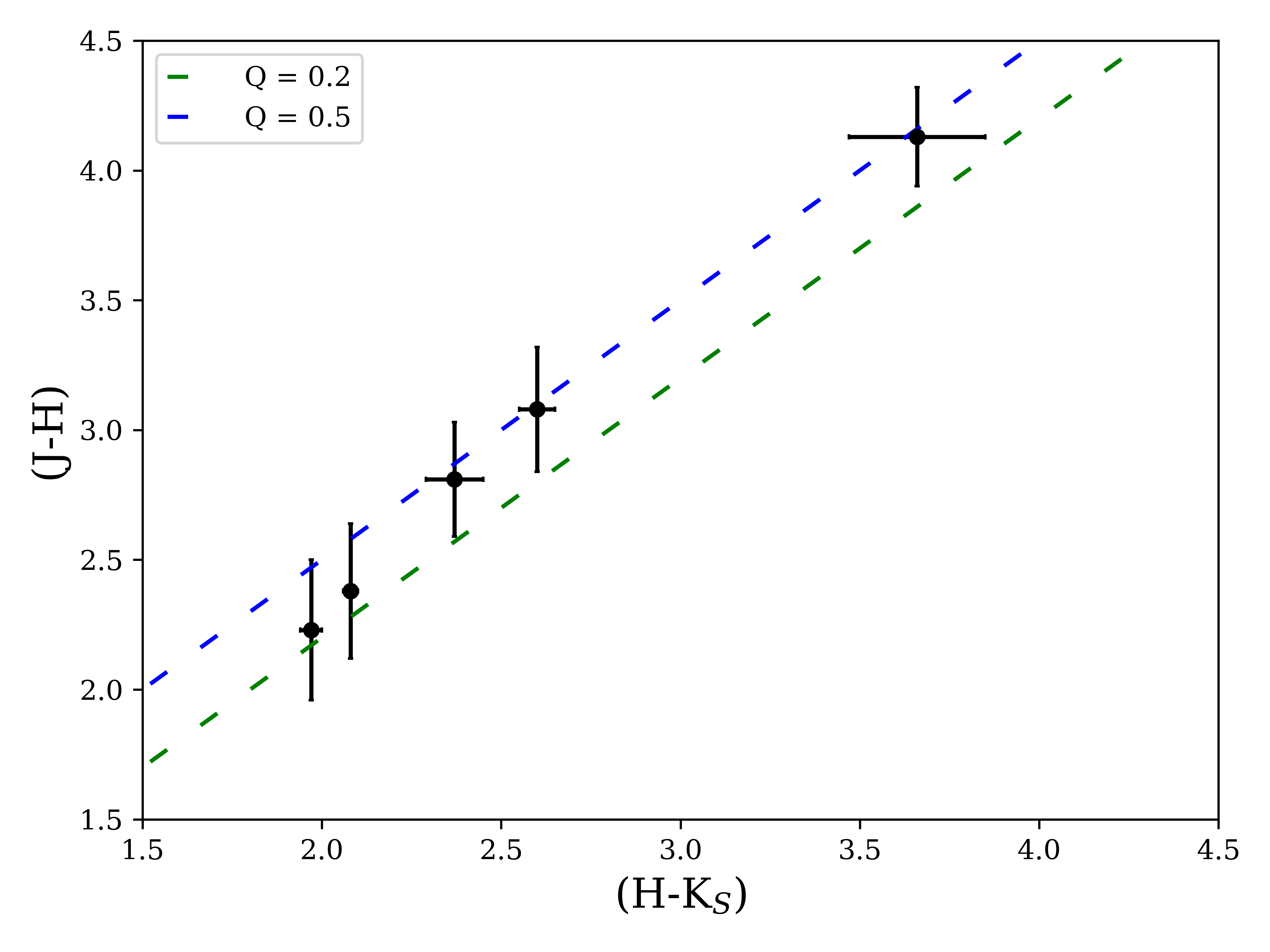}
	\caption{Color-color diagram of IRS 2L, E1, E2, E4, and E7. Since we find no emission above the noise for E3, we exclude it. The green and blue dashed line represent the Q values 0.2 and 0.5, respectively. All sources inside this range are most likely cluster members \citep[see][]{Comeron2005}.}
\label{fig:irs13_qfactor}
\end{figure}

In Paper I, we estimated a Hill radius of 22 mpc from the velocity dispersion based on the proper motion analysis of the cluster members. The enclosed projected region remarkably matches the dimensions of the core region of IRS 13 with E3 at its center. However, parts of the proposed {\it ridge} that connects IRS 13 with IRS 2L (Fig. \ref{fig:irs13_dimensions}) are not inside the Hill radius which suggests that the southern part of the cluster will evolve into a gravitational detached region from the core.
In the following, we will explore tracers that imply a connection between IRS 2L (Fig. \ref{fig:irs13_dimensions}) and the core region of IRS 13. Considering the colors of IRS 2L (Paper I and Table \ref{tab:cluster_membership}) in combination with the early-type classification \citep[B-type, see][]{Moultaka2009}, the star belongs to the dissolved and elongated part of IRS 13, i.e., the tail region. Interestingly, \cite{Paumard2006} excluded the idea of a dissolving cluster due to missing and expelled high-mass stars. We will elaborate on this point in the next section. Furthermore, the polarization observations \citep{Witzel2011} presented in \citet{Buchholz2013} strongly suggest that IRS 2L is indeed an embedded source in the dust feature that also encloses the IRS 13 cluster \citep[the results of Buchholz et al. are independently confirmed by][]{Roche2018}\footnote{The authors of Roche et al. investigate the orientation of the magnetic field lines in the {\it inner parsec}.}. The correlation of the polarization angle with the alignment of the magnetic field lines is a strong sign that IRS 2L is a (former) cluster member of IRS 13 which is underlined by the Q-factor introduced by \cite{Comeron2005} that is defined as $Q\,=\,(J-H)\cdot a\,-\,(H-K)$. The equation to determine the Q-factor is inspired by \cite{Rieke1985} where $a=1.7$ is related to the reddening vector. Inspecting the $J$-band data of the GC observed with NACO\footnote{Programme ID: 091.B-0081(B)} reveals E1, E2, E4, and E7 as prominent and bright sources above the confusion level (see Paper I).\newline
In contrast, we find no $J$-band counterpart of E3 above the noise level, but a faint emission from IRS 2L at its corresponding K-band location, indicating weak tracers at $1.27\mu m$ of the B-type star. 
Based on the photometric analysis performed by \citet{Blum1996}, we adapt a $J$-band magnitude of $13.8\,\pm\,0.6$ mag for the used reference star IRS7 \citep[see also][]{Cutri2003}. The uncertainty of $\pm$0.6 mag for the estimated $J$-band magnitude reflects the variability of IRS 7. Furthermore, we adapt the H and H-K values from Paper I to estimate the J-H colors (Table \ref{tab:cluster_membership}).
\begin{table}[hbt!]
    \centering
    \begin{tabular}{|ccc|}
         \hline 
         \hline
           Star ID & J-H & H-K  \\
         \hline
         IRS 2L  &  4.13$\pm$0.19  & 3.66$\pm$0.19  \\
         E1      &  2.23$\pm$0.27 & 1.97$\pm$0.03 \\         
         E2      &  2.38$\pm$0.26 & 2.08$\pm$0.02 \\         
         E4      &  2.81$\pm$0.22 & 2.37$\pm$0.08 \\     
         E7      &  3.08$\pm$0.24 & 2.60$\pm$0.05 \\
       \hline
    \end{tabular}
    \caption{Derredened colors for stars observed in the J, H, K band. The H and K magnitudes and uncertainties are listed in Paper I. For the reference star IRS 2L, we adapt the uncertainties from \cite{Blum1996}. Regarding the E-stars, we scale the uncertainties listed in this table to IRS 2L by using the mean values of the standard deviation of the individual magnitudes to achieve reasonable comparability.}
    \label{tab:cluster_membership}
\end{table}
With the estimated derredened J-H and H-K colors, we incorporate the Q factor, where we set $a=1.0$ due to the obliterated reddening. In Fig. \ref{fig:irs13_qfactor}, we show the resulting correlation where $\rm Q\,\in{0.2,0.5}$ defines the cluster membership of the investigated sources. The presented correlation is not surprising because of the studies mentioned before by \cite{Buchholz2013}, where the authors already state a connection between the enveloping local dust structures displayed in Fig. \ref{fig:irs13_dimensions} and IRS 2L.
Since the dust around E3 is associated with IRS 13, we find no compelling reason (other than a historical one) why the southern part of the cluster should be excluded. In addition, CO, H30$\alpha$, and [FeIII] are enveloping the same region as the dust, implying a {connection between the gas emission lines and the grains}. Due to the nature of the radio emission lines, they should be destroyed close to the massive WR stars in the cluster unless sufficient gas-dust coupling and shielding is present \citep{Morris2020}.\newline
Therefore, we propose that the nature of the IRS 13 core can be described as the tidally stable remnant of a massive inspiralling cluster that is suffering from ongoing evaporation due to the tidal forces of Sgr~A* underlined by the gas and dust distribution as shown, for example, in Fig. \ref{fig:irs13_dimensions}. Although we present strong indications for the IRS 13 cluster membership of IRS 2L, the star could be a chance association due to its location. We want to emphasize that the cluster membership of IRS 2L has no impact on the proposed dissolving nature of IRS 13, as other high-mass candidates could also serve as tracers of expelled stars \citep[][]{Yusef-Zadeh2013}.  

\subsection{IRS 13, an in-spiraling cluster}
\label{sec:discuss_disturber}

{In Sec. \ref{sec:simulations}, we have shown that a cluster can migrate from a few parsecs in the {\it inner parsec}. For these naive N-body simulations, we neglected stellar winds or the ambient medium, which should impact the simulations \citep{Kaczmarek2011, Pfalzner2014}. Another limitation is the same mass of all particles used. Since we wanted to analyze the migration timescale of a cluster with a given mass, we expect a comparable reduced impact on the outcome.}\newline 
Independent of the particle number used in the simulations, we find that a dense concentration of stars is indeed able to migrate towards the inner parsec (indicated with a red circle in Fig. \ref{fig:irs13_evo}). This takes place on very short timescales of $t\lesssim 0.3$Myrs, agreeing with the simulations of \cite{Portegies-Zwart2003} and \cite{Jalali2014}. For every simulation, the cluster enters the inner parsec (red circle in Fig. \ref{fig:irs13_evo}) from south with a net direction towards north. Furthermore, we notice a tendency for high-mass stars to migrate to the inner parsec. Since the mass of the cluster is distributed among the stellar members, the cluster with fewer particles exhibits a higher mass/object ratio. We do find that for $N=64$, $>\,50\%$ of stars are able to migrate to the {\it inner parsec}. In contrast, only $\lesssim\,30\%$ of stellar members of the cluster with $N=248$ and $N=512$ enter the {\it inner parsec} which implies a migration preference for high-mass stars. This relation could serve as an explanation for the exceptionally high stellar mass density in the core region of IRS 13.\newline
It is also evident that the cluster seems to dissolve but does not leave a star trail. Stars need an escape velocity of v$_{esc}\,>\,2\bar{v}$ to get kicked out of a cluster, where $\bar{v}$ defines the mean velocity of all the stellar cluster members. As argued in the literature \citep[see, for example,][]{Paumard2006}, one should find star trails to classify IRS 13 as an in-spiraling cluster. This argument is weak simply because it assumes that the cluster consists of non-interacting stars that lose orbital energy due to dynamical friction. 
However, we can safely assume that identifying former and expelled IRS 13 members is challenging because they are not following a simple star trail trajectory motivated by general cluster dynamics. In detail, the interaction of a multi-component system is the dominant parameter for the evolution of stellar systems in an embedded cluster \citep[][]{Zwart2010}. In Appendix \ref{sec:star_trail_app}, we will elaborate on this particular point.\newline

\subsection{IRS 13 and the (C)CWS}

\begin{figure*}%
	\centering
	\includegraphics[width=1.\textwidth]{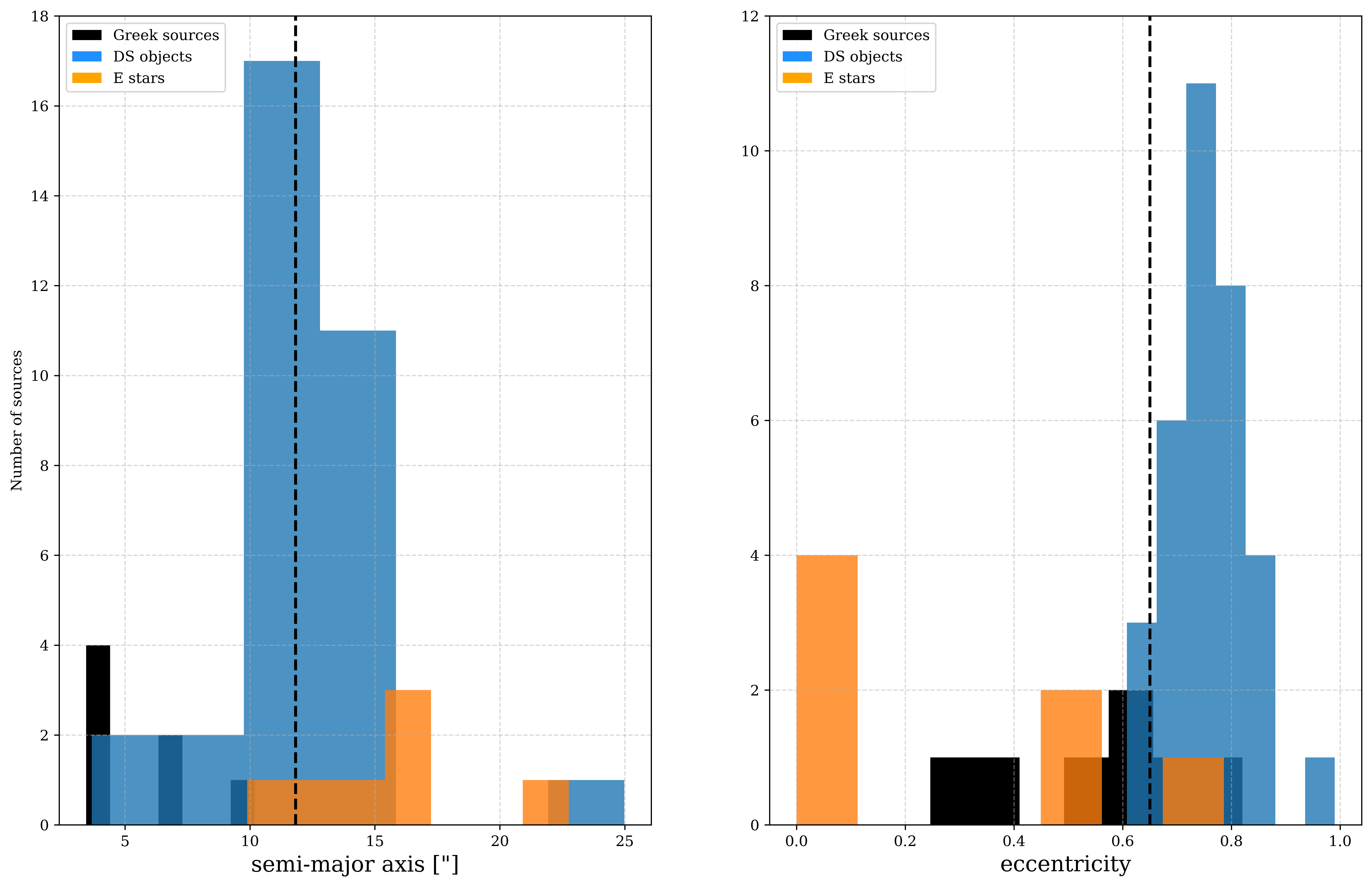}
	\caption{Semi-major axis $a$ and eccentricity $e$ of the IRS 13 cluster members analyzed in Paper I and this work. Both quantities exhibit a non-randomized distribution with an average value of $a=(11.82\pm 4.39)"$ and $e=(0.65\pm 0.23)$ which is reflected by the black dashed line. Compared to the kinematic features shown in \cite{Fellenberg2022} (see their Fig. 18), the distribution presented here does not correspond to any known structure. The bin width of the shown histogram is defined by the square root of the sample number.}
\label{fig:irs13_eccentricity_semimajor}
\end{figure*}

The composition of the {\it inner parsec} exhibits various sub-regions containing different stellar types and ages. Due to migration time scales, the presence of late-type stars is expected \citep{Morris1993}. In contrast, the young age of the S-cluster stars resulted in the formulation of the {\it Paradox of Youth} by \cite{Ghez2003}. Taking into account the age of the E-stars and the here investigated dusty sources in the IRS 13 cluster, the mentioned paradox seems to be valid for other regions in addition to the S-cluster. While this claim still needs to be verified, a detailed analysis of the IRS 13 cluster already bears fruitful results. From a larger point of view, the shape of the cluster suggests a dynamic interaction with its environment. For example, IRS 13 exhibits a prominent footprint from the mini-cavity on its eastern edge. However, it is not yet clear to determine the reason for the creation of the mini-cavity. So far, the question about the nature and history of the mini-cavity is only approached by speculations about a possible wind that originates at the position of Sgr~A*. {For a detailed discussion of the wind that might interfere with its environment, we refer the interested reader to Appendix \ref{sec:wind_app}.}\newline 
However, due to the high-mass of the cluster members inside of the Hill radius (Fig. \ref{fig:irs13_dimensions} {and Paper I}), it is suggested that IRS 13 is a core remnant of a massive cluster \citep[for a schematic diagram, see Fig. 8 in][]{Motte2018}\footnote{Please also consider \cite{Tige2017}.}. This claim is strengthened by the theoretical work of \citet{Bonnell2008} \citep[see also][]{Wardle2008} where the authors proposed inspiralling molecular clouds that form stars with a top-heavy mass function at projected distances of up to 0.25 pc. With the estimated distance of about 0.1 pc for IRS 13 from Sgr~A*, the idea of inspiralling clouds that undergo star-formation process can be applied to the cluster. Therefore, we consider IRS 13 as the prime example of a new star-formation channel that promises to explain at least partially the paradox of youth. Moreover, it is interesting to note that the non-randomized distribution of stellar objects in the cluster shown in Fig. \ref{fig:cluster_structure} and Fig. \ref{fig:cluster_structure_pa} can also be found for the members of the S-cluster \citep{Ali2020}. Both clusters are part of the NSC for which it is established that the stars at various distances are arranged in the clockwise and counterclockwise disks which are located at different distances from Sgr~A* \citep{Paumard2006, Lu2009}. Due to the distance of IRS 13 from Sgr~A* of about 0.1 pc, only the clockwise disk is considered to have a connection to the cluster because it resides between 0.05-0.5 pc with a mean inclination of 115$^{\circ}$. Inspecting Fig. \ref{fig:cluster_structure} (right plot) directly shows that the mean inclination of the clockwise disk is slightly larger than the main peak that corresponds with the DS-sources. Argumentatively, we can infer two different options for the DS sources:
\begin{itemize}
    \item[a)] The DS sources will keep their distinctive inclination distribution,
    \item[b)] Due to relaxation, the DS sources will (partially) migrate into the clockwise disk.
\end{itemize}
Addressing both options exceeds the scope of this work, but it is important to note that the dusty sources already follow an ordered distribution despite their young age. Hence, it is suggested that the larger structure, in which the DS sources are embedded, i.e. IRS 13, forces a distributive imprint on the cluster members. Interestingly, the arrangement of an inspiraling cluster in a disk-like structure is exactly what was proposed by the simulations of \cite{Bonnell2008}. The only difference is the nature of the modeled object, i.e., IRS 13 harbors evolved stars with an age of several 10$^6$yrs whereas the initial molecular cloud simulated by \cite{Bonnell2008} lacks any stellar components. Ignoring these obvious differences, the quantitative statement of the investigations outlined by \cite{Bonnell2008} is surely not the inspiraling nature of a molecular cloud, as it is also investigated by, for example, \cite{Jalali2014}. It is rather the fact that the interplay between a supermassive black hole and an object, which is prone to the formation of stars, arranges itself in a non-randomized stellar distribution as is shown for the S-cluster \citep{Ali2020}, the NSC \citep{Paumard2006, Lu2009, Fellenberg2022}, and the IRS 13 cluster (Paper I and this work). We strengthen the statement about a non-randomized distribution in a cluster that gravitationally interacts with an SMBH by inspecting the semi-major axis $a$ and the eccentricity $e$ in Fig. \ref{fig:irs13_eccentricity_semimajor}.
Based on this figure, it becomes clear that the IRS 13 population forms an individual arrangement that does not follow any known stellar density structures, such as the CWS, CCWS/F1, F2, and F3 \citep[][]{Paumard2006, Yelda2014, Fellenberg2022}. {Recently,} \cite{Jia2023} mentioned the reduced significance of the large-scale features F1-F3 following the argumentation of the disputed CCWS by \cite{Lu2006, Lu2009}. In addition, the reduced significance of the CCWS was already mentioned by \cite{Bartko2008} but without excluding this feature. However, we do not comment on the significance of these large-scale features since we consider IRS 13 as an individual structure with unique properties. Taking into account the absence of additional features in Fig. \ref{fig:irs13_eccentricity_semimajor}, we do not find traces of a second kinematic structure, which challenges the proposed classification of the IRS 13 cluster members into two groups (N1 and N2, see Table \ref{tab:final_orbit_table}-\ref{tab:orbit_table_estars}). A possible explanation for the arrangement of the LOAN, inclination, and PA (Figure \ref{fig:cluster_structure} and Fig. \ref{fig:cluster_structure_pa}) could be projection effects that result in a warped disk.\newline
Despite the manifold theoretical and observational attempts that target the star formation history of the {\it inner parsec}, there are still unresolved issues due to the limitations of the respective models. Although we are in favor of models that describe the infall of massive molecular clouds or a cluster \citep[][]{Portegies-Zwart2003, Maillard2004, Bonnell2008, Jalali2014} in order to explain the star formation history of the {\it inner parsec}, it is still questionable how these structures could have migrated from the host habitat into the NSC. {In the following, we will explore a possible mechanism.}

\subsubsection{Cluster migration}

One proposed idea is described with cloud-cloud collisions which cause the initial and required decrease in the angular momentum of the material \citep{Moser2017, Hsieh2017, Tsuboi2021}. This hypothesis has recently been challenged by the work of \cite{Salas2021}, who introduced turbulence as an additional parameter to explain the infall of the material required for star formation. The authors used a Fourier forced module \citep{MacLow1999} to mimic the {various} origins of turbulence. This approach seems reasonable because of several sub-regions and complex background emissions resulting from stellar density fluctuations implied by the simulations outlined in \cite{Dinh2021}. However, the high-velocity dispersion \citep{Moser2017, Hsieh2021} that is observed for the gas material {requires} extended models. The complexity of molecular clouds that interact tidally with the gravitational potential of Sgr~A* was recently underlined by observations of \cite{Paumard2022}. The authors find different clumps in the CND that exhibit a wide range of densities or filling factors. However, the general process of the creation of these clumps should not allow a huge variety of defining parameters as, for example, shown in \cite{Dinh2021}. It is, therefore, tempting to overcome some of the complex and challenging problems by assuming a gravitationally stable cluster that migrates towards the {\it inner parsec} to explain its star-formation history.\newline
Nonetheless, we propose that the high-mass stars (i.e. the first generation) of IRS 13 formed outside the {\it inner parsec} where the parent cloud was located inside the CND a few million years ago. Despite the mentioned and debated cloud-cloud collision, {this violent event could be substituted with} fragmentation {that resulted in a decreased} angular momentum of the initial cluster. For example, \cite{Misugi2023} finds a {decrease in} angular momentum caused by the collapse of filaments inside a molecular cloud. A transfer of the angular momentum to protostellar cores is plausible \citep{Misugi2019} but should not be confused with the orbital angular momentum parameter. Based on general cluster dynamics, an energy transfer between the inner and outer cluster regions seems at least plausible and should be considered for a detailed approach studying a possible link between the core and cluster angular momentum \citep{Makino1998, Boily1999, Zwart2010}.\newline

\subsection{IRS 13E3, a colliding wind product?}

\begin{figure*}%
	\centering
	\includegraphics[width=1.\textwidth]{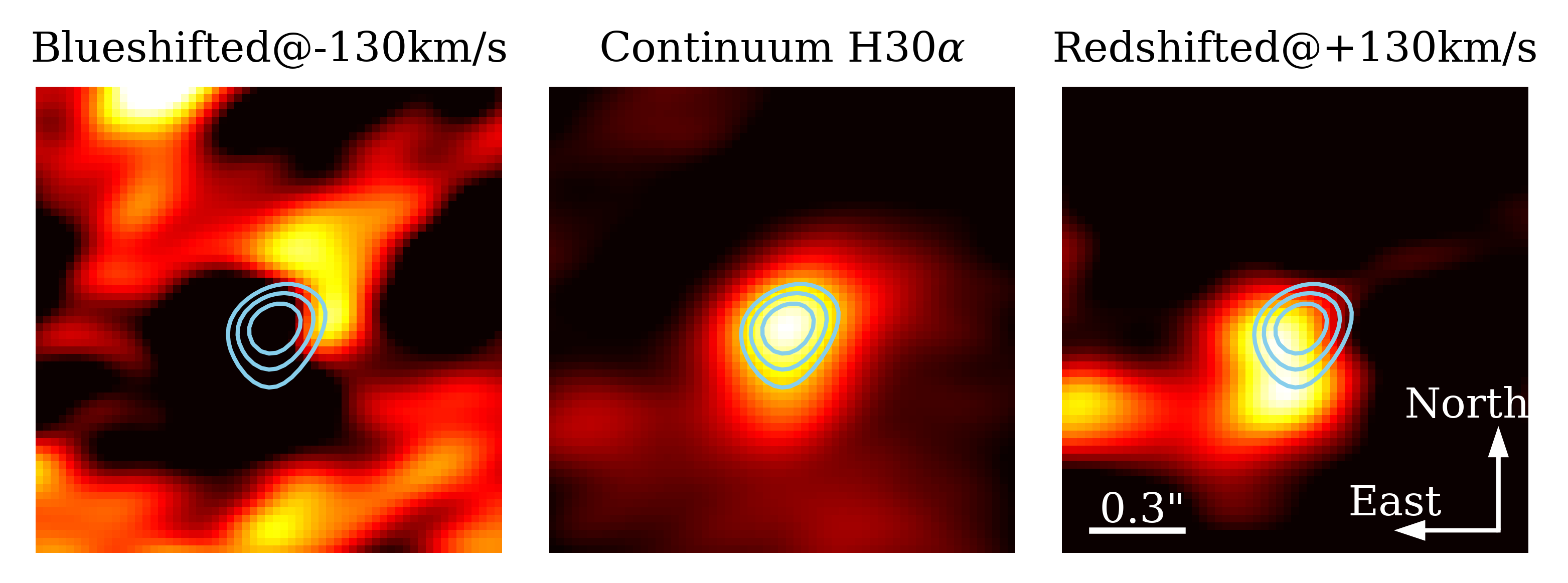}
	\caption{Ionized gas around the E-star E3 observed with ALMA at $231.9\,$GHz. The contour lines represent 70$\%$, 80$\%$, 90$\%$, and 100$\%$ of the normalized flux distribution of E3. For the blue-shifted velocity map, we select the H30$\alpha$ line at $231.8\,$GHz resulting in a velocity of $-130\pm30\,$km/s. The red-shifted velocity map is associated with $232.0\,$GHz which can be transferred to $+130\pm30\,$km/s. From the Doppler-shifted velocity maps, it is evident that the H30$\alpha$ emission with a rest-wavelength at $231.9\,$GHz is not centered on E3 but rather shows a ring-like distribution. We find an approximate diameter of this ring of about $(0.20\pm0.02)"$. The detection of an ionized ring around E3 was independently reported by \cite{tsuboi2019}.}
\label{fig:irs13e3_velocity_map}
\end{figure*}

Until now, the presence of an IMBH embedded in the IRS 13 cluster has been debated \citep{Maillard2004,Schoedel2005, Wang2020, Zhu2020}. For example, \cite{Zhu2020} and \cite{Wang2020} argued that colliding winds are responsible for the bright X-ray emission that is associated with the IRS 13 cluster. Given the nature of the dusty sources as high-mass YSOs which could be associated with Herbig Ae/Be stars, the region should in principle exhibit multiple stellar-wind interactions. Motivated by the observations by \citet{Fritz2010}, the authors of \citet{Zhu2020} concluded that the E3 source is rather a hot blob and the product of colliding winds which explains the prominent X-ray emission observed with \textit{Chandra}. In contrast, \citet{Tsuboi2017, tsuboi2019} showed radio/submm data observed with ALMA, which revealed a rotating ring of gas (H30$\alpha$) around E3 with velocities between -200 km/s and +200 km/s. By examining the same wavelength coverage used by Tsuboi et al. and observed with ALMA\footnote{PI Lena Murchikova \citep[project code: 2016.1.00870. S, see][]{Murchikova2019}}, we independently confirm the results of \cite{tsuboi2019, Tsuboi2021} by choosing a different representation of the ionized ring of the H30$\alpha$ line distribution around the position of the E star E3 (Fig. \ref{fig:irs13e3_velocity_map}).
The blue- and red-shifted motion of gas at the position of E3 was also shown by \citet{Wang2020} (see their Appendix) but not further commented on. Nevertheless, the rotation of ionized gas is the strongest indication of the presence of a massive body or an IMBH. \citet{Murchikova2019} {claimed to uncover} a comparable distribution of H30$\alpha$ around Sgr~A*. We note that this pattern of rotating gas is also reported in \citet{Lutz1993}\footnote{Please see the channel maps in their Fig. 5.} where the authors investigated the Doppler-shifted [FeIII] line centered on the core region of IRS 13.\newline
In principle, these two scenarios can coexist with each other, although we note that the collision of stellar winds to explain the nature of E3 is a stationary solution that relies on the distance of E2 and E4. Nonetheless, the proper motion vectors for these two stars show large deviations in the literature, including this work. If colliding winds can explain the nature of E3, we should see strong variations in the flux if the distance is variable. However, \cite{Zhu2020} presents an almost constant K-band and X-ray flux, underlined by a matching proper motion vector for E2 and E4. To explain the nature of E3 with colliding stellar winds, the constant distance of E2 and E4 is a necessary condition that is fulfilled \citep[see also][]{Wang2020}. However, we observed an increase in the distance between E3 and E4 from 0.1" in 2002 to 0.15" in 2018 using K- and L-band NACO data, resulting in $\Delta d\,=\,50$ mas. Consequently, the distance between E2 to E3 also increases due to their different proper motion directions. We find an increase in distance of $\Delta d\,=\,40$ mas between 2002 and 2018 for E2 and E3. The varying distance of E3 to E2 and E4 should be reflected in a significant flux variability that is described by the radiative braking point \citep[][]{Gayley1997, Owoki2002}. This breaking point of colliding winds is defined as
\begin{equation}
    r_B\,=\,D/(1+\sqrt{P_{WR/\nu}})
    \label{eq:breacking-point}
\end{equation}
where $D$ defines the distance and $P_{WR/\nu}$ {stands for} the mass-loss-dependent momentum flux of the WR star. Since E2 and E4 are classified and simulated as WR stars with the comparable mass-loss rate, $D$ must be constant to produce a constant flux \citep[][]{Zhu2020}. As we mentioned, the projected distances of E2 and E4 are indeed constant, but the position of E3 {with respect to} both stars is not. It is evident that Eq.~\ref{eq:breacking-point} can be written as
\begin{equation}
    P_{WR/\nu}\,=\,\bigl(D/r_B\,-\,1\bigr)^2
    \label{eq:momentum-flux}
\end{equation}
where $P_{WR/\nu}$ fluctuates as a function of the distance. Nevertheless, we note that \cite{Lepine1999} report significant wind inhomogeneities for WR stars that could serve as a naive explanation for the increasing distance between E3 and E2 as well as E4 but not for the reported constant flux.

\subsection{IRS 13E3, an IMBH?}
\label{sec:IMBH}

To verify whether the hypothetical IMBH is consistent with the multi-wavelength properties of the region close to IRS 13E3, we calculate its spectral energy distribution. {We underline the importance of this analysis since it has a major impact on the cluster dynamics discussed in this work.}
To this goal, we assume that the IMBH accretes material mostly from stellar winds of the surrounding WR stars and hence the accretion flow can be described as radiatively inefficient, advection-dominated, similar to Sgr~A* \citep{2010ApJ...716..504S,2018MNRAS.478.3544R}, but corresponding to a smaller black-hole mass and a different relative accretion rate.

The IMBH mass can be estimated from the rotation of the gas traced by H30$\alpha$ line. The gas forms a ring-like feature around the E3 infrared source, see Fig.~\ref{fig:irs13e3_velocity_map}, that is located within the region of $R_{\rm gas}=0.10 \pm 0.01''\sim 825 \pm 83\,{\rm AU}$. The line-of-sight velocity traced by the line is $v_{\rm R}=130 \pm 30\,{\rm km\,s^{-1}}$. Assuming that the gas is fully virialized, the IMBH mass is 
\begin{equation}
    M_{\rm IMBH}=\frac{R_{\rm gas}v_{\rm R}^2}{G}\sim 31434 \pm 7422\,M_{\odot}\,,
    \label{eq_virial_mass}
\end{equation}
where the uncertainty is given by the propagation of errors. This is rather a lower limit since the 3D velocity includes both the line-of-sight as well as the tangential velocity component.

Within the model, we assume that the IMBH gravitationally dominates within the region given by the tidal (Hill) radius, where we assume that its distance from Sgr~A* is $d_{\rm IRS13}\sim 0.1\,{\rm pc}$,

\begin{align}
    r_{\rm t}&\sim d_{\rm IRS13} \left(\frac{m_{\rm IMBH}}{3M_{\rm SgrA*}} \right)^{1/3}\,\notag\\
    &\sim 0.01 \left(\frac{d_{\rm IRS13}}{0.1\,{\rm pc}} \right)\left(\frac{m_{\rm IMBH}}{3\times 10^4\,M_{\odot}} \right)^{1/3}\,{\rm pc}\,,
\end{align}
where the IMBH mass $m_{\rm IMBH}$ is scaled to $3\times 10^4\,M_{\odot}$, which is given by the virial mass in Eq.~\ref{eq_virial_mass} using the velocity-resolved H30$\alpha$ line. The IMBH can effectively accrete the hot, X-ray emitting material from the region given by the Bondi-radius,
\begin{align}
    r_{\rm Bondi}&=\frac{2Gm_{\rm IMBH}}{c_{\rm s}^2}\\
    &\simeq 146 \left(\frac{m_{\rm IMBH}}{3\times 10^4\,M_{\odot}} \right)\left(\frac{T}{2.2\times 10^7\,{\rm K}} \right)^{-1}\,{\rm AU},
\end{align}
where the temperature is scaled to the best-fit value by \citet{Zhu2020}. The rotating ring structure corresponding to the recombination line H30$\alpha$ (see Fig.~\ref{fig:irs13e3_velocity_map}) represents a colder material that is captured from larger distances (e.g. stellar-wind material) and that circularizes on the scale of $R_{\rm gas}$.   

In analogy to Sgr~A*, we assume that the accretion flow onto the putative IMBH located close to the E3 region is dominated by advection, i.e. it is a hot, radiatively inefficient advection-dominated accretion flow \citep[ADAF; ][]{Narayan1998,2014ARA&A..52..529Y}. This assumption will be verified a posteriori from the comparison between the ADAF model and the broad-band continuum luminosities corresponding to the E3 region.

\begin{figure}
    \centering
    \includegraphics[width=\columnwidth]{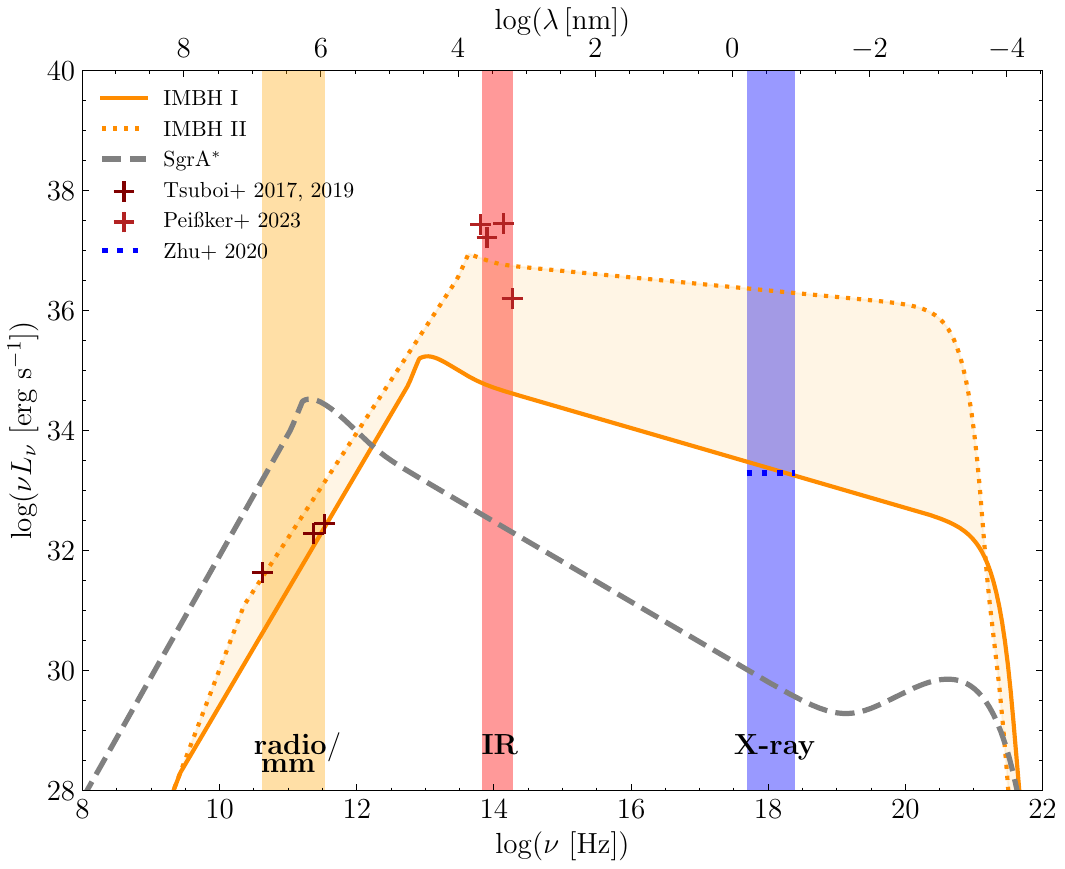}
    \caption{Broad-band spectral energy distribution (SED) of IRS 13 E3 source (data in the legend) in comparison with the SED of the advection-dominated accretion flow (ADAF) around the putative IMBH of $3\times 10^4\,M_{\odot}$. The solid orange line (IMBH I) corresponds to the hot accretion flow (ADAF) around the IMBH with the relative accretion rate of $\dot{m}=2\times 10^{-6}$, which is in agreement with the mm data at 232 and 340 GHz as well as the X-ray data in the 2-10 keV band. The dotted orange line (IMBH II) corresponds to the ADAF with $\dot{m}=10^{-4}$ that can capture better the radio data at 42 GHz as well as the infrared data (from H to M band); however, this model overpredicts the X-ray luminosity by three orders of magnitude. For comparison, we also show the SED of Sgr~A* (dashed gray line), whose peak is shifted to longer wavelengths in the mm domain. The X-ray measurement in the 2-10 keV band was adopted from \citet{Zhu2020}, infrared measurements were taken from \citet{peissker2023c}, and the mm and radio data are adopted according to \citet{tsuboi2019} and \citet{Tsuboi2017b}. Observed regions are highlighted in orange, red, and blue for radio/mm, infrared, and X-ray bands respectively. }
    \label{fig_IRS13E3_ADAF}
\end{figure}

We adopt the X-ray luminosity constraint in the 2-10 keV band from  \citet{Zhu2020}, while the infrared flux densities in the H, K, L, and M bands are taken from Paper I \citep{peissker2023c} {for which we used the peak intensity of IRS 13E3}. The low-frequency mm/radio flux densities at 340, 232, and 42 GHz are adopted from \citet{Tsuboi2017b} and \citet{tsuboi2019}. When combined, the broad-band continuum SED of E3 has a characteristic bump in the infrared domain, see Fig.~\ref{fig_IRS13E3_ADAF}.
To model the ADAF SED, we use the self-similar model of \citet{2021ApJ...923..260P} \citep[see also][for details]{1995ApJ...452..710N,1997ApJ...477..585M} that takes into account cooling by advection and radiation due to synchrotron, inverse Compton effect, and thermal bremsstrahlung radiation\footnote{The python script \texttt{SEDmodel.py} is available at \url{https://github.com/dpesce/LLAGNSED}.}. The solved energy-balance equation is as follows,
\begin{equation}
     \delta Q_{\rm vis}^{+}+Q^{ie}=Q_{\rm rad}^{-}+Q^{\rm adv,e}\,,
     \label{eq_energy_balance}
\end{equation}
where on the left side, $ Q_{\rm vis}^{+}$ stands for the total viscous heating rate, $\delta Q_{\rm vis}^{+}$ represents the fraction of the viscous heating rate that goes into electrons, and $Q^{ie}$ stands for the energy transfer from ions to electrons due to Coulomb interaction. On the right side, $Q_{\rm rad}^{-}$ is the radiative cooling and $Q^{\rm adv,e}$ is the advective rate of the electron energy into the black hole. The radiative cooling term $Q_{\rm rad}^{-}=P_{\rm synch}+P_{\rm Comp}+P_{\rm brems}$ consists of the radiative power due to synchrotron, inverse Compton, and bremsstrahlung processes.

\begin{figure}
    \centering
    \includegraphics[width=\columnwidth]{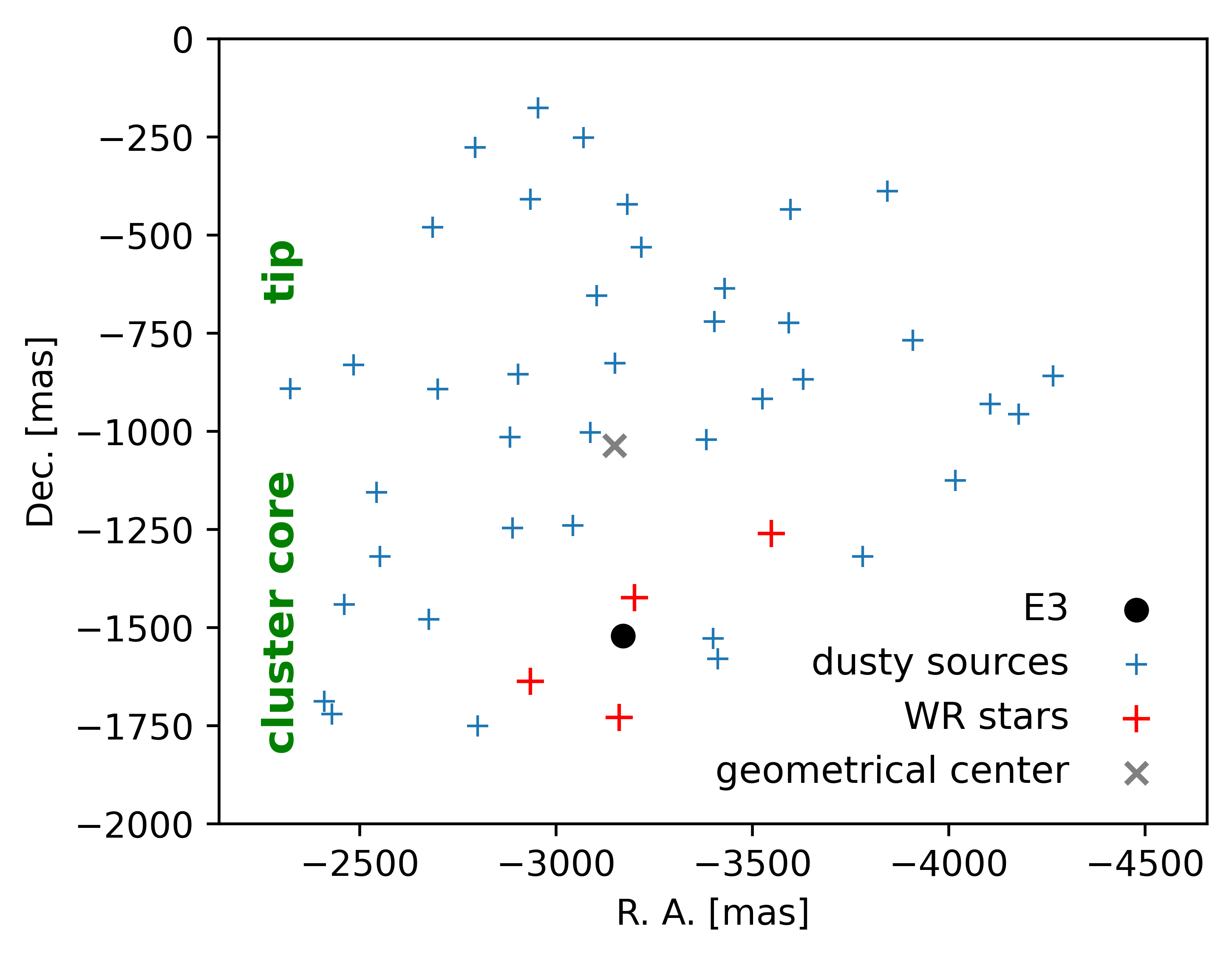}
    \caption{On-sky projected scatter map of the resolved sources of the IRS 13 cluster showing the position of the E3 star (black dot). The figure also indicates the positions of all other known E stars (red +), which are projected around the position of E3. Blue symbols mark the position of the dusty objects, while the geometric center of all cluster members is represented by a black $\times$. If the putative IMBH coincides with the position of the E3 source, we would expect a homogeneous distribution of the massive WR stars around its location. As shown, this is the case for the cluster members of IRS 13. However, the apparent distance between the geometrical center of all cluster members and the proposed position of a putative IMBH implies a complex setup that will be addressed in Labaj et al. (in prep.). The different components (core region and tip) of the northern part of the cluster are indicated and adapted from Fig. \ref{fig:irs13_dimensions} and Paper I.}
    \label{fig:scattermap}
\end{figure}

The X-ray emission as well as the mm emission is reproduced by the ADAF model (IMBH I, see the solid line in Fig.~\ref{fig_IRS13E3_ADAF}) with the IMBH mass of $3\times 10^4\,M_{\odot}$ and the relative accretion rate of $\dot{m}=2\times 10^{-6}$, which is defined as $\dot{m}=\dot{m}_{\rm acc}/\dot{m}_{\rm Edd}$. For the radiative efficiency of $\epsilon=0.1$ and $m_{\rm IMBH}=3\times 10^4\,M_{\odot}$, the Eddington accretion rate is $\dot{m}_{\rm Edd}=4 \pi G m_{\rm IMBH} m_{\rm p}/(\epsilon \sigma_{\rm T}c)\sim 6.6 \times 10^{-4}\,M_{\odot}\,{\rm yr^{-1}}$, where $m_{\rm p}$ denotes the proton mass, $\sigma_{\rm T}$ is the Thomson cross-section, and $c$ is the light speed. The IMBH I case does not reproduce well the prominent infrared emission. To reach the infrared luminosities as well as the radio luminosity at 42 GHz, the accretion rate would have to reach $\dot{m}=10^{-4}$ (IMBH II; see the dotted line in Fig.~\ref{fig_IRS13E3_ADAF}), which, however, overpredicts the X-ray luminosity by three orders of magnitude. Hence, we can constrain the relative accretion rate of the putative E3 IMBH to $10^{-4}\lesssim \dot{m}\lesssim 2\times 10^{-6}$, which is low enough for ADAF, and hence the initial assumption is justified. The case IMBH I with the lower accretion rate appears more plausible since the infrared emission is likely enhanced by dust and stellar emission; the accreting IMBH just contributes to it at a lower level. In that case, the actual accretion rate is $\dot{m}_{\rm acc}=\dot{m}\dot{m}_{\rm Edd}\sim 1.3\times 10^{-9}M_{\odot}\,{\rm yr^{-1}}$. Hence, if we have $N_{\rm WR}\sim 4$ prominent WR stars around the E3 source (Fig. \ref{fig:scattermap}) with each having a mass-loss rate of $\dot{m}_{\rm w}\sim 10^{-5}M_{\odot}\,{\rm yr^{-1}}$, only the fraction $\eta_{\rm acc}=\dot{m}_{\rm acc}/(N_{\rm WR}\dot{m}_{\rm w})\sim 3.3 \times 10^{-5}$ is eventually accreted by the putative IMBH. The rest is contributing to an outflow. Besides the quantitative agreement with the SED constraints, we also stress the qualitative characteristic of the IMBH ADAF SED, which peaks in the mid-IR domain, i.e. at shorter wavelengths than the SED of Sgr~A* that peaks in the mm domain, see the dashed gray line in Fig.~\ref{fig_IRS13E3_ADAF}. Specifically for $m_{\rm IMBH}=3\times 10^4\,M_{\odot}$ and $\dot{m}=2\times 10^{-6}$, the SED has a luminosity peak at $\lambda\sim 28\,{\rm \mu m}$, which is due to the equal contribution of the optically thin synchrotron emission and the inverse Compton radiation. 
A qualitative as well as an approximately quantitative agreement between the model of the accreting IMBH with the broad-band spectral data at the position of the E3 source provides tentative support for the hypothetical IMBH \citep{Fragione2022}. However, for the confirmation of the IMBH presence in IRS 13, a more detailed multiwavelength timing and spectral analysis is required in combination with the study of the rotating ionized material, as demonstrated by the gas ring traced by the H30$\alpha$ line in Fig.~\ref{fig:irs13e3_velocity_map} and by \citet{Tsuboi2017b} and \citet{tsuboi2019}. {It should be noted that the expected variability timescale of the proposed IMBH is on the order of a minute. No current X-ray instrument is sensitive enough to provide access to the necessary timescales. However, we will expand on these points in a forthcoming publication (Labaj et al., in prep.). }



\subsection{Limitations of the analysis}
\label{sec:discuss_limitations}

The Keplerian solutions for the projected on-sky positions of the investigated sources lack radial velocities. While comparable studies that investigate the IRS 13 sources suffer from the same limitation \citep{Schoedel2005, muzic2008, Eckart2013}, we want to discuss a possible impact on the results presented here. 
Naturally, an additional observational parameter, such as the LOS velocity, decreases the uncertainty for the Keplerian elements (especially the periapse passage) for stellar objects such as the S-stars in the S-cluster. However, this comparison is weak, since some stars in the S-cluster show proper motions of several 1000 km/s \citep{Peissker2022}, the IRS 13 cluster members exhibit velocities of a few tens-hundreds km/s (Paper I).
In Fig. \ref{fig:cluster_structure}, we present the results of the LOAN and the inclination for all investigated IRS 13 cluster members. It is evident that the impact of the missing radial velocity on the inclination is negligible because the absolute value is expected to be in the same order as the values listed in Table  \ref{tab:orbit_table_all_ds}-\ref{tab:final_orbit_table}. 
For the LOAN, the situation is more complex, since it defines the alignment of a given orbit, which results in a retrograde or prograde motion of the investigated object. However, we can extract the retrograde or prograde motion of the cluster members from the fits shown in Fig. \ref{fig:orbits_1} where we indicate the direction of the investigated trajectories with arrows. {Inspecting Table \ref{tab:orbit_table} suggests, that the LOAN is not affected by choosing an orbit with a poor $\chi^2$ Keplerian solution. From the distribution of the reduced $\chi^2$, we find a similar trend for the majority of the investigated sources.}
\begin{figure}%
	\centering
	\includegraphics[width=.5\textwidth]{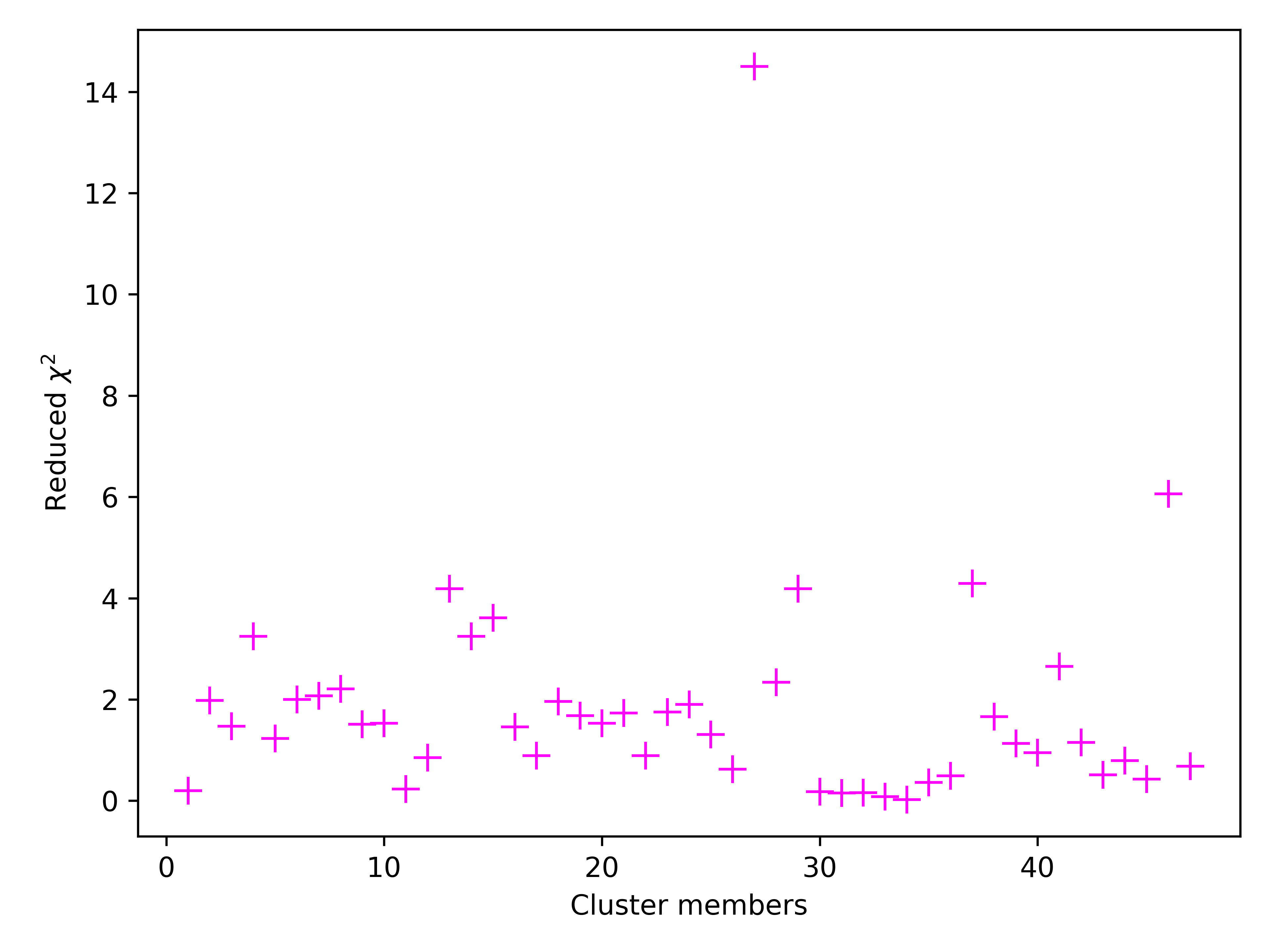}
	\caption{Distribution of the reduced $\chi^2$. The corresponding values for the cluster members are adapted from Tables \ref{tab:orbit_table}-\ref{tab:orbit_table_estars}.}
\label{fig:chi_distribution}
\end{figure}
{In addition to the above discussion, we guide the interested reader to Appendix \ref{sec:parameter_space_e_a_app} for a short discussion about possible values for the semi-major axis a and the eccentricity e.
To check the validity of the plots discussed in this work}, we refer to the orbit plots of all investigated sources of the IRS 13 cluster that are released along with this manuscript (Appendix \ref{sec:sup_mat_app}).\newline
Based on this discussion, we expect that neither the absolute value of the inclination or the retrograde and prograde motion of the investigated cluster members is affected by the radial velocity. In addition, we do not expect a significantly different result for the position angle as shown in Fig. \ref{fig:cluster_structure_pa} because the orientation relative to Sgr~A* is independent of the inclination or the retrograde and prograde motion.

\subsection{Alternative interpretations}

Here, we want to discuss another possible interpretation of the presented data. While we favor the scenario of an inspiralling cluster based on the comprehensive analysis presented in Paper I and this work, we cannot exclude the possibility of rejuvenation. This process describes an apparent young age of older stars that underwent merging events, as is proposed for stars in the inner parsec \citep{stephan2019}. As discussed, for example, in \cite{Genzel2003}, the presence of super-blue stragglers may be plausible and explain the relatively young age of the stars in the S-cluster \citep{Habibi2017}. This scenario could also be transferred to the IRS 13 cluster, although some of the E-stars are categorized as WR stars. In particular, WN8 and WC9 type stars \citep{Paumard2006} that can be associated with stellar masses between 10-15 M$_{\odot}$ which is in the same order as the S-stars. Hence, rejunivated stars in the IRS 13 cluster cannot be ruled out, which impacts our classification as an inspiralling cluster in the first place.
Furthermore, \cite{Zajacek2020} argues that the depletion of red giants through a possible jet interaction with Sgr~A* leads to stars, that mimic the emission of bluer and younger stellar sources. Although this hypothesis is limited to the S-cluster, it cannot be ruled out that the IRS 13 cluster may have been closer ($\sim 40 mpc$) to Sgr~A*. However, this scenario again demands an inspiralling cluster that undergoes a dynamical migration.
Still, an exact solution is challenging, but it could be addressed in future high-resolution observations by identifying close binary systems to qualitatively derive a relation between the multiplicity fraction and the relaxed stars \citep{Alexander2014}. With this knowledge, one could estimate the exact stellar age of the S-cluster but also IRS 13 without the chance of "fake" young stars that have been rejuvenated by merging processes \citep{stephan2019}. So far, theoretical models show that the relaxation timescales for possible binary systems are in the range of $10^8-10^{10}$ yr \citep{Alexander2014} whereas rejuvenation can take place in the same time frame \citep{Genzel2003}. However, \citep{Pfuhl2014} identified a massive binary system at a distance of 0.1 pc to Sgr~A* and and estimated age of $~10^6$yr which is 2-3 magnitudes lower than the time frame for possible rejuvenation processes. Although this binary is not part of IRS 13, it is an example of how to distinguish between young stars and older stars that appear younger. Future surveys with ERIS (VLT), MIRI (JWST), GRAVITY (VLTI), MATISSE (VLTI) and METIS (ELT) will put tight constraints on the age of the stars and their habitat by identifying possible binary systems. {In addition to the classification of IRS 13 displayed in Fig. \ref{fig:irs13_dimensions}, a comprehensive global point of view on the dynamical behavior of the cluster is required to confirm the scenario proposed here of an inspiralling cluster.} Speculatively, the recent observation of the globular cluster VVV CL002 could serve as a link between the inner parsec region and distances of about 300 pc \citep{Minniti2023}.\newline
In addition to these scenarios, an even more profound possibility of an alternative explanation for IRS 13 should be discussed. The chance association that would affect the classification of IRS 13 in the first place. Although \cite{muzic2008}, \cite{EckartAA2013}, and \cite{peissker2023b} discussed these scenarios, we want to revise the idea of a chance association for at least some IRS 13 cluster members. Especially when it comes to binaries, this problem is eminent and often arises when the data baseline is limited \citep[][]{Oudmaijer2010, Longhitano2010, Martayan2016}. In the following, we will distinguish between back- and foreground contamination.

\subsubsection{Background stars}

\begin{figure}%
	\centering
	\includegraphics[width=.5\textwidth]{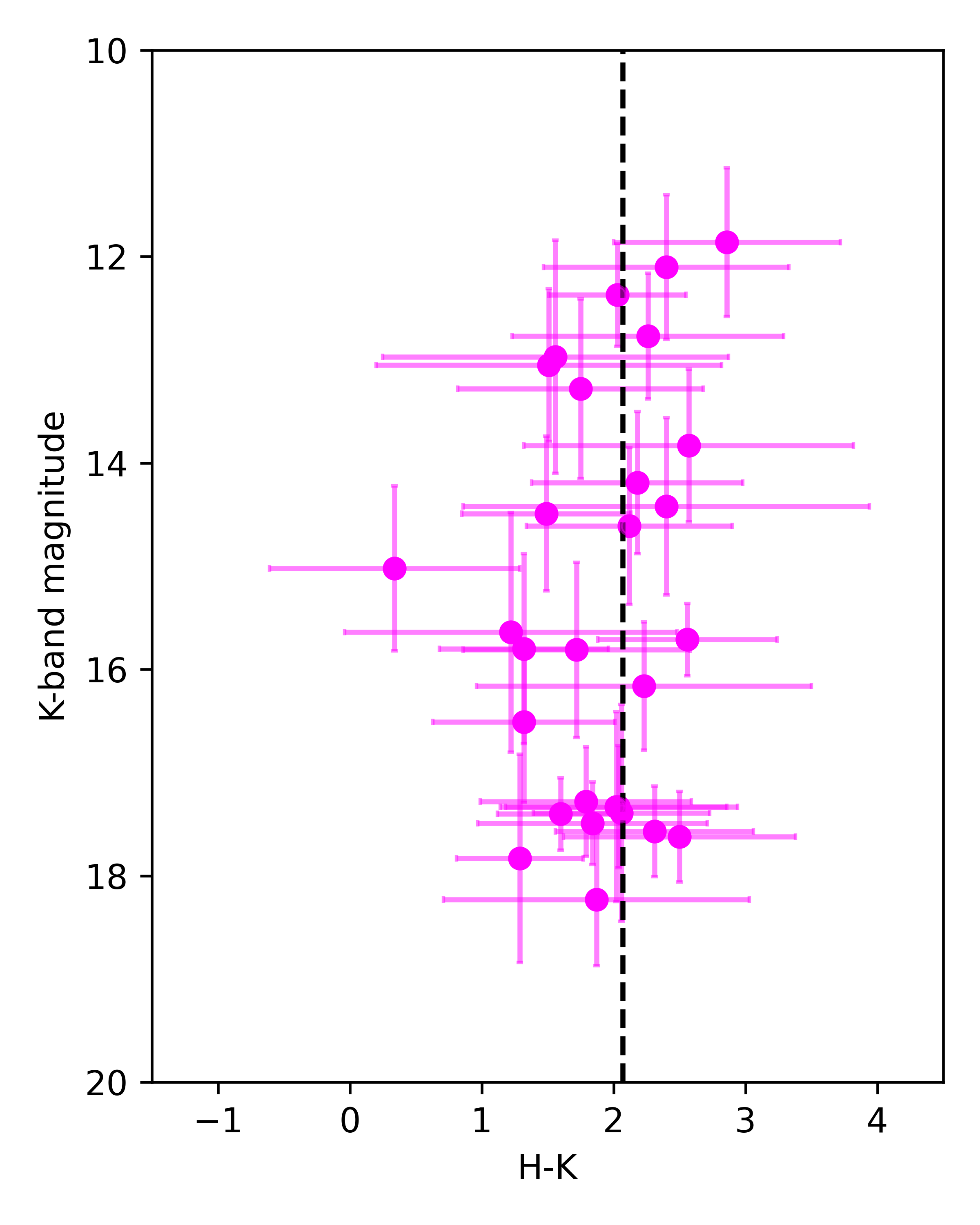}
	\caption{{Near-infrared K-band magnitudes as a function of H-K colors of the new IRS 13 cluster members analyzed in Paper I and this work. The dashed line indicates the mean H-K color of all stars in the NSC and is adapted from \cite{Schoedel2010}. Considering the uncertainty range of the estimated colors (see Paper I), we do not find any significant outliers.}}
\label{fig:cluster_membership}
\end{figure}

{Taking into account the extinction maps presented in \cite{Schoedel2010}, we find a rather constant reddening along IRS 13 in agreement with the analysis of \cite{Fritz2010, Fritz2011} who estimates $\rm A_K\,=\,3.6\pm 1.6$ for the K-band. Hence, the chance of stellar contamination is reduced \citep{Comeron2007}. In addition, the authors of \cite{Fritz2010} find that the probability of background stars at the position of IRS 13 is less than 0.1$\%$. Using the magnitudes of Paper I, we construct a K-band magnitude and H-K color diagram for the new IRS 13 members shown in Fig. \ref{fig:cluster_membership}.}
{In this diagram, we indicate the mean H-K color based on the analysis of \cite{Schoedel2010} who investigates all stars in the NSC using the same NACO set as discussed in this work. On the basis of the current data set and the necessary condition for contamination for the H-K colors of $\ll\,2.07$, we did not find photometric outliers that would suggest a population of background stars that significantly impact the analysis. Despite the possibility of a bright background star, we conclude that the majority of objects in the IRS 13 cluster are most plausible not background stars, in agreement with the literature \citep{muzic2008, Fritz2010}. A possible superposition of random stars in the line of sight will be discussed in the next section.}

\subsubsection{Foreground stars}

\begin{figure}%
	\centering
	\includegraphics[width=.5\textwidth]{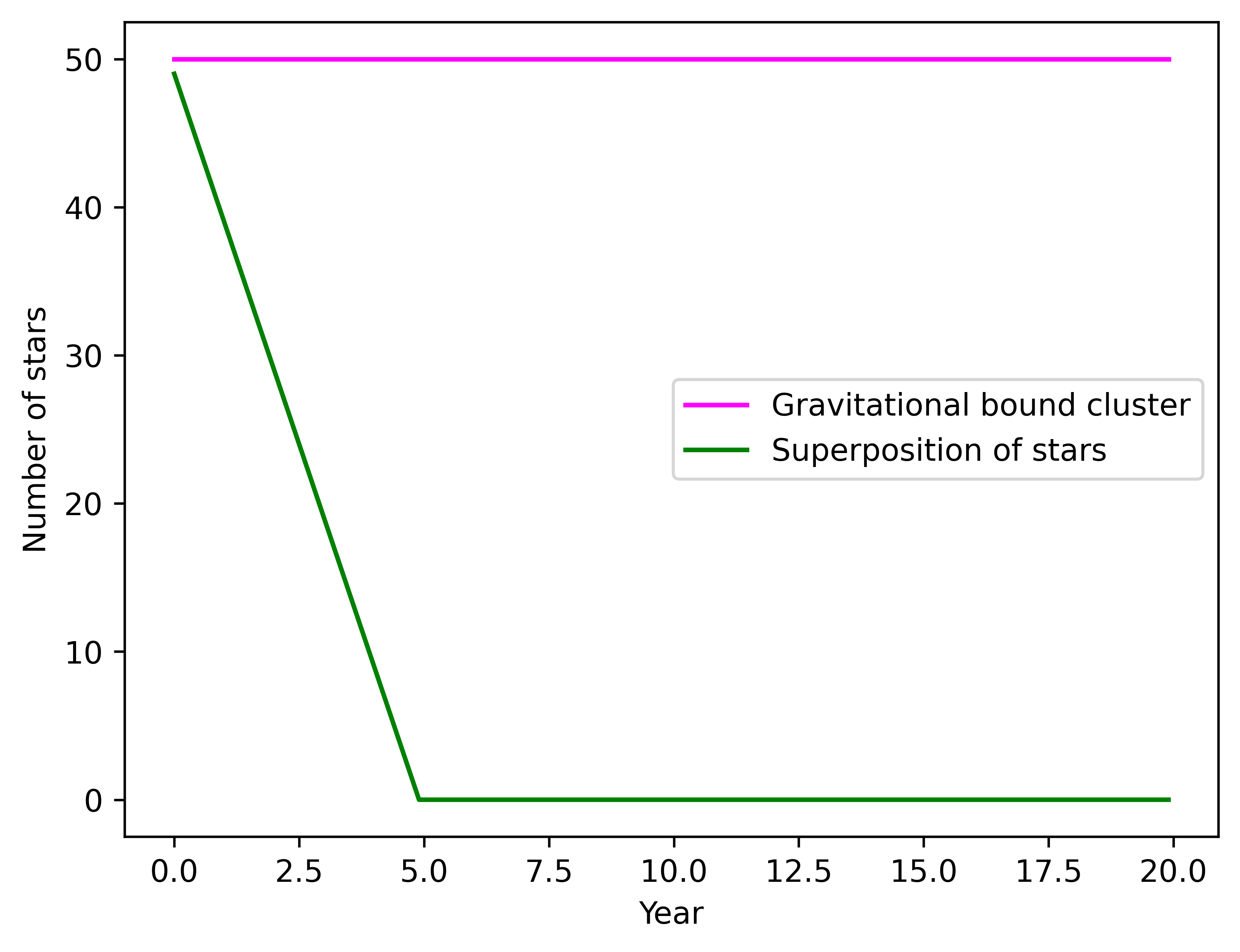}
	\caption{Monte Carlo simulations of a gravitational bound cluster and a group of stars that are in superposition with IRS 13. These stars act as a cluster that dissolves completely after 5 years.}
\label{fig:superpostion}
\end{figure}

{The data set investigated in this work covers almost two decades, which reduces the possibility of contamination by a foreground star. However, we want to take a closer look at the possibility of a superposition of several stars that may act as a fake cluster. For this, we model a foreground fake cluster that consists of stars with a randomized velocity and direction. Due to the decreased distance of this fake cluster towards earth, the cluster is smaller than the measured size of the core region of IRS 13 which we set to 1 arcsec (Fig. \ref{fig:irs13_finding_chart}). The "real" size of the fake cluster is 0.75 arcsec at a distance of 2 kpc away from the GC. Assuming a distance of 8 kpc for the GC, this fake cluster would be the size of IRS 13. In Fig. \ref{fig:superpostion}, we show Monte Carlo simulations that compare a gravitational bound cluster with a group of stars that are in superpositon with IRS 13.}
{Based on the presented figure, it is evident that the fake cluster dissolves after 5 years in agreement with the confusion arguments for two stars outlined in \cite{Sabha2012} and \cite{Eckart2013}. Both authors estimate that stellar confusion due to background or foreground stars is on the order of a few years.}
\section{Conclusions} 
\label{sec:conclusion}

In addition to the manifold results shown in Paper I, we investigate the possibility of a Keplerian solution that represents the trajectory of the cluster members of IRS 13. Assuming a circular motion of the cluster, we find with the proper motion and the LOS velocity a 3d distance of about 0.12 pc for IRS 13 to Sgr~A*. With this distance estimate, we can safely assume that the IRS 13 cluster is gravitationally bound to Sgr~A*, i.e. it is inside the sphere of gravitational influence of Sgr~A* ($\sim 2$ pc). The hot plasma as well as colder gas and dust of IRS 13 can also provide material for the accretion flow of Sgr A* since they lie inside the Sgr~A* Bondi radius of $\sim 0.2\,{\rm pc}$.
Therefore, we assume that all cluster members are gravitationally bound to Sgr~A* which justifies the inference of Keplerian orbits that cover a data baseline of about 20 years (Paper I). {Despite the lack of radial velocities and the uncertainties that arise from this, we find that contamination by fore and background stars is highly unlikely. Based on our analysis, we formulate} the following key results:
\begin{itemize}
    \item We argue that the true dimensions of IRS 13 are larger than it was previously known in the literature and the cluster expands to the south in projection,
    \item With these derived dimensions, the nature of IRS 13 can be described as an evaporating cluster that suffers from the tidal interaction with Sgr~A*, resulting in a stretched tail,
    \item IRS 13 is the remnant of a massive star-formation cluster, in which the first generation of stars started to form most likely in the region of the current CND,
    \item The second generation of stellar objects, i.e., the investigated dusty sources, were formed during the inspiral motion of IRS 13,
    \item Our simulations show, that a cluster with the derived properties for IRS 13 in Paper I and this work can migrate into the {\it inner parsec},
    \item The simulations suggest that especially embedded high-mass stars can undergo a migration from the CND to the NSC,
    \item {Independent of the number of particles used in the simulations, the migration timescales from the CND in the NSC is $\sim\,0.3$Myr,}
    \item In agreement with the literature, we propose a fragmenting dense bow-shock shell as {a potential birthplace for the candidate YSOs},
    \item The orbital distribution of the DS sources is not randomized as it is proposed for other regions in the NSC,
    \item The majority of the investigated cluster members of IRS 13 are part of at least one disk that is not related to the clockwise disk or the counter-clockwise disk,
    \item The massive and embedded cluster IRS 13 forms an individual structure on the basis of about 50 individual stellar orbits.
    \item Using the H30$\alpha$-line emitting gas kinematics and the spectral energy model of an ADAF around the IMBH of $3\times 10^4\,M_{\odot}$, we provide tentative support for the hypothetical presence of the IMBH in the IRS 13 E3 region.
\end{itemize}
With the upcoming James Webb Space Telescope (JWST) and the Enhanced Resolution Imaging Spectrograph (ERIS) observations, we will test our hypothesis of the putative presence of an IMBH and further verify the nature of the cluster members. In addtion, we will further decrease the limitations of the Keplerian elements by incorporating the radial velocity for the dusty sources. {Due to the 3d distance of the IRS 13 cluster to Sgr~A*, we estimated that these limitations on the eccentricity e and semi-major axis a are in a reasonable uncertainty range of about 5-10$\%$.} As for the H30$\alpha$ line distribution presented in this work, the IFU data provided by JWST and ERIS will be another cornerstone for studying the various components of the IRS 13 cluster which we addressed in Paper I and this work.

\begin{acknowledgments}
FP and LL gratefully acknowledges the Collaborative Research Center 1601 funded by the Deutsche Forschungsgemeinschaft (DFG, German Research Foundation) – SFB 1601 [sub-project A3] - 500700252. 
We acknowledge support for the Article Processing Charge from the DFG (German Research Foundation, 491454339).
MZ is grateful to the GA\v{C}R EXPRO grant no. GX21-13491X for financial support while working on the project. ML acknowledges the GA\v{C}R-LA grant No. GF23-04053L for financial support. Part of this work was supported by fruitful discussions with members of the European Union funded COST Action MP0905: Black Holes in a Violent Universe and the Czech Science Foundation (No.\ 21-06825X). VK has been partially supported by the Czech Ministry of Education, Youth and Sports Research Infrastructure (LM2023047). AP, SE, JC, and GB contributed useful points to the discussion. We also would like to thank the members of the SINFONI/NACO/VISIR and ESO's Paranal/Chile team for their support and collaboration. This paper makes use of the following ALMA data: ADS/JAO.ALMA$\#$2016.1.00870.S and ADS/JAO.ALMA$\#$2015.1.01080.S. ALMA is a partnership of ESO (representing its member states), NSF (USA) and NINS (Japan), together with NRC (Canada), MOST and ASIAA (Taiwan), and KASI (Republic of Korea), in cooperation with the Republic of Chile. The Joint ALMA Observatory is operated by ESO, AUI/NRAO and NAOJ.
\end{acknowledgments}
\vspace{5mm}
\facilities{VLT (SINFONI and NACO), ALMA (Band 7)}

\software{astropy \citep{2013A&A...558A..33A,2018AJ....156..123A},
          SciPy \citep{SciPy2020},
          Hyperion \citep{Robitaille2011, Robitaille2017},
          DPuser \citep{Ott2013}
          }

\bibliography{bib}{}

\begin{thebibliography}{}
\expandafter\ifx\csname natexlab\endcsname\relax\def\natexlab#1{#1}\fi
\providecommand{\url}[1]{\href{#1}{#1}}
\providecommand{\dodoi}[1]{doi:~\href{http://doi.org/#1}{\nolinkurl{#1}}}
\providecommand{\doeprint}[1]{\href{http://ascl.net/#1}{\nolinkurl{http://ascl.net/#1}}}
\providecommand{\doarXiv}[1]{\href{https://arxiv.org/abs/#1}{\nolinkurl{https://arxiv.org/abs/#1}}}

\bibitem[{{Alexander} \& {Pfuhl}(2014)}]{Alexander2014}
{Alexander}, T., \& {Pfuhl}, O. 2014, \apj, 780, 148, \dodoi{10.1088/0004-637X/780/2/148}

\bibitem[{{Ali} {et~al.}(2020){Ali}, {Paul}, {Eckart}, {Parsa}, {Zajacek}, {Pei{\ss}ker}, {Subroweit}, {Valencia-S.}, {Thomkins}, \& {Witzel}}]{Ali2020}
{Ali}, B., {Paul}, D., {Eckart}, A., {et~al.} 2020, \apj, 896, 100, \dodoi{10.3847/1538-4357/ab93ae}

\bibitem[{{Astropy Collaboration} {et~al.}(2013){Astropy Collaboration}, {Robitaille}, {Tollerud}, {Greenfield}, {Droettboom}, {Bray}, {Aldcroft}, {Davis}, {Ginsburg}, {Price-Whelan}, {Kerzendorf}, {Conley}, {Crighton}, {Barbary}, {Muna}, {Ferguson}, {Grollier}, {Parikh}, {Nair}, {Unther}, {Deil}, {Woillez}, {Conseil}, {Kramer}, {Turner}, {Singer}, {Fox}, {Weaver}, {Zabalza}, {Edwards}, {Azalee Bostroem}, {Burke}, {Casey}, {Crawford}, {Dencheva}, {Ely}, {Jenness}, {Labrie}, {Lim}, {Pierfederici}, {Pontzen}, {Ptak}, {Refsdal}, {Servillat}, \& {Streicher}}]{2013A&A...558A..33A}
{Astropy Collaboration}, {Robitaille}, T.~P., {Tollerud}, E.~J., {et~al.} 2013, \aap, 558, A33, \dodoi{10.1051/0004-6361/201322068}

\bibitem[{{Astropy Collaboration} {et~al.}(2018){Astropy Collaboration}, {Price-Whelan}, {Sip{\H{o}}cz}, {G{\"u}nther}, {Lim}, {Crawford}, {Conseil}, {Shupe}, {Craig}, {Dencheva}, {Ginsburg}, {VanderPlas}, {Bradley}, {P{\'e}rez-Su{\'a}rez}, {de Val-Borro}, {Aldcroft}, {Cruz}, {Robitaille}, {Tollerud}, {Ardelean}, {Babej}, {Bach}, {Bachetti}, {Bakanov}, {Bamford}, {Barentsen}, {Barmby}, {Baumbach}, {Berry}, {Biscani}, {Boquien}, {Bostroem}, {Bouma}, {Brammer}, {Bray}, {Breytenbach}, {Buddelmeijer}, {Burke}, {Calderone}, {Cano Rodr{\'\i}guez}, {Cara}, {Cardoso}, {Cheedella}, {Copin}, {Corrales}, {Crichton}, {D'Avella}, {Deil}, {Depagne}, {Dietrich}, {Donath}, {Droettboom}, {Earl}, {Erben}, {Fabbro}, {Ferreira}, {Finethy}, {Fox}, {Garrison}, {Gibbons}, {Goldstein}, {Gommers}, {Greco}, {Greenfield}, {Groener}, {Grollier}, {Hagen}, {Hirst}, {Homeier}, {Horton}, {Hosseinzadeh}, {Hu}, {Hunkeler}, {Ivezi{\'c}}, {Jain}, {Jenness}, {Kanarek}, {Kendrew}, {Kern}, {Kerzendorf}, {Khvalko}, {King}, {Kirkby}, {Kulkarni},
  {Kumar}, {Lee}, {Lenz}, {Littlefair}, {Ma}, {Macleod}, {Mastropietro}, {McCully}, {Montagnac}, {Morris}, {Mueller}, {Mumford}, {Muna}, {Murphy}, {Nelson}, {Nguyen}, {Ninan}, {N{\"o}the}, {Ogaz}, {Oh}, {Parejko}, {Parley}, {Pascual}, {Patil}, {Patil}, {Plunkett}, {Prochaska}, {Rastogi}, {Reddy Janga}, {Sabater}, {Sakurikar}, {Seifert}, {Sherbert}, {Sherwood-Taylor}, {Shih}, {Sick}, {Silbiger}, {Singanamalla}, {Singer}, {Sladen}, {Sooley}, {Sornarajah}, {Streicher}, {Teuben}, {Thomas}, {Tremblay}, {Turner}, {Terr{\'o}n}, {van Kerkwijk}, {de la Vega}, {Watkins}, {Weaver}, {Whitmore}, {Woillez}, {Zabalza}, \& {Astropy Contributors}}]{2018AJ....156..123A}
{Astropy Collaboration}, {Price-Whelan}, A.~M., {Sip{\H{o}}cz}, B.~M., {et~al.} 2018, \aj, 156, 123, \dodoi{10.3847/1538-3881/aabc4f}

\bibitem[{{Barnes} \& {Hut}(1986)}]{Barnes1986}
{Barnes}, J., \& {Hut}, P. 1986, \nat, 324, 446, \dodoi{10.1038/324446a0}

\bibitem[{{Bartko} {et~al.}(2008){Bartko}, {Eisenhauer}, {Fritz}, {Genzel}, {Gillessen}, {Martins}, {Ott}, {Paumard}, {Pfuhl}, \& {Trippe}}]{Bartko2008}
{Bartko}, H., {Eisenhauer}, F., {Fritz}, T., {et~al.} 2008, in Journal of Physics Conference Series, Vol. 131, Journal of Physics Conference Series, 012010, \dodoi{10.1088/1742-6596/131/1/012010}

\bibitem[{{Bartko} {et~al.}(2009){Bartko}, {Martins}, {Fritz}, {Genzel}, {Levin}, {Perets}, {Paumard}, {Nayakshin}, {Gerhard}, {Alexander}, {Dodds-Eden}, {Eisenhauer}, {Gillessen}, {Mascetti}, {Ott}, {Perrin}, {Pfuhl}, {Reid}, {Rouan}, {Sternberg}, \& {Trippe}}]{2009ApJ...697.1741B}
{Bartko}, H., {Martins}, F., {Fritz}, T.~K., {et~al.} 2009, \apj, 697, 1741, \dodoi{10.1088/0004-637X/697/2/1741}

\bibitem[{{Becklin} {et~al.}(1978){Becklin}, {Matthews}, {Neugebauer}, \& {Willner}}]{Becklin1978}
{Becklin}, E.~E., {Matthews}, K., {Neugebauer}, G., \& {Willner}, S.~P. 1978, \apj, 219, 121, \dodoi{10.1086/155761}

\bibitem[{{Blum} {et~al.}(1996){Blum}, {Sellgren}, \& {Depoy}}]{Blum1996}
{Blum}, R.~D., {Sellgren}, K., \& {Depoy}, D.~L. 1996, \apj, 470, 864, \dodoi{10.1086/177917}

\bibitem[{{Boily} \& {Spurzem}(1999)}]{Boily1999}
{Boily}, C.~M., \& {Spurzem}, R. 1999, in Astronomische Gesellschaft Abstract Series, Vol.~15, Astronomische Gesellschaft Abstract Series, 32

\bibitem[{{Bonnell} \& {Rice}(2008)}]{Bonnell2008}
{Bonnell}, I.~A., \& {Rice}, W.~K.~M. 2008, Science, 321, 1060, \dodoi{10.1126/science.1160653}

\bibitem[{{Buchholz} {et~al.}(2013){Buchholz}, {Witzel}, {Sch{\"o}del}, \& {Eckart}}]{Buchholz2013}
{Buchholz}, R.~M., {Witzel}, G., {Sch{\"o}del}, R., \& {Eckart}, A. 2013, \aap, 557, A82, \dodoi{10.1051/0004-6361/201220338}

\bibitem[{{Chernoff} \& {Shapiro}(1987)}]{Chernoff1987}
{Chernoff}, D.~F., \& {Shapiro}, S.~L. 1987, \apj, 322, 113, \dodoi{10.1086/165708}

\bibitem[{{Ciurlo} {et~al.}(2016){Ciurlo}, {Paumard}, {Rouan}, \& {Cl{\'e}net}}]{Ciurlo2016}
{Ciurlo}, A., {Paumard}, T., {Rouan}, D., \& {Cl{\'e}net}, Y. 2016, \aap, 594, A113, \dodoi{10.1051/0004-6361/201527173}

\bibitem[{{Ciurlo} {et~al.}(2023){Ciurlo}, {Campbell}, {Morris}, {Do}, {Ghez}, {Becklin}, {Bentley}, {Chu}, {Gautam}, {Gursahani}, {Hees}, {O'Neil}, {Lu}, {Martinez}, {Naoz}, {Sakai}, \& {Sch{\"o}del}}]{Ciurlo2023}
{Ciurlo}, A., {Campbell}, R.~D., {Morris}, M.~R., {et~al.} 2023, \apj, 944, 136, \dodoi{10.3847/1538-4357/acb344}

\bibitem[{{Comer{\'o}n} \& {Schneider}(2007)}]{Comeron2007}
{Comer{\'o}n}, F., \& {Schneider}, N. 2007, \aap, 473, 149, \dodoi{10.1051/0004-6361:20077733}

\bibitem[{{Comer{\'o}n} {et~al.}(2005){Comer{\'o}n}, {Schneider}, \& {Russeil}}]{Comeron2005}
{Comer{\'o}n}, F., {Schneider}, N., \& {Russeil}, D. 2005, \aap, 433, 955, \dodoi{10.1051/0004-6361:20041586}

\bibitem[{{Cutri} {et~al.}(2003){Cutri}, {Skrutskie}, {van Dyk}, {Beichman}, {Carpenter}, {Chester}, {Cambresy}, {Evans}, {Fowler}, {Gizis}, {Howard}, {Huchra}, {Jarrett}, {Kopan}, {Kirkpatrick}, {Light}, {Marsh}, {McCallon}, {Schneider}, {Stiening}, {Sykes}, {Weinberg}, {Wheaton}, {Wheelock}, \& {Zacarias}}]{Cutri2003}
{Cutri}, R.~M., {Skrutskie}, M.~F., {van Dyk}, S., {et~al.} 2003, VizieR Online Data Catalog, II/246

\bibitem[{{Dinh} {et~al.}(2021){Dinh}, {Salas}, {Morris}, \& {Naoz}}]{Dinh2021}
{Dinh}, C.~K., {Salas}, J.~M., {Morris}, M.~R., \& {Naoz}, S. 2021, \apj, 920, 79, \dodoi{10.3847/1538-4357/ac185b}

\bibitem[{{Do} {et~al.}(2019){Do}, {Hees}, {Ghez}, {Martinez}, {Chu}, {Jia}, {Sakai}, {Lu}, {Gautam}, {O'Neil}, {Becklin}, {Morris}, {Matthews}, {Nishiyama}, {Campbell}, {Chappell}, {Chen}, {Ciurlo}, {Dehghanfar}, {Gallego-Cano}, {Kerzendorf}, {Lyke}, {Naoz}, {Saida}, {Sch{\"o}del}, {Takahashi}, {Takamori}, {Witzel}, \& {Wizinowich}}]{Do2019S2}
{Do}, T., {Hees}, A., {Ghez}, A., {et~al.} 2019, Science, 365, 664, \dodoi{10.1126/science.aav8137}

\bibitem[{{Eckart} {et~al.}(2004){Eckart}, {Moultaka}, {Viehmann}, {Straubmeier}, \& {Mouawad}}]{Eckart2004a}
{Eckart}, A., {Moultaka}, J., {Viehmann}, T., {Straubmeier}, C., \& {Mouawad}, N. 2004, \apj, 602, 760, \dodoi{10.1086/381178}

\bibitem[{{Eckart} {et~al.}(2013){Eckart}, {Mu{\v z}i{\'c}}, {Yazici}, {Sabha}, {Shahzamanian}, {Witzel}, {Moser}, {Garcia-Marin}, {Valencia-S.}, {Jalali}, {Bremer}, {Straubmeier}, {Rauch}, {Buchholz}, {Kunneriath}, \& {Moultaka}}]{Eckart2013}
{Eckart}, A., {Mu{\v z}i{\'c}}, K., {Yazici}, S., {et~al.} 2013, aap, 551, A18, \dodoi{10.1051/0004-6361/201219994}

\bibitem[{{Eckart, A.} {et~al.}(2013){Eckart, A.}, {Muzi\'{}c, K.}, {Yazici, S.}, {Sabha, N.}, {Shahzamanian, B.}, {Witzel, G.}, {Moser, L.}, {Garcia-Marin, M.}, {, M. Valencia-S.}, {Jalali, B.}, {Bremer, M.}, {Straubmeier, C.}, {Rauch, C.}, {Buchholz, R.}, {Kunneriath, D.}, \& {Moultaka, J.}}]{EckartAA2013}
{Eckart, A.}, {Muzi\'{}c, K.}, {Yazici, S.}, {et~al.} 2013, A\&A, 551, A18, \dodoi{10.1051/0004-6361/201219994}

\bibitem[{{Event Horizon Telescope Collaboration} {et~al.}(2022){Event Horizon Telescope Collaboration}, {Akiyama}, {Alberdi}, {Alef}, {Algaba}, {Anantua}, {Asada}, {Azulay}, {Bach}, {Baczko}, {Ball}, {Balokovi{\'c}}, {Barrett}, {Baub{\"o}ck}, {Benson}, {Bintley}, {Blackburn}, {Blundell}, {Bouman}, {Bower}, {Boyce}, {Bremer}, {Brinkerink}, {Brissenden}, {Britzen}, {Broderick}, {Broguiere}, {Bronzwaer}, {Bustamante}, {Byun}, {Carlstrom}, {Ceccobello}, {Chael}, {Chan}, {Chatterjee}, {Chatterjee}, {Chen}, {Chen}, {Cheng}, {Cho}, {Christian}, {Conroy}, {Conway}, {Cordes}, {Crawford}, {Crew}, {Cruz-Osorio}, {Cui}, {Davelaar}, {Laurentis}, {Deane}, {Dempsey}, {Desvignes}, {Dexter}, {Dhruv}, {Doeleman}, {Dougal}, {Dzib}, {Eatough}, {Emami}, {Falcke}, {Farah}, {Fish}, {Fomalont}, {Ford}, {Fraga-Encinas}, {Freeman}, {Friberg}, {Fromm}, {Fuentes}, {Galison}, {Gammie}, {Garc{\'\i}a}, {Gentaz}, {Georgiev}, {Goddi}, {Gold}, {G{\'o}mez-Ruiz}, {G{\'o}mez}, {Gu}, {Gurwell}, {Hada}, {Haggard}, {Haworth}, {Hecht}, {Hesper},
  {Heumann}, {Ho}, {Ho}, {Honma}, {Huang}, {Huang}, {Hughes}, {Ikeda}, {Impellizzeri}, {Inoue}, {Issaoun}, {James}, {Jannuzi}, {Janssen}, {Jeter}, {Jiang}, {Jim{\'e}nez-Rosales}, {Johnson}, {Jorstad}, {Joshi}, {Jung}, {Karami}, {Karuppusamy}, {Kawashima}, {Keating}, {Kettenis}, {Kim}, {Kim}, {Kim}, {Kim}, {Kino}, {Koay}, {Kocherlakota}, {Kofuji}, {Koch}, {Koyama}, {Kramer}, {Kramer}, {Krichbaum}, {Kuo}, {Bella}, {Lauer}, {Lee}, {Lee}, {Leung}, {Levis}, {Li}, {Lico}, {Lindahl}, {Lindqvist}, {Lisakov}, {Liu}, {Liu}, {Liuzzo}, {Lo}, {Lobanov}, {Loinard}, {Lonsdale}, {Lu}, {Mao}, {Marchili}, {Markoff}, {Marrone}, {Marscher}, {Mart{\'\i}-Vidal}, {Matsushita}, {Matthews}, {Medeiros}, {Menten}, {Michalik}, {Mizuno}, {Mizuno}, {Moran}, {Moriyama}, {Moscibrodzka}, {M{\"u}ller}, {Mus}, {Musoke}, {Myserlis}, {Nadolski}, {Nagai}, {Nagar}, {Nakamura}, {Narayan}, {Narayanan}, {Natarajan}, {Nathanail}, {Fuentes}, {Neilsen}, {Neri}, {Ni}, {Noutsos}, {Nowak}, {Oh}, {Okino}, {Olivares}, {Ortiz-Le{\'o}n}, {Oyama}, {{\"O}zel},
  {Palumbo}, {Paraschos}, {Park}, {Parsons}, {Patel}, {Pen}, {Pesce}, {Pi{\'e}tu}, {Plambeck}, {PopStefanija}, {Porth}, {P{\"o}tzl}, {Prather}, {Preciado-L{\'o}pez}, {Psaltis}, {Pu}, {Ramakrishnan}, {Rao}, {Rawlings}, {Raymond}, {Rezzolla}, {Ricarte}, {Ripperda}, {Roelofs}, {Rogers}, {Ros}, {Romero-Ca{\~n}izales}, {Roshanineshat}, {Rottmann}, {Roy}, {Ruiz}, {Ruszczyk}, {Rygl}, {S{\'a}nchez}, {S{\'a}nchez-Arg{\"u}elles}, {S{\'a}nchez-Portal}, {Sasada}, {Satapathy}, {Savolainen}, {Schloerb}, {Schonfeld}, {Schuster}, {Shao}, {Shen}, {Small}, {Sohn}, {SooHoo}, {Souccar}, {Sun}, {Tazaki}, {Tetarenko}, {Tiede}, {Tilanus}, {Titus}, {Torne}, {Traianou}, {Trent}, {Trippe}, {Turk}, {van Bemmel}, {van Langevelde}, {van Rossum}, {Vos}, {Wagner}, {Ward-Thompson}, {Wardle}, {Weintroub}, {Wex}, {Wharton}, {Wielgus}, {Wiik}, {Witzel}, {Wondrak}, {Wong}, {Wu}, {Yamaguchi}, {Yoon}, {Young}, {Young}, {Younsi}, {Yuan}, {Yuan}, {Zensus}, {Zhang}, {Zhao}, {Zhao}, {Agurto}, {Allardi}, {Amestica}, {Araneda}, {Arriagada}, {Berghuis},
  {Bertarini}, {Berthold}, {Blanchard}, {Brown}, {C{\'a}rdenas}, {Cantzler}, {Caro}, {Castillo-Dom{\'\i}nguez}, {Chan}, {Chang}, {Chang}, {Chang}, {Chang}, {Chen}, {Chilson}, {Chuter}, {Ciechanowicz}, {Colin-Beltran}, {Coulson}, {Crowley}, {Degenaar}, {Dornbusch}, {Dur{\'a}n}, {Everett}, {Faber}, {Forster}, {Fuchs}, {Gale}, {Geertsema}, {Gonz{\'a}lez}, {Graham}, {Gueth}, {Halverson}, {Han}, {Han}, {Hasegawa}, {Hern{\'a}ndez-Rebollar}, {Herrera}, {Herrero-Illana}, {Heyminck}, {Hirota}, {Hoge}, {Hostler Schimpf}, {Howie}, {Huang}, {Jiang}, {Jinchi}, {John}, {Kimura}, {Klein}, {Kubo}, {Kuroda}, {Kwon}, {Lacasse}, {Laing}, {Leitch}, {Li}, {Liu}, {Liu}, {Lin}, {Lu}, {Mac-Auliffe}, {Martin-Cocher}, {Matulonis}, {Maute}, {Messias}, {Meyer-Zhao}, {Monta{\~n}a}, {Montenegro-Montes}, {Montgomerie}, {Moreno Nolasco}, {Muders}, {Nishioka}, {Norton}, {Nystrom}, {Ogawa}, {Olivares}, {Oshiro}, {P{\'e}rez-Beaupuits}, {Parra}, {Phillips}, {Poirier}, {Pradel}, {Qiu}, {Raffin}, {Rahlin}, {Ram{\'\i}rez}, {Ressler}, {Reynolds},
  {Rodr{\'\i}guez-Montoya}, {Saez-Madain}, {Santana}, {Shaw}, {Shirkey}, {Silva}, {Snow}, {Sousa}, {Sridharan}, {Stahm}, {Stark}, {Test}, {Torstensson}, {Venegas}, {Walther}, {Wei}, {White}, {Wieching}, {Wijnands}, {Wouterloot}, {Yu}, {Yu (于威)}, \& {Zeballos}}]{eht2022}
{Event Horizon Telescope Collaboration}, {Akiyama}, K., {Alberdi}, A., {et~al.} 2022, \apjl, 930, L12, \dodoi{10.3847/2041-8213/ac6674}

\bibitem[{{Fragione} {et~al.}(2022){Fragione}, {Kocsis}, {Rasio}, \& {Silk}}]{Fragione2022}
{Fragione}, G., {Kocsis}, B., {Rasio}, F.~A., \& {Silk}, J. 2022, \apj, 927, 231, \dodoi{10.3847/1538-4357/ac5026}

\bibitem[{{Fritz} {et~al.}(2010){Fritz}, {Gillessen}, {Dodds-Eden}, {Martins}, {Bartko}, {Genzel}, {Paumard}, {Ott}, {Pfuhl}, {Trippe}, {Eisenhauer}, \& {Gratadour}}]{Fritz2010}
{Fritz}, T.~K., {Gillessen}, S., {Dodds-Eden}, K., {et~al.} 2010, \apj, 721, 395, \dodoi{10.1088/0004-637X/721/1/395}

\bibitem[{{Fritz} {et~al.}(2011){Fritz}, {Gillessen}, {Dodds-Eden}, {Lutz}, {Genzel}, {Raab}, {Ott}, {Pfuhl}, {Eisenhauer}, \& {Yusef-Zadeh}}]{Fritz2011}
---. 2011, ApJ, 737, 73, \dodoi{10.1088/0004-637X/737/2/73}

\bibitem[{{Gayley} {et~al.}(1997){Gayley}, {Owocki}, \& {Cranmer}}]{Gayley1997}
{Gayley}, K.~G., {Owocki}, S.~P., \& {Cranmer}, S.~R. 1997, \apj, 475, 786, \dodoi{10.1086/303573}

\bibitem[{{Genzel} {et~al.}(1996){Genzel}, {Thatte}, {Krabbe}, {Kroker}, \& {Tacconi-Garman}}]{Genzel1996}
{Genzel}, R., {Thatte}, N., {Krabbe}, A., {Kroker}, H., \& {Tacconi-Garman}, L.~E. 1996, \apj, 472, 153, \dodoi{10.1086/178051}

\bibitem[{{Genzel} {et~al.}(2003){Genzel}, {Sch{\"o}del}, {Ott}, {Eisenhauer}, {Hofmann}, {Lehnert}, {Eckart}, {Alexander}, {Sternberg}, {Lenzen}, {Cl{\'e}net}, {Lacombe}, {Rouan}, {Renzini}, \& {Tacconi-Garman}}]{Genzel2003}
{Genzel}, R., {Sch{\"o}del}, R., {Ott}, T., {et~al.} 2003, \apj, 594, 812, \dodoi{10.1086/377127}

\bibitem[{{Ghez} {et~al.}(2003){Ghez}, {Duch{\^e}ne}, {Matthews}, {Hornstein}, {Tanner}, {Larkin}, {Morris}, {Becklin}, {Salim}, {Kremenek}, {Thompson}, {Soifer}, {Neugebauer}, \& {McLean}}]{Ghez2003}
{Ghez}, A.~M., {Duch{\^e}ne}, G., {Matthews}, K., {et~al.} 2003, \apjl, 586, L127, \dodoi{10.1086/374804}

\bibitem[{{Gillessen} {et~al.}(2009){Gillessen}, {Eisenhauer}, {Trippe}, {Alexander}, {Genzel}, {Martins}, \& {Ott}}]{Gillessen2009}
{Gillessen}, S., {Eisenhauer}, F., {Trippe}, S., {et~al.} 2009, \apj, 692, 1075, \dodoi{10.1088/0004-637X/692/2/1075}

\bibitem[{{Gravity Collaboration} {et~al.}(2018){Gravity Collaboration}, {Abuter}, {Amorim}, {Anugu}, {Baub{\"o}ck}, {Benisty}, {Berger}, {Blind}, {Bonnet}, {Brandner}, {Buron}, {Collin}, {Chapron}, {Cl{\'e}net}, {Coud{\'e} Du Foresto}, {de Zeeuw}, {Deen}, {Delplancke-Str{\"o}bele}, {Dembet}, {Dexter}, {Duvert}, {Eckart}, {Eisenhauer}, {Finger}, {F{\"o}rster Schreiber}, {F{\'e}dou}, {Garcia}, {Garcia Lopez}, {Gao}, {Gendron}, {Genzel}, {Gillessen}, {Gordo}, {Habibi}, {Haubois}, {Haug}, {Hau{\ss}mann}, {Henning}, {Hippler}, {Horrobin}, {Hubert}, {Hubin}, {Jimenez Rosales}, {Jochum}, {Jocou}, {Kaufer}, {Kellner}, {Kendrew}, {Kervella}, {Kok}, {Kulas}, {Lacour}, {Lapeyr{\`e}re}, {Lazareff}, {Le Bouquin}, {L{\'e}na}, {Lippa}, {Lenzen}, {M{\'e}rand}, {M{\"u}ler}, {Neumann}, {Ott}, {Palanca}, {Paumard}, {Pasquini}, {Perraut}, {Perrin}, {Pfuhl}, {Plewa}, {Rabien}, {Ram{\'\i}rez}, {Ramos}, {Rau}, {Rodr{\'\i}guez-Coira}, {Rohloff}, {Rousset}, {Sanchez-Bermudez}, {Scheithauer}, {Sch{\"o}ller}, {Schuler}, {Spyromilio},
  {Straub}, {Straubmeier}, {Sturm}, {Tacconi}, {Tristram}, {Vincent}, {von Fellenberg}, {Wank}, {Waisberg}, {Widmann}, {Wieprecht}, {Wiest}, {Wiezorrek}, {Woillez}, {Yazici}, {Ziegler}, \& {Zins}}]{gravity2018}
{Gravity Collaboration}, {Abuter}, R., {Amorim}, A., {et~al.} 2018, \aap, 615, L15, \dodoi{10.1051/0004-6361/201833718}

\bibitem[{{Gravity Collaboration} {et~al.}(2023){Gravity Collaboration}, {Wojtczak}, {Labadie}, {Perraut}, {Tessore}, {Soulain}, {Ganci}, {Bouvier}, {Dougados}, {Al{\'e}cian}, {Nowacki}, {Cozzo}, {Brandner}, {Caratti O Garatti}, {Garcia}, {Garcia Lopez}, {Sanchez-Bermudez}, {Amorim}, {Benisty}, {Berger}, {Bourdarot}, {Caselli}, {Cl{\'e}net}, {de Zeeuw}, {Davies}, {Drescher}, {Duvert}, {Eckart}, {Eisenhauer}, {Eupen}, {F{\"o}rster-Schreiber}, {Gendron}, {Gillessen}, {Grant}, {Grellmann}, {Hei{\ss}el}, {Henning}, {Hippler}, {Horrobin}, {Hubert}, {Jocou}, {Kervella}, {Lacour}, {Lapeyr{\`e}re}, {Le Bouquin}, {L{\'e}na}, {Lutz}, {Mang}, {Ott}, {Paumard}, {Perrin}, {Scheithauer}, {Shangguan}, {Shimizu}, {Spezzano}, {Straub}, {Straubmeier}, {Sturm}, {van Dishoeck}, {Vincent}, \& {Widmann}}]{gravitycollaboration2023}
{Gravity Collaboration}, {Wojtczak}, J.~A., {Labadie}, L., {et~al.} 2023, \aap, 669, A59, \dodoi{10.1051/0004-6361/202244675}

\bibitem[{{Habibi} {et~al.}(2017){Habibi}, {Gillessen}, {Martins}, {Eisenhauer}, {Plewa}, {Pfuhl}, {George}, {Dexter}, {Waisberg}, {Ott}, {von Fellenberg}, {Baub{\"o}ck}, {Jimenez-Rosales}, \& {Genzel}}]{Habibi2017}
{Habibi}, M., {Gillessen}, S., {Martins}, F., {et~al.} 2017, \apj, 847, 120, \dodoi{10.3847/1538-4357/aa876f}

\bibitem[{{Harfst} {et~al.}(2007){Harfst}, {Gualandris}, {Merritt}, {Spurzem}, {Portegies Zwart}, \& {Berczik}}]{Harfst2007}
{Harfst}, S., {Gualandris}, A., {Merritt}, D., {et~al.} 2007, \na, 12, 357, \dodoi{10.1016/j.newast.2006.11.003}

\bibitem[{{Heggie}(1975)}]{Heggie1975}
{Heggie}, D.~C. 1975, \mnras, 173, 729, \dodoi{10.1093/mnras/173.3.729}

\bibitem[{{Hei{\ss}el} {et~al.}(2022){Hei{\ss}el}, {Paumard}, {Perrin}, \& {Vincent}}]{Heissel2022}
{Hei{\ss}el}, G., {Paumard}, T., {Perrin}, G., \& {Vincent}, F. 2022, \aap, 660, A13, \dodoi{10.1051/0004-6361/202142114}

\bibitem[{{Hills}(1975{\natexlab{a}})}]{Hills1975a}
{Hills}, J.~G. 1975{\natexlab{a}}, \aj, 80, 809, \dodoi{10.1086/111815}

\bibitem[{{Hills}(1975{\natexlab{b}})}]{Hills1975b}
---. 1975{\natexlab{b}}, \aj, 80, 1075, \dodoi{10.1086/111842}

\bibitem[{{Hills}(1988)}]{Hills1988}
---. 1988, \nat, 331, 687, \dodoi{10.1038/331687a0}

\bibitem[{{Hobbs} \& {Nayakshin}(2009)}]{Hobbs2009}
{Hobbs}, A., \& {Nayakshin}, S. 2009, \mnras, 394, 191, \dodoi{10.1111/j.1365-2966.2008.14359.x}

\bibitem[{{Hsieh} {et~al.}(2017){Hsieh}, {Koch}, {Ho}, {Kim}, {Tang}, {Wang}, {Yen}, \& {Hwang}}]{Hsieh2017}
{Hsieh}, P.-Y., {Koch}, P.~M., {Ho}, P. T.~P., {et~al.} 2017, \apj, 847, 3, \dodoi{10.3847/1538-4357/aa8329}

\bibitem[{{Hsieh} {et~al.}(2021){Hsieh}, {Koch}, {Kim}, {Mart{\'\i}n}, {Yen}, {Carpenter}, {Harada}, {Turner}, {Ho}, {Tang}, \& {Beck}}]{Hsieh2021}
{Hsieh}, P.-Y., {Koch}, P.~M., {Kim}, W.-T., {et~al.} 2021, \apj, 913, 94, \dodoi{10.3847/1538-4357/abf4cd}

\bibitem[{{Hurley} {et~al.}(2002){Hurley}, {Tout}, \& {Pols}}]{Hurley2002}
{Hurley}, J.~R., {Tout}, C.~A., \& {Pols}, O.~R. 2002, \mnras, 329, 897, \dodoi{10.1046/j.1365-8711.2002.05038.x}

\bibitem[{{Jalali} {et~al.}(2014){Jalali}, {Pelupessy}, {Eckart}, {Portegies Zwart}, {Sabha}, {Borkar}, {Moultaka}, {Mu{\v z}i{\'c}}, \& {Moser}}]{Jalali2014}
{Jalali}, B., {Pelupessy}, F.~I., {Eckart}, A., {et~al.} 2014, \mnras, 444, 1205, \dodoi{10.1093/mnras/stu1483}

\bibitem[{{Jia} {et~al.}(2023){Jia}, {Xu}, {Lu}, {Chu}, {O'Neil}, {Drechsler}, {Hosek}, {Sakai}, {Do}, {Ciurlo}, {Gautam}, {Ghez}, {Becklin}, {Morris}, \& {Bentley}}]{Jia2023}
{Jia}, S., {Xu}, N., {Lu}, J.~R., {et~al.} 2023, \apj, 949, 18, \dodoi{10.3847/1538-4357/acb939}

\bibitem[{{Joshi} {et~al.}(2001){Joshi}, {Nave}, \& {Rasio}}]{Joshi2001}
{Joshi}, K.~J., {Nave}, C.~P., \& {Rasio}, F.~A. 2001, \apj, 550, 691, \dodoi{10.1086/319771}

\bibitem[{{Kaczmarek} {et~al.}(2011){Kaczmarek}, {Olczak}, \& {Pfalzner}}]{Kaczmarek2011}
{Kaczmarek}, T., {Olczak}, C., \& {Pfalzner}, S. 2011, \aap, 528, A144, \dodoi{10.1051/0004-6361/201015233}

\bibitem[{{King}(1966)}]{King1966}
{King}, I.~R. 1966, \aj, 71, 64, \dodoi{10.1086/109857}

\bibitem[{{Krabbe} {et~al.}(1995){Krabbe}, {Genzel}, {Eckart}, {Najarro}, {Lutz}, {Cameron}, {Kroker}, {Tacconi-Garman}, {Thatte}, {Weitzel}, {Drapatz}, {Geballe}, {Sternberg}, \& {Kudritzki}}]{Krabbe1995}
{Krabbe}, A., {Genzel}, R., {Eckart}, A., {et~al.} 1995, \apjl, 447, L95, \dodoi{10.1086/309579}

\bibitem[{Kraft(1988)}]{Kraft1988}
Kraft, D. 1988, A software package for sequential quadratic programming, Tech. Rep. DFVLR-FB 88-2, Institut fuer Dynamik der Flugsysteme, Oberpfaffenhofen

\bibitem[{{Kroupa}(2001)}]{Kroupa2001}
{Kroupa}, P. 2001, \mnras, 322, 231, \dodoi{10.1046/j.1365-8711.2001.04022.x10.48550/arXiv.astro-ph/0009005}

\bibitem[{{L{\'e}pine} \& {Moffat}(1999)}]{Lepine1999}
{L{\'e}pine}, S., \& {Moffat}, A. F.~J. 1999, \apj, 514, 909, \dodoi{10.1086/306958}

\bibitem[{{Longhitano} \& {Binggeli}(2010)}]{Longhitano2010}
{Longhitano}, M., \& {Binggeli}, B. 2010, \aap, 509, A46, \dodoi{10.1051/0004-6361/200913109}

\bibitem[{{Lu} {et~al.}(2006){Lu}, {Ghez}, {Hornstein}, {Morris}, {Matthews}, {Thompson}, \& {Becklin}}]{Lu2006}
{Lu}, J.~R., {Ghez}, A.~M., {Hornstein}, S.~D., {et~al.} 2006, in Journal of Physics Conference Series, Vol.~54, Journal of Physics Conference Series, 279--287, \dodoi{10.1088/1742-6596/54/1/044}

\bibitem[{{Lu} {et~al.}(2009){Lu}, {Ghez}, {Hornstein}, {Morris}, {Becklin}, \& {Matthews}}]{Lu2009}
{Lu}, J.~R., {Ghez}, A.~M., {Hornstein}, S.~D., {et~al.} 2009, \apj, 690, 1463, \dodoi{10.1088/0004-637X/690/2/1463}

\bibitem[{{Lutz} {et~al.}(1993){Lutz}, {Krabbe}, \& {Genzel}}]{Lutz1993}
{Lutz}, D., {Krabbe}, A., \& {Genzel}, R. 1993, \apj, 418, 244, \dodoi{10.1086/173386}

\bibitem[{{L{\"u}tzgendorf} {et~al.}(2016){L{\"u}tzgendorf}, {Helm}, {Pelupessy}, \& {Portegies Zwart}}]{Luetzgendorf2016}
{L{\"u}tzgendorf}, N., {Helm}, E. v.~d., {Pelupessy}, F.~I., \& {Portegies Zwart}, S. 2016, \mnras, 456, 3645, \dodoi{10.1093/mnras/stv2918}

\bibitem[{{Mac Low}(1999)}]{MacLow1999}
{Mac Low}, M.-M. 1999, \apj, 524, 169, \dodoi{10.1086/307784}

\bibitem[{{Mahadevan}(1997)}]{1997ApJ...477..585M}
{Mahadevan}, R. 1997, \apj, 477, 585, \dodoi{10.1086/303727}

\bibitem[{{Maillard} {et~al.}(2004){Maillard}, {Paumard}, {Stolovy}, \& {Rigaut}}]{Maillard2004}
{Maillard}, J.~P., {Paumard}, T., {Stolovy}, S.~R., \& {Rigaut}, F. 2004, \aap, 423, 155, \dodoi{10.1051/0004-6361:20034147}

\bibitem[{{Makino} \& {Taiji}(1998)}]{Makino1998}
{Makino}, J., \& {Taiji}, M. 1998, {Scientific Simulations with Special-Purpose Computers--the GRAPE Systems}

\bibitem[{{Martayan} {et~al.}(2016){Martayan}, {Lobel}, {Baade}, {Mehner}, {Rivinius}, {Boffin}, {Girard}, {Mawet}, {Montagnier}, {Blomme}, {Kervella}, {Sana}, {{\v{S}}tefl}, {Zorec}, {Lacour}, {Le Bouquin}, {Martins}, {M{\'e}rand}, {Patru}, {Selman}, \& {Fr{\'e}mat}}]{Martayan2016}
{Martayan}, C., {Lobel}, A., {Baade}, D., {et~al.} 2016, \aap, 587, A115, \dodoi{10.1051/0004-6361/201526578}

\bibitem[{{Massey} \& {Hunter}(1998)}]{Massey1998}
{Massey}, P., \& {Hunter}, D.~A. 1998, \apj, 493, 180, \dodoi{10.1086/305126}

\bibitem[{{Menten} {et~al.}(1997){Menten}, {Reid}, {Eckart}, \& {Genzel}}]{Menten1997}
{Menten}, K.~M., {Reid}, M.~J., {Eckart}, A., \& {Genzel}, R. 1997, \apjl, 475, L111, \dodoi{10.1086/310472}

\bibitem[{{Mezger} {et~al.}(1996){Mezger}, {Duschl}, \& {Zylka}}]{Mezger1996}
{Mezger}, P.~G., {Duschl}, W.~J., \& {Zylka}, R. 1996, \aapr, 7, 289, \dodoi{10.1007/s001590050007}

\bibitem[{{Minniti} {et~al.}(2023){Minniti}, {Matsunaga}, {Fernandez-Trincado}, {Otsubo}, {Sarugaku}, {Takeuchi}, {Katoh}, {Hamano}, {Ikeda}, {Kawakita}, {Lucas}, {Smith}, {Petralia}, {Garro}, {Saito}, {Alonso-Garcia}, {Gomez}, \& {Navarro}}]{Minniti2023}
{Minniti}, D., {Matsunaga}, N., {Fernandez-Trincado}, J.~G., {et~al.} 2023, arXiv e-prints, arXiv:2312.16028, \dodoi{10.48550/arXiv.2312.16028}

\bibitem[{{Misugi} {et~al.}(2019){Misugi}, {Inutsuka}, \& {Arzoumanian}}]{Misugi2019}
{Misugi}, Y., {Inutsuka}, S.-i., \& {Arzoumanian}, D. 2019, \apj, 881, 11, \dodoi{10.3847/1538-4357/ab2382}

\bibitem[{{Misugi} {et~al.}(2023){Misugi}, {Inutsuka}, \& {Arzoumanian}}]{Misugi2023}
---. 2023, \apj, 943, 76, \dodoi{10.3847/1538-4357/aca88d}

\bibitem[{{Morris}(1993)}]{Morris1993}
{Morris}, M. 1993, \apj, 408, 496, \dodoi{10.1086/172607}

\bibitem[{{Morris} {et~al.}(2020){Morris}, {Charnley}, {Corcoran}, {Cordiner}, {Damineli}, {Groh}, {Gull}, {Loinard}, {Madura}, {Mehner}, {Moffat}, {Palmer}, {Rau}, {Richardson}, \& {Weigelt}}]{Morris2020}
{Morris}, P.~W., {Charnley}, S.~B., {Corcoran}, M., {et~al.} 2020, \apjl, 892, L23, \dodoi{10.3847/2041-8213/ab784a}

\bibitem[{{Moser} {et~al.}(2017){Moser}, {S{\'a}nchez-Monge}, {Eckart}, {Requena-Torres}, {Garc{\'\i}a-Marin}, {Kunneriath}, {Zensus}, {Britzen}, {Sabha}, {Shahzamanian}, {Borkar}, \& {Fischer}}]{Moser2017}
{Moser}, L., {S{\'a}nchez-Monge}, {\'A}., {Eckart}, A., {et~al.} 2017, \aap, 603, A68, \dodoi{10.1051/0004-6361/201628385}

\bibitem[{{Motte} {et~al.}(2018){Motte}, {Bontemps}, \& {Louvet}}]{Motte2018}
{Motte}, F., {Bontemps}, S., \& {Louvet}, F. 2018, \araa, 56, 41, \dodoi{10.1146/annurev-astro-091916-055235}

\bibitem[{{Moultaka} {et~al.}(2009){Moultaka}, {Eckart}, \& {Sch{\"o}del}}]{Moultaka2009}
{Moultaka}, J., {Eckart}, A., \& {Sch{\"o}del}, R. 2009, \apj, 703, 1635, \dodoi{10.1088/0004-637X/703/2/1635}

\bibitem[{{Murchikova} {et~al.}(2019){Murchikova}, {Phinney}, {Pancoast}, \& {Blandford}}]{Murchikova2019}
{Murchikova}, E.~M., {Phinney}, E.~S., {Pancoast}, A., \& {Blandford}, R.~D. 2019, \nat, 570, 83, \dodoi{10.1038/s41586-019-1242-z}

\bibitem[{{Mu{\v{z}}i{\'c}} {et~al.}(2008){Mu{\v{z}}i{\'c}}, {Sch{\"o}del}, {Eckart}, {Meyer}, \& {Zensus}}]{muzic2008}
{Mu{\v{z}}i{\'c}}, K., {Sch{\"o}del}, R., {Eckart}, A., {Meyer}, L., \& {Zensus}, A. 2008, \aap, 482, 173, \dodoi{10.1051/0004-6361:20078352}

\bibitem[{{Narayan} {et~al.}(1998){Narayan}, {Mahadevan}, {Grindlay}, {Popham}, \& {Gammie}}]{Narayan1998}
{Narayan}, R., {Mahadevan}, R., {Grindlay}, J.~E., {Popham}, R.~G., \& {Gammie}, C. 1998, apj, 492, 554, \dodoi{10.1086/305070}

\bibitem[{{Narayan} \& {Yi}(1995)}]{1995ApJ...452..710N}
{Narayan}, R., \& {Yi}, I. 1995, \apj, 452, 710, \dodoi{10.1086/176343}

\bibitem[{{Ott}(2013)}]{Ott2013}
{Ott}, T. 2013, {DPUSER: Interactive language for image analysis}.
\newblock \doeprint{1303.025}

\bibitem[{{Oudmaijer} \& {Parr}(2010)}]{Oudmaijer2010}
{Oudmaijer}, R.~D., \& {Parr}, A.~M. 2010, \mnras, 405, 2439, \dodoi{10.1111/j.1365-2966.2010.16609.x}

\bibitem[{{Owocki}(2002)}]{Owoki2002}
{Owocki}, S. 2002, in Astronomical Society of the Pacific Conference Series, Vol. 260, Interacting Winds from Massive Stars, ed. A.~F.~J. {Moffat} \& N.~{St-Louis}, 15

\bibitem[{{Parsa} {et~al.}(2017){Parsa}, {Eckart}, {Shahzamanian}, {Karas}, {Zaja{\v c}ek}, {Zensus}, \& {Straubmeier}}]{Parsa2017}
{Parsa}, M., {Eckart}, A., {Shahzamanian}, B., {et~al.} 2017, \apj, 845, 22, \dodoi{10.3847/1538-4357/aa7bf0}

\bibitem[{{Paumard} {et~al.}(2022){Paumard}, {Ciurlo}, {Morris}, {Do}, \& {Ghez}}]{Paumard2022}
{Paumard}, T., {Ciurlo}, A., {Morris}, M.~R., {Do}, T., \& {Ghez}, A.~M. 2022, \aap, 664, A97, \dodoi{10.1051/0004-6361/202243228}

\bibitem[{Paumard {et~al.}(2006)Paumard, Genzel, Martins, Nayakshin, Beloborodov, Levin, Trippe, Eisenhauer, Ott, Gillessen, Abuter, Cuadra, Alexander, \& Sternberg}]{Paumard2006}
Paumard, T., Genzel, R., Martins, F., {et~al.} 2006, The Astrophysical Journal, 643, 1011.
\newblock \url{http://stacks.iop.org/0004-637X/643/i=2/a=1011}

\bibitem[{{Pei{\ss}ker} {et~al.}(2020a){Pei{\ss}ker}, {Eckart}, \& {Parsa}}]{peissker2020a}
{Pei{\ss}ker}, F., {Eckart}, A., \& {Parsa}, M. 2020a, \apj, 889, 61, \dodoi{10.3847/1538-4357/ab5afd}

\bibitem[{{Pei{\ss}ker} {et~al.}(2020c){Pei{\ss}ker}, {Eckart}, {Sabha}, {Zaja{\v{c}}ek}, \& {Bhat}}]{Peissker2020c}
{Pei{\ss}ker}, F., {Eckart}, A., {Sabha}, N.~B., {Zaja{\v{c}}ek}, M., \& {Bhat}, H. 2020c, \apj, 897, 28, \dodoi{10.3847/1538-4357/ab9826}

\bibitem[{{Pei{\ss}ker} {et~al.}(2020d){Pei{\ss}ker}, {Eckart}, {Zaja{\v{c}}ek}, {Ali}, \& {Parsa}}]{Peissker2020d}
{Pei{\ss}ker}, F., {Eckart}, A., {Zaja{\v{c}}ek}, M., {Ali}, B., \& {Parsa}, M. 2020d, The Astrophysical Journal, 899, 50, \dodoi{10.3847/1538-4357/ab9c1c}

\bibitem[{{Pei{\ss}ker} {et~al.}(2022){Pei{\ss}ker}, {Eckart}, {Zaja{\v{c}}ek}, \& {Britzen}}]{Peissker2022}
{Pei{\ss}ker}, F., {Eckart}, A., {Zaja{\v{c}}ek}, M., \& {Britzen}, S. 2022, \apj, 933, 49, \dodoi{10.3847/1538-4357/ac752f}

\bibitem[{{Pei{\ss}ker} {et~al.}(2020b){Pei{\ss}ker}, {Hosseini}, {Zaja{\v{c}}ek}, {Eckart}, {Saalfeld}, {Valencia-S.}, {Parsa}, \& {Karas}}]{Peissker2020b}
{Pei{\ss}ker}, F., {Hosseini}, S.~E., {Zaja{\v{c}}ek}, M., {et~al.} 2020b, \aap, 634, A35, \dodoi{10.1051/0004-6361/201935953}

\bibitem[{{Pei{\ss}ker} {et~al.}(2021a){Pei{\ss}ker}, {Ali}, {Zaja{\v{c}}ek}, {Eckart}, {Hosseini}, {Karas}, {Cl{\'e}net}, {Sabha}, {Labadie}, \& {Subroweit}}]{peissker2021}
{Pei{\ss}ker}, F., {Ali}, B., {Zaja{\v{c}}ek}, M., {et~al.} 2021a, \apj, 909, 62, \dodoi{10.3847/1538-4357/abd9c6}

\bibitem[{{Pei{\ss}ker} {et~al.}(2021c){Pei{\ss}ker}, {Zaja{\v{c}}ek}, {Eckart}, {Ali}, {Karas}, {Sabha}, {Grellmann}, {Labadie}, \& {Shahzamanian}}]{peissker2021c}
{Pei{\ss}ker}, F., {Zaja{\v{c}}ek}, M., {Eckart}, A., {et~al.} 2021c, \apj, 923, 69, \dodoi{10.3847/1538-4357/ac23df}

\bibitem[{{Pei{\ss}ker} {et~al.}(2023a){Pei{\ss}ker}, {Zaja{\v{c}}ek}, {Eckart}, {Ali}, {Karas}, {Sabha}, {Grellmann}, {Labadie}, \& {Shahzamanian}}]{peissker2023a}
---. 2023a, \apj, 943, 183, \dodoi{10.3847/1538-4357/acb435}

\bibitem[{{Pei{\ss}ker} {et~al.}(2023b){Pei{\ss}ker}, {Zaja{\v{c}}ek}, {Sabha}, {Tsuboi}, {Moultaka}, {Labadie}, {Eckart}, {Karas}, {Steiniger}, {Subroweit}, {Suresh}, {Melamed}, \& {Cl{\'e}net}}]{peissker2023b}
{Pei{\ss}ker}, F., {Zaja{\v{c}}ek}, M., {Sabha}, N.~B., {et~al.} 2023b, \apj, 944, 231, \dodoi{10.3847/1538-4357/aca977}

\bibitem[{{Pei{\ss}ker} {et~al.}(2023c){Pei{\ss}ker}, {Zaja{\v{c}}ek}, {Thomkins}, {Eckart}, {Labadie}, {Karas}, {Sabha}, {Steiniger}, \& {Melamed}}]{peissker2023c}
{Pei{\ss}ker}, F., {Zaja{\v{c}}ek}, M., {Thomkins}, L., {et~al.} 2023c, \apj, 956, 70, \dodoi{10.3847/1538-4357/acf6b5}

\bibitem[{{Pelupessy} {et~al.}(2013){Pelupessy}, {van Elteren}, {de Vries}, {McMillan}, {Drost}, \& {Portegies Zwart}}]{Pelupessy2013}
{Pelupessy}, F.~I., {van Elteren}, A., {de Vries}, N., {et~al.} 2013, \aap, 557, A84, \dodoi{10.1051/0004-6361/201321252}

\bibitem[{{Pesce} {et~al.}(2021){Pesce}, {Palumbo}, {Narayan}, {Blackburn}, {Doeleman}, {Johnson}, {Ma}, {Nagar}, {Natarajan}, \& {Ricarte}}]{2021ApJ...923..260P}
{Pesce}, D.~W., {Palumbo}, D. C.~M., {Narayan}, R., {et~al.} 2021, \apj, 923, 260, \dodoi{10.3847/1538-4357/ac2eb5}

\bibitem[{{Petts} \& {Gualandris}(2017)}]{Petts2017}
{Petts}, J.~A., \& {Gualandris}, A. 2017, \mnras, 467, 3775, \dodoi{10.1093/mnras/stx296}

\bibitem[{{Pfalzner} {et~al.}(2014){Pfalzner}, {Parmentier}, {Steinhausen}, {Vincke}, \& {Menten}}]{Pfalzner2014}
{Pfalzner}, S., {Parmentier}, G., {Steinhausen}, M., {Vincke}, K., \& {Menten}, K. 2014, \apj, 794, 147, \dodoi{10.1088/0004-637X/794/2/147}

\bibitem[{{Pfuhl} {et~al.}(2014){Pfuhl}, {Alexander}, {Gillessen}, {Martins}, {Genzel}, {Eisenhauer}, {Fritz}, \& {Ott}}]{Pfuhl2014}
{Pfuhl}, O., {Alexander}, T., {Gillessen}, S., {et~al.} 2014, \apj, 782, 101, \dodoi{10.1088/0004-637X/782/2/101}

\bibitem[{{Plewa} {et~al.}(2015){Plewa}, {Gillessen}, {Eisenhauer}, {Ott}, {Pfuhl}, {George}, {Dexter}, {Habibi}, {Genzel}, {Reid}, \& {Menten}}]{Plewa2015}
{Plewa}, P.~M., {Gillessen}, S., {Eisenhauer}, F., {et~al.} 2015, \mnras, 453, 3234, \dodoi{10.1093/mnras/stv1910}

\bibitem[{{Plewa} {et~al.}(2018){Plewa}, {Gillessen}, {Baub{\"o}ck}, {Dexter}, {Eisenhauer}, {von Fellenberg}, {Gao}, {Genzel}, {Habibi}, {Jimenez-Rosales}, {Ott}, {Pfuhl}, {Waisberg}, \& {Widmann}}]{Plewa2018}
{Plewa}, P.~M., {Gillessen}, S., {Baub{\"o}ck}, M., {et~al.} 2018, Research Notes of the American Astronomical Society, 2, 35, \dodoi{10.3847/2515-5172/aab3df}

\bibitem[{{Portegies Zwart} \& {McMillan}(2018)}]{Portegies-Wart2018}
{Portegies Zwart}, S., \& {McMillan}, S. 2018, {Astrophysical Recipes; The art of AMUSE} ({Bristol, UK: IOP Publishing}), \dodoi{10.1088/978-0-7503-1320-9}

\bibitem[{{Portegies Zwart} {et~al.}(2013){Portegies Zwart}, {McMillan}, {van Elteren}, {Pelupessy}, \& {de Vries}}]{Portegies-Wart2013}
{Portegies Zwart}, S., {McMillan}, S.~L.~W., {van Elteren}, E., {Pelupessy}, I., \& {de Vries}, N. 2013, Computer Physics Communications, 184, 456, \dodoi{10.1016/j.cpc.2012.09.024}

\bibitem[{{Portegies Zwart} {et~al.}(2009){Portegies Zwart}, {McMillan}, {Harfst}, {Groen}, {Fujii}, {Nuall{\'a}in}, {Glebbeek}, {Heggie}, {Lombardi}, {Hut}, {Angelou}, {Banerjee}, {Belkus}, {Fragos}, {Fregeau}, {Gaburov}, {Izzard}, {Juri{\'c}}, {Justham}, {Sottoriva}, {Teuben}, {van Bever}, {Yaron}, \& {Zemp}}]{Portegies-Wart2009}
{Portegies Zwart}, S., {McMillan}, S., {Harfst}, S., {et~al.} 2009, \na, 14, 369, \dodoi{10.1016/j.newast.2008.10.006}

\bibitem[{{Portegies Zwart} {et~al.}(2023){Portegies Zwart}, {Boekholt}, \& {Heggie}}]{Portegies-Zwart2023}
{Portegies Zwart}, S.~F., {Boekholt}, T. C.~N., \& {Heggie}, D.~C. 2023, \mnras, \dodoi{10.1093/mnras/stad2654}

\bibitem[{{Portegies Zwart} {et~al.}(2002){Portegies Zwart}, {Makino}, {McMillan}, \& {Hut}}]{Portegies-Zwart2002}
{Portegies Zwart}, S.~F., {Makino}, J., {McMillan}, S. L.~W., \& {Hut}, P. 2002, \apj, 565, 265, \dodoi{10.1086/324141}

\bibitem[{{Portegies Zwart} {et~al.}(2003){Portegies Zwart}, {McMillan}, \& {Gerhard}}]{Portegies-Zwart2003}
{Portegies Zwart}, S.~F., {McMillan}, S. L.~W., \& {Gerhard}, O. 2003, apj, 593, 352, \dodoi{10.1086/376439}

\bibitem[{{Portegies Zwart} {et~al.}(2010){Portegies Zwart}, {McMillan}, \& {Gieles}}]{Zwart2010}
{Portegies Zwart}, S.~F., {McMillan}, S. L.~W., \& {Gieles}, M. 2010, \araa, 48, 431, \dodoi{10.1146/annurev-astro-081309-130834}

\bibitem[{{Requena-Torres} {et~al.}(2012){Requena-Torres}, {G{\"u}sten}, {Wei{\ss}}, {Harris}, {Mart{\'\i}n-Pintado}, {Stutzki}, {Klein}, {Heyminck}, \& {Risacher}}]{Requena-Torres2012}
{Requena-Torres}, M.~A., {G{\"u}sten}, R., {Wei{\ss}}, A., {et~al.} 2012, \aap, 542, L21, \dodoi{10.1051/0004-6361/201219068}

\bibitem[{{Ressler} {et~al.}(2018){Ressler}, {Quataert}, \& {Stone}}]{2018MNRAS.478.3544R}
{Ressler}, S.~M., {Quataert}, E., \& {Stone}, J.~M. 2018, \mnras, 478, 3544, \dodoi{10.1093/mnras/sty1146}

\bibitem[{{Rieke} \& {Lebofsky}(1985)}]{Rieke1985}
{Rieke}, G.~H., \& {Lebofsky}, M.~J. 1985, \apj, 288, 618, \dodoi{10.1086/162827}

\bibitem[{{Rieke} {et~al.}(1978){Rieke}, {Telesco}, \& {Harper}}]{Rieke1978}
{Rieke}, G.~H., {Telesco}, C.~M., \& {Harper}, D.~A. 1978, \apj, 220, 556, \dodoi{10.1086/155936}

\bibitem[{{Robitaille}(2011)}]{Robitaille2011}
{Robitaille}, T.~P. 2011, \aap, 536, A79, \dodoi{10.1051/0004-6361/201117150}

\bibitem[{{Robitaille}(2017)}]{Robitaille2017}
---. 2017, \aap, 600, A11, \dodoi{10.1051/0004-6361/201425486}

\bibitem[{{Roche} {et~al.}(2018){Roche}, {Lopez-Rodriguez}, {Telesco}, {Sch{\"o}del}, \& {Packham}}]{Roche2018}
{Roche}, P.~F., {Lopez-Rodriguez}, E., {Telesco}, C.~M., {Sch{\"o}del}, R., \& {Packham}, C. 2018, \mnras, 476, 235, \dodoi{10.1093/mnras/sty129}

\bibitem[{{Sabha} {et~al.}(2012){Sabha}, {Eckart}, {Merritt}, {Zamaninasab}, {Witzel}, {Garc{\'{\i}}a-Mar{\'{\i}}n}, {Jalali}, {Valencia-S.}, {Yazici}, {Buchholz}, {Shahzamanian}, {Rauch}, {Horrobin}, \& {Straubmeier}}]{Sabha2012}
{Sabha}, N., {Eckart}, A., {Merritt}, D., {et~al.} 2012, \aap, 545, A70, \dodoi{10.1051/0004-6361/201219203}

\bibitem[{{Salas} {et~al.}(2021){Salas}, {Morris}, \& {Naoz}}]{Salas2021}
{Salas}, J.~M., {Morris}, M.~R., \& {Naoz}, S. 2021, \aj, 161, 243, \dodoi{10.3847/1538-3881/abefd3}

\bibitem[{{Sana} {et~al.}(2012){Sana}, {de Mink}, {de Koter}, {Langer}, {Evans}, {Gieles}, {Gosset}, {Izzard}, {Le Bouquin}, \& {Schneider}}]{Sana2012}
{Sana}, H., {de Mink}, S.~E., {de Koter}, A., {et~al.} 2012, Science, 337, 444, \dodoi{10.1126/science.1223344}

\bibitem[{{Sch{\"o}del} {et~al.}(2005){Sch{\"o}del}, {Eckart}, {Iserlohe}, {Genzel}, \& {Ott}}]{Schoedel2005}
{Sch{\"o}del}, R., {Eckart}, A., {Iserlohe}, C., {Genzel}, R., \& {Ott}, T. 2005, \apjl, 625, L111, \dodoi{10.1086/431307}

\bibitem[{{Sch{\"o}del} {et~al.}(2010){Sch{\"o}del}, {Najarro}, {Muzic}, \& {Eckart}}]{Schoedel2010}
{Sch{\"o}del}, R., {Najarro}, F., {Muzic}, K., \& {Eckart}, A. 2010, \aap, 511, A18, \dodoi{10.1051/0004-6361/200913183}

\bibitem[{{Shcherbakov} \& {Baganoff}(2010)}]{2010ApJ...716..504S}
{Shcherbakov}, R.~V., \& {Baganoff}, F.~K. 2010, \apj, 716, 504, \dodoi{10.1088/0004-637X/716/1/504}

\bibitem[{{Simon} {et~al.}(1990){Simon}, {Chen}, {Forrest}, {Garnett}, {Longmore}, {Gauer}, \& {Dixon}}]{Simon1990}
{Simon}, M., {Chen}, W.~P., {Forrest}, W.~J., {et~al.} 1990, \apj, 360, 95, \dodoi{10.1086/169099}

\bibitem[{{Smith} {et~al.}(1990){Smith}, {Aitken}, \& {Roche}}]{Smith1990}
{Smith}, C.~H., {Aitken}, D.~K., \& {Roche}, P.~F. 1990, \mnras, 246, 1

\bibitem[{{Spitzer}(1987)}]{Spitzer1987}
{Spitzer}, L. 1987, {Dynamical evolution of globular clusters}

\bibitem[{{Stephan} {et~al.}(2019){Stephan}, {Naoz}, {Ghez}, {Morris}, {Ciurlo}, {Do}, {Breivik}, {Coughlin}, \& {Rodriguez}}]{stephan2019}
{Stephan}, A.~P., {Naoz}, S., {Ghez}, A.~M., {et~al.} 2019, \apj, 878, 58, \dodoi{10.3847/1538-4357/ab1e4d}

\bibitem[{{Tan}(2000)}]{2000ApJ...536..173T}
{Tan}, J.~C. 2000, \apj, 536, 173, \dodoi{10.1086/308905}

\bibitem[{{Tig{\'e}} {et~al.}(2017){Tig{\'e}}, {Motte}, {Russeil}, {Zavagno}, {Hennemann}, {Schneider}, {Hill}, {Nguyen Luong}, {Di Francesco}, {Bontemps}, {Louvet}, {Didelon}, {K{\"o}nyves}, {Andr{\'e}}, {Leuleu}, {Bardagi}, {Anderson}, {Arzoumanian}, {Benedettini}, {Bernard}, {Elia}, {Figueira}, {Kirk}, {Martin}, {Minier}, {Molinari}, {Nony}, {Persi}, {Pezzuto}, {Polychroni}, {Rayner}, {Rivera-Ingraham}, {Roussel}, {Rygl}, {Spinoglio}, \& {White}}]{Tige2017}
{Tig{\'e}}, J., {Motte}, F., {Russeil}, D., {et~al.} 2017, \aap, 602, A77, \dodoi{10.1051/0004-6361/201628989}

\bibitem[{{Tsuboi} {et~al.}(2019){Tsuboi}, {Kitamura}, {Tsutsumi}, {Miyawaki}, {Miyoshi}, \& {Miyazaki}}]{tsuboi2019}
{Tsuboi}, M., {Kitamura}, Y., {Tsutsumi}, T., {et~al.} 2019, \pasj, 71, 105, \dodoi{10.1093/pasj/psz089}

\bibitem[{{Tsuboi} {et~al.}(2020){Tsuboi}, {Kitamura}, {Tsutsumi}, {Miyawaki}, {Miyoshi}, \& {Miyazaki}}]{Tsuboi2020b}
---. 2020, \pasj, 72, L5, \dodoi{10.1093/pasj/psaa016}

\bibitem[{{Tsuboi} {et~al.}(2021){Tsuboi}, {Kitamura}, {Tsutsumi}, {Miyawaki}, {Miyoshi}, \& {Miyazaki}}]{Tsuboi2021}
{Tsuboi}, M., {Kitamura}, Y., {Tsutsumi}, T., {et~al.} 2021, in Astronomical Society of the Pacific Conference Series, Vol. 528, New Horizons in Galactic Center Astronomy and Beyond, ed. M.~{Tsuboi} \& T.~{Oka}, 155

\bibitem[{{Tsuboi} {et~al.}(2017{\natexlab{a}}){Tsuboi}, {Kitamura}, {Tsutsumi}, {Uehara}, {Miyoshi}, {Miyawaki}, \& {Miyazaki}}]{Tsuboi2017b}
---. 2017{\natexlab{a}}, \apjl, 850, L5, \dodoi{10.3847/2041-8213/aa97d3}

\bibitem[{{Tsuboi} {et~al.}(2017{\natexlab{b}}){Tsuboi}, {Kitamura}, {Uehara}, {Miyawaki}, {Tsutsumi}, {Miyazaki}, \& {Miyoshi}}]{Tsuboi2017}
{Tsuboi}, M., {Kitamura}, Y., {Uehara}, K., {et~al.} 2017{\natexlab{b}}, \apj, 842, 94, \dodoi{10.3847/1538-4357/aa74e3}

\bibitem[{Virtanen {et~al.}(2020)Virtanen, Gommers, Oliphant, Haberland, Reddy, Cournapeau, Burovski, Peterson, Weckesser, Bright, {van der Walt}, Brett, Wilson, Millman, Mayorov, Nelson, Jones, Kern, Larson, Carey, Polat, Feng, Moore, {VanderPlas}, Laxalde, Perktold, Cimrman, Henriksen, Quintero, Harris, Archibald, Ribeiro, Pedregosa, {van Mulbregt}, \& {SciPy 1.0 Contributors}}]{SciPy2020}
Virtanen, P., Gommers, R., Oliphant, T.~E., {et~al.} 2020, Nature Methods, 17, 261, \dodoi{10.1038/s41592-019-0686-2}

\bibitem[{{Vollmer} {et~al.}(2022){Vollmer}, {Davies}, {Gratier}, {Liz{\'e}e}, {Imanishi}, {Gallimore}, {Impellizzeri}, {Garc{\'\i}a-Burillo}, \& {Le Petit}}]{Vollmer2022}
{Vollmer}, B., {Davies}, R.~I., {Gratier}, P., {et~al.} 2022, \aap, 665, A102, \dodoi{10.1051/0004-6361/202141684}

\bibitem[{{von Fellenberg} {et~al.}(2022){von Fellenberg}, {Gillessen}, {Stadler}, {Baub{\"o}ck}, {Genzel}, {de Zeeuw}, {Pfuhl}, {Amaro Seoane}, {Drescher}, {Eisenhauer}, {Habibi}, {Ott}, {Widmann}, \& {Young}}]{Fellenberg2022}
{von Fellenberg}, S.~D., {Gillessen}, S., {Stadler}, J., {et~al.} 2022, \apjl, 932, L6, \dodoi{10.3847/2041-8213/ac68ef}

\bibitem[{{Wang} {et~al.}(2020){Wang}, {Li}, {Russell}, \& {Cuadra}}]{Wang2020}
{Wang}, Q.~D., {Li}, J., {Russell}, C. M.~P., \& {Cuadra}, J. 2020, \mnras, 492, 2481, \dodoi{10.1093/mnras/stz3624}

\bibitem[{{Wardle} \& {Yusef-Zadeh}(1992)}]{Wardle1992}
{Wardle}, M., \& {Yusef-Zadeh}, F. 1992, \nat, 357, 308, \dodoi{10.1038/357308a0}

\bibitem[{{Wardle} \& {Yusef-Zadeh}(2008)}]{Wardle2008}
---. 2008, \apjl, 683, L37, \dodoi{10.1086/591471}

\bibitem[{{Whitney} \& {Hartmann}(1993)}]{Whitney1993}
{Whitney}, B.~A., \& {Hartmann}, L. 1993, \apj, 402, 605, \dodoi{10.1086/172163}

\bibitem[{{Wilson} {et~al.}(1987){Wilson}, {Franza}, \& {Noethe}}]{Wilson1987}
{Wilson}, R.~N., {Franza}, F., \& {Noethe}, L. 1987, Journal of Modern Optics, 34, 485, \dodoi{10.1080/09500348714550501}

\bibitem[{{Witzel} {et~al.}(2011){Witzel}, {Eckart}, {Buchholz}, {Zamaninasab}, {Lenzen}, {Sch{\"o}del}, {Araujo}, {Sabha}, {Bremer}, {Karas}, {Straubmeier}, \& {Muzic}}]{Witzel2011}
{Witzel}, G., {Eckart}, A., {Buchholz}, R.~M., {et~al.} 2011, \aap, 525, A130, \dodoi{10.1051/0004-6361/201015009}

\bibitem[{{Witzel} {et~al.}(2017){Witzel}, {Sitarski}, {Ghez}, {Morris}, {Hees}, {Do}, {Lu}, {Naoz}, {Boehle}, {Martinez}, {Chappell}, {Sch{\"o}del}, {Meyer}, {Yelda}, {Becklin}, \& {Matthews}}]{Witzel2017}
{Witzel}, G., {Sitarski}, B.~N., {Ghez}, A.~M., {et~al.} 2017, \apj, 847, 80, \dodoi{10.3847/1538-4357/aa80ea}

\bibitem[{{Wolk} {et~al.}(2008){Wolk}, {Spitzbart}, {Bourke}, {Gutermuth}, {Vigil}, \& {Comer{\'o}n}}]{Wolk2008}
{Wolk}, S.~J., {Spitzbart}, B.~D., {Bourke}, T.~L., {et~al.} 2008, \aj, 135, 693, \dodoi{10.1088/0004-6256/135/2/693}

\bibitem[{{Yelda} {et~al.}(2014){Yelda}, {Ghez}, {Lu}, {Do}, {Meyer}, {Morris}, \& {Matthews}}]{Yelda2014}
{Yelda}, S., {Ghez}, A.~M., {Lu}, J.~R., {et~al.} 2014, \apj, 783, 131, \dodoi{10.1088/0004-637X/783/2/131}

\bibitem[{{Yuan} \& {Narayan}(2014)}]{2014ARA&A..52..529Y}
{Yuan}, F., \& {Narayan}, R. 2014, \araa, 52, 529, \dodoi{10.1146/annurev-astro-082812-141003}

\bibitem[{{Yusef-Zadeh} {et~al.}(2013){Yusef-Zadeh}, {Royster}, {Wardle}, {Arendt}, {Bushouse}, {Lis}, {Pound}, {Roberts}, {Whitney}, \& {Wootten}}]{Yusef-Zadeh2013}
{Yusef-Zadeh}, F., {Royster}, M., {Wardle}, M., {et~al.} 2013, \apjl, 767, L32, \dodoi{10.1088/2041-8205/767/2/L32}

\bibitem[{{Yusef-Zadeh} {et~al.}(2017){Yusef-Zadeh}, {Sch{\"o}del}, {Wardle}, {Bushouse}, {Cotton}, {Royster}, {Kunneriath}, {Roberts}, \& {Gallego-Cano}}]{Yusef-Zadeh2017-ALMAVLA}
{Yusef-Zadeh}, F., {Sch{\"o}del}, R., {Wardle}, M., {et~al.} 2017, \mnras, 470, 4209, \dodoi{10.1093/mnras/stx1439}

\bibitem[{{Zaja{\v c}ek} {et~al.}(2017){Zaja{\v c}ek}, {Britzen}, {Eckart}, {Shahzamanian}, {Busch}, {Karas}, {Parsa}, {Peissker}, {Dov{\v c}iak}, {Subroweit}, {Dinnbier}, \& {Zensus}}]{Zajacek2017}
{Zaja{\v c}ek}, M., {Britzen}, S., {Eckart}, A., {et~al.} 2017, \aap, 602, A121, \dodoi{10.1051/0004-6361/201730532}

\bibitem[{{Zaja{\v{c}}ek} {et~al.}(2020){Zaja{\v{c}}ek}, {Araudo}, {Karas}, {Czerny}, \& {Eckart}}]{Zajacek2020}
{Zaja{\v{c}}ek}, M., {Araudo}, A., {Karas}, V., {Czerny}, B., \& {Eckart}, A. 2020, \apj, 903, 140, \dodoi{10.3847/1538-4357/abbd94}

\bibitem[{{Zhu} {et~al.}(2020){Zhu}, {Li}, {Ciurlo}, {Morris}, {Zhang}, {Do}, \& {Ghez}}]{Zhu2020}
{Zhu}, Z., {Li}, Z., {Ciurlo}, A., {et~al.} 2020, \apj, 897, 135, \dodoi{10.3847/1538-4357/ab980d}

\end{thebibliography}
\bibliographystyle{aasjournal}

\appendix

In this Appendix, we show the results of the MCMC simulations used for the analysis and provide additional information regarding the analysis. Furthermore, we expand the results and discussion of Sec. \ref{sec:results} and Sec. \ref{sec:discuss} to cover some background aspects such as the wind topic which will be covered in an upcoming publication.

\section{Parameter space of the eccentricity and semi-major axis}
\label{sec:parameter_space_e_a_app}

\begin{figure}%
	\centering
	\includegraphics[width=.5\textwidth]{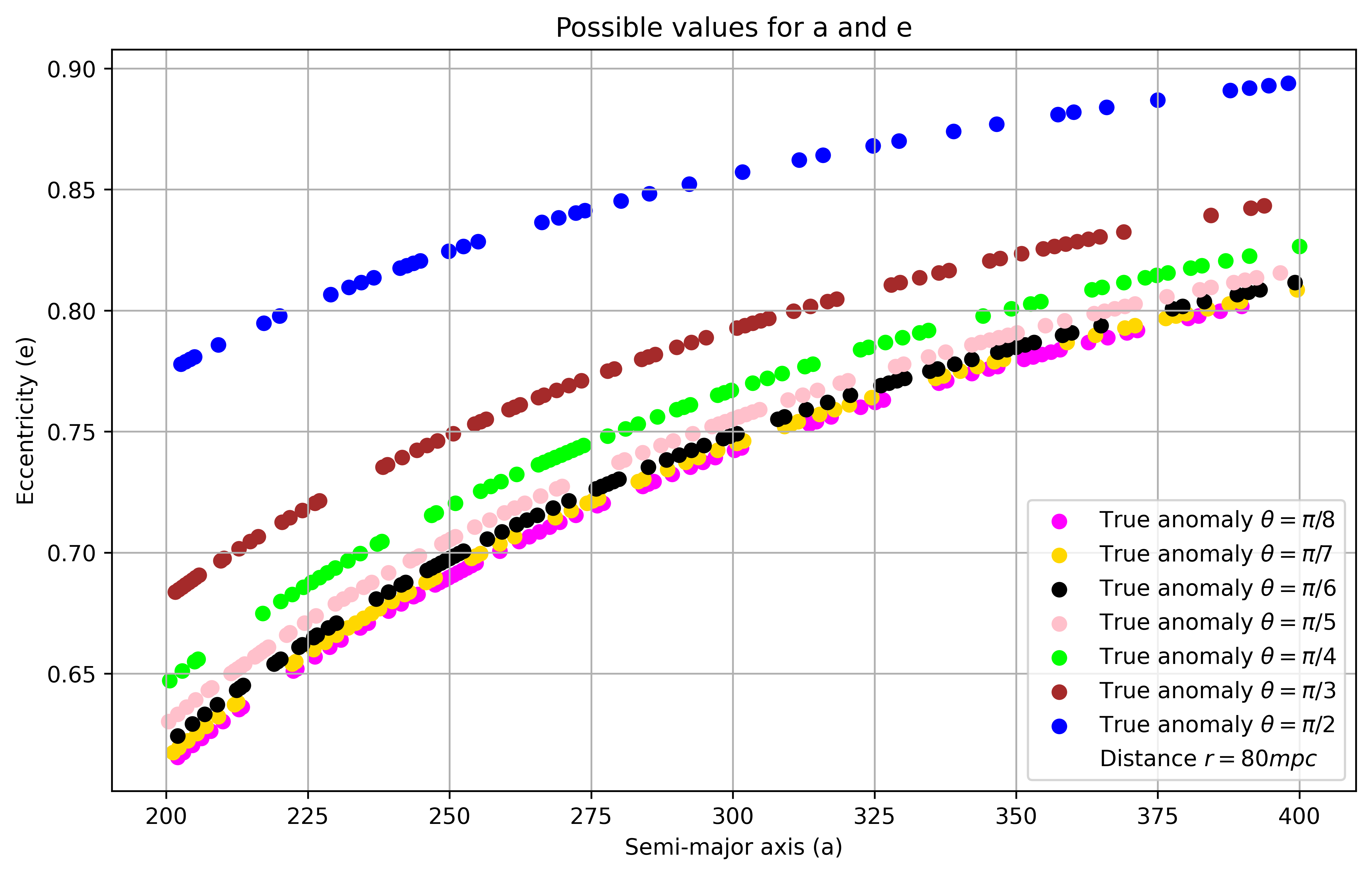}
	\caption{{All possible values for the semi-major axis and eccentricity assuming a constant r and $\theta$. For r, we use the estimated cluster distance of 80 mpc (Sec. \ref{sec:results}). Different values of $\theta$ are indicated. We find a possible eccentricity range between $\pm\,5\%$ and $\pm\,10\%$ with a corresponding uncertainty for the semi-major axis of $300\,\pm\,100$ mpc. As implied by this figure, the uncertainty of the eccentricity decreases for higher values of a.}}
\label{fig:eccentricity_semimajoraxis_parameterspace}
\end{figure}

{Since we discussed already possible values for the inclination and the LOAN in Sec. \ref{sec:discuss_limitations}, we want to complete this discussion by inspecting possible values for the eccentricity e and the semi-major axis a. For this, we refer to Eq. \ref{eq:kepler} in Sec. \ref{sec:results}. Because the projected distance of the data points to Sgr~A* is fixed, we can assume a constant r and true anomaly $\theta$. With this, we plot the possible parameter space for these quantities (Fig. \ref{fig:eccentricity_semimajoraxis_parameterspace}).}
{For this example, we used the estimated 3d distance for IRS 13 of r=80 mpc (Sec. \ref{sec:results}) with an angle for the true anomaly between $\theta=\pi/2$ and $\theta=\pi/8$. In the shown example, we used semi-major axis values between 200-400 mpc and find that for large uncertainties of $\pm\,100$, the resulting eccentricity range is 0.1-0.2. It furthermore demonstrates that the derived orbital parameter e is inside a reasonable uncertainty range of $\pm\,10\%$ which is an upper limit. However, this result has no impact on the discussion about the inclination i and the LOAN (Sec. \ref{sec:discuss_limitations}) and the result presented in Fig. \ref{fig:cluster_structure} and Fig. \ref{fig:cluster_structure_pa}.}

\section{Missing star trails}
\label{sec:star_trail_app}

Assuming a gas-to-dust ratio of 100 and 1000, we will use the massive, young, and embedded cluster RCW 108 as an example for the mentioned cluster dynamics and star trails \citep{Comeron2005, Comeron2007}. Employing the indicated gas-to-dust ratio, we get a total stellar mass of $2.3\,\times\,10^3\,-\,2.3\,\times\,10^4\,M_{\odot}$ for RCW 108 \citep{Wolk2008} which is in a comparable range of the IRS 13 estimate of $1.5\,\times\,10^4\,M_{\odot}$ \citep{Paumard2006}. \cite{Comeron2007} estimate a multiplicity fraction of $\sim\,0.43$ which serves as an quantitative argument. A more qualitative approach exceeds the scope of this work. However, a closer look at the derived multiplicity fraction of RCW 108 reveals that the triplet fraction is about $\sim\,0.07$ whereas higher-order systems show about $\sim\,0.10$. This example implies that the young cluster tends to destroy high-order companion systems. A possible explanation could be hard and soft binary systems \citep[][]{Heggie1975, Hills1975a, Hills1975b}. Assuming a random stellar encounter, a soft binary gets destroyed, resulting in at least one singlet whereas the companion may become a runaway star \citep[][]{Hills1988}. In contrast, a hard binary can be described as a system with higher (binding) energy and increased mean stellar kinetic energy compared to the soft binary. Using the definition from \cite{Heggie1975}, it is
\begin{equation}
    E_B\,>\,\frac{1}{2}\langle mv^2\rangle
    \label{eq:binding_energy}
\end{equation}
where E$_B$ defines the binding energy of the system. It is obvious that the mass of the multiplet system is the main contributor, which explains why most massive stars are accompanied by at least one stellar companion \citep{Sana2012}. Furthermore, Eq.~\ref{eq:binding_energy} implies that single components of a triplet (or higher order) system suffer from a decreased binding energy because of their distance to the primary. This is reflected by
\begin{equation}
    E_B\,=\,-G\frac{m_1 m_2}{2a_B}
\end{equation}
where $G$ is the gravitational constant, $m_1$ and $m_2$ are the masses of the binary system, and $a_B$ is the semi-major axis of the system. For simplicity, we set $m_1=m_2=m_3$ where $m_3$ is the mass of the triplet. From these definitions, it is evident that
\begin{equation}
    E_B\,=\,-G\frac{m_1m_2}{2a_B}\,>\,-G\frac{(m_1 + m_2)\cdot m_3}{2a_T}\,=\,E_T
\end{equation}
where E$_T$ is the binding energy of the triplet system with a$_T\,\gg$ a$_B$ \citep[see][]{Heggie1975}. The above equation demonstrates that high-order multiplet systems are prone to dissolve in the presence of a hard binary. Hence, a system tends to create a higher fraction of binaries.\newline
Although RCW 108 serves just as an example, it is suggested that this dynamic behavior of the cluster can be transferred to IRS 13. Because the trajectory of the expelled component of a triplet (or soft binary) system does not rely on the interaction of IRS 13 with Sgr~A*, we conclude a random direction for this cluster member. Hence, no star trail is needed to classify IRS 13 as an evaporating cluster. A detailed overview of the destruction and creation of multiplets, a common process for young cluster timescales of a few $10^4$ yr \citep{Hurley2002}, is described in \cite{Zwart2010}. 

\section{Winds in the inner parsec}
\label{sec:wind_app}
{It is assumed that the black hole feeding is driven by stellar winds that are orginating from the S-stars \citep{Luetzgendorf2016}. While this idea promises many different interesting scientific aspects, it is limited by the exact knowledge of the stellar type of the S-stars and the related wind velocities. However, \cite{Wardle1992} formulated the idea that stellar winds originating at the IRS 16 cluster are responsible for the mini-cavity to describe large scale imprints of the massive stars in the inner parsec. This is followed by the analysis of \cite{Krabbe1995} who estimated wind velocities of several hundred km/s for the massive stars in the IRS 16 cluster. Recently, \cite{peissker2021} followed up on this speculation to explain the bow-shock shape of X7. Furthermore, \cite{Ciurlo2023} independently confirmed the misalignment of X7 and Sgr~A*.
Both Pei{\ss}ker and Ciurlo et al. conclude that elongated and dusty sources, such as X7, pointing in the direction of the IRS 16 cluster exclude a speculative nuclear wind that is launched at the position of Sgr~A*. If the IRS 16 cluster is responsible for the mini-cavity, it can be argued that the possible interaction of stellar winds with X7 could also be the reason for the bow shock of X3 \citep{peissker2023b}. Hence, it is implied that the mini-cavity is neither created by a central speculative star or the interaction IRS 13 with the northern arm. From this discussion, we conclude that the mini-cavity is a distinct feature that is created by the interaction with the surrounding and the ambient medium. We will address this point in more detail in an upcoming publication.}

\section{Supplementary material}
\label{sec:sup_mat_app}

{Along with this article, we publish supplementary material to present the analysis done for each individual source listed in Table \ref{tab:orbit_table_all_ds}-\ref{tab:final_orbit_table}. The associated uncertainties for the orbital elements are extracted from the provided corner plots. Furthermore, we publish the RA and DEC fits for all investigated sources. Finally, we release .txt files with the derived Keplerian orbit for all cluster members, arranged as follows. The first column represents the epoch, and the second and third columns give the RA and DEC coordinates of each source in arcseconds.}

\section{MCMC simulations}
\label{sec:mcmc-app}

\begin{figure*}%
	\centering
	\includegraphics[width=1.\textwidth]{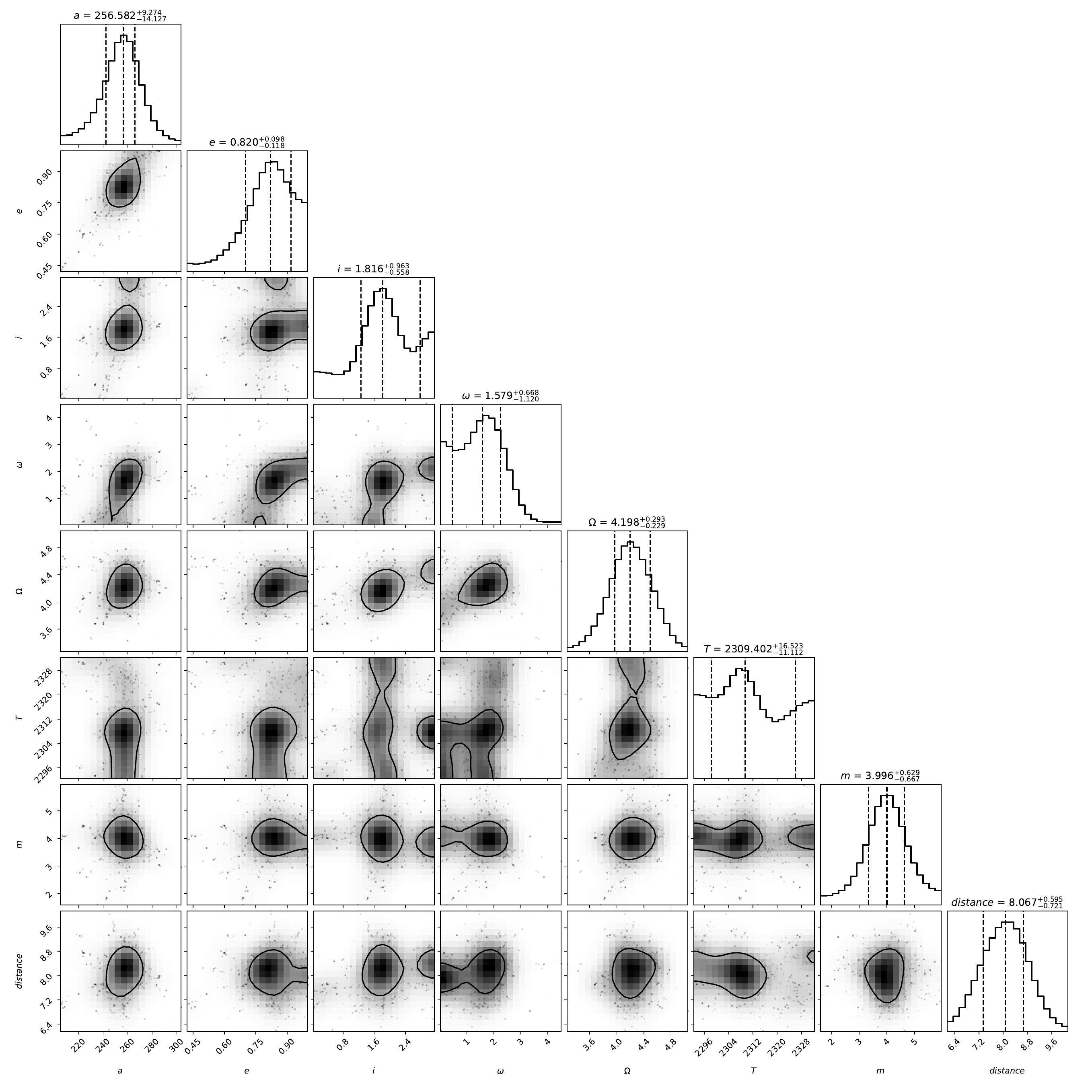}
	\caption{MCMC simulation of the Keplerian orbit of $\alpha$ (Fig. \ref{fig:orbits_1}).}
\label{fig:mcmc_alpha}
\end{figure*}

\begin{figure*}%
	\centering
	\includegraphics[width=1.\textwidth]{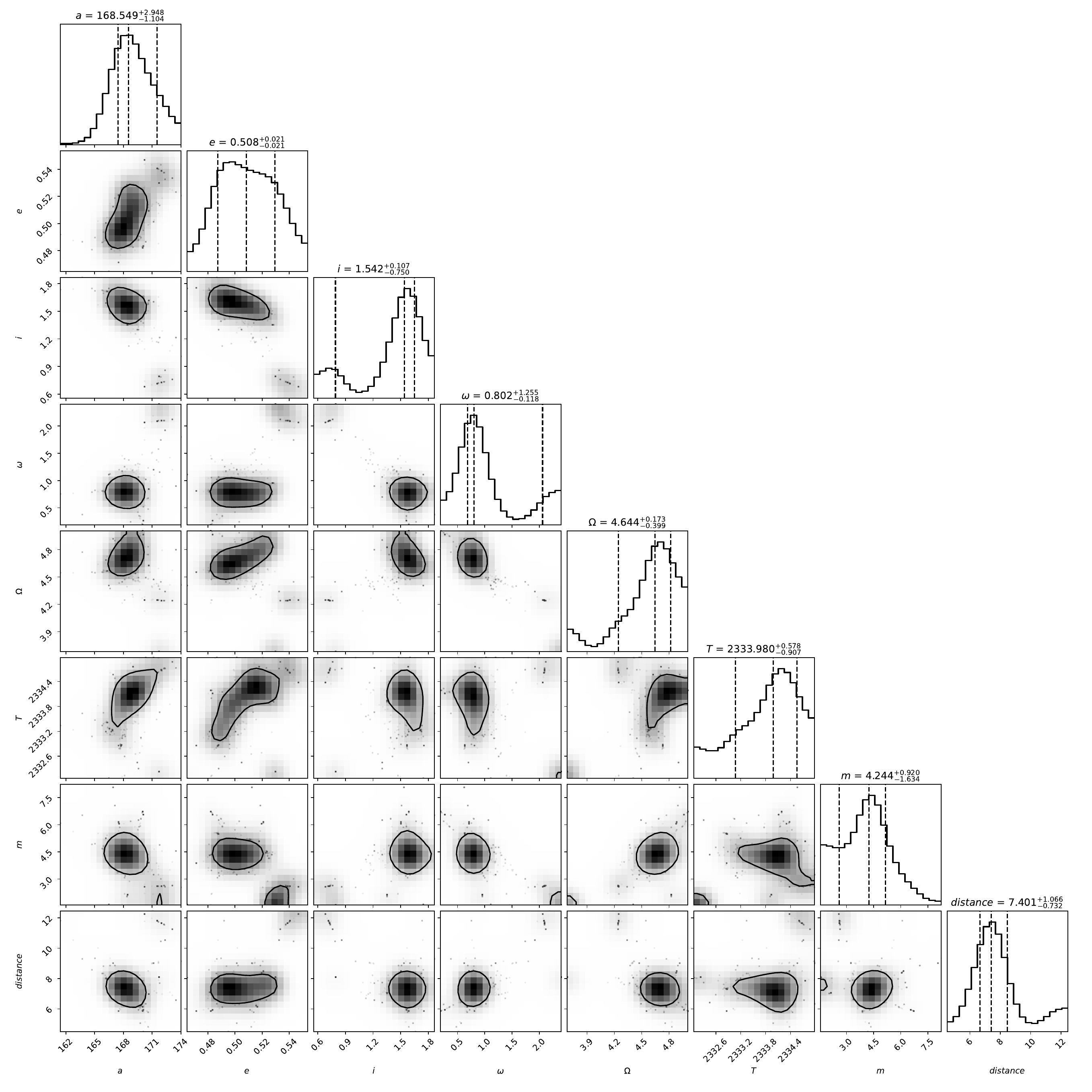}
	\caption{MCMC simulation of the Keplerian orbit of $\beta$ (Fig. \ref{fig:orbits_1}).}
\label{fig:mcmc_beta}
\end{figure*}

\begin{figure*}%
	\centering
	\includegraphics[width=1.\textwidth]{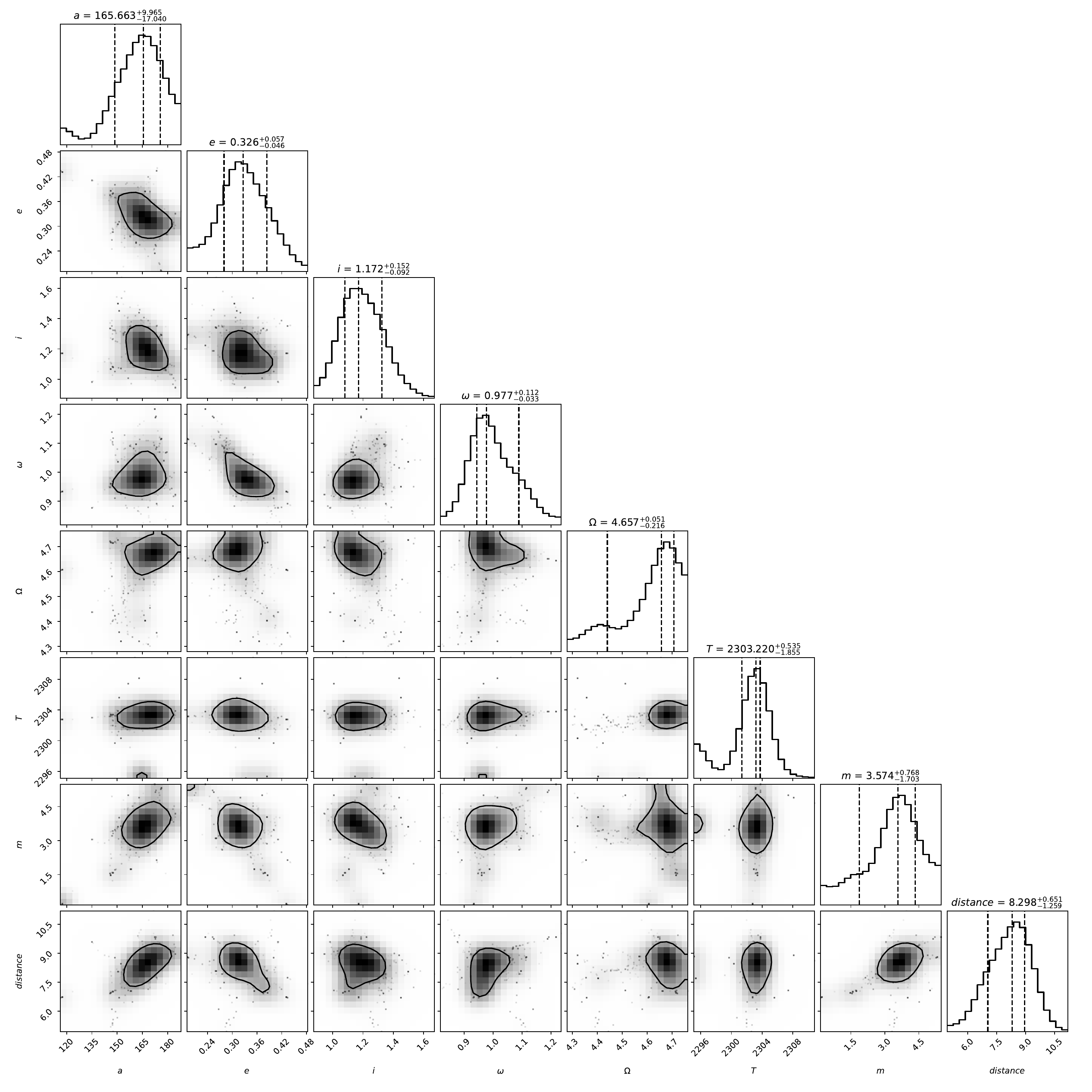}
	\caption{MCMC simulation of the Keplerian orbit of $\delta$ (Fig. \ref{fig:orbits_1}).}
\label{fig:mcmc_delta}
\end{figure*}

\begin{figure*}%
	\centering
	\includegraphics[width=1.\textwidth]{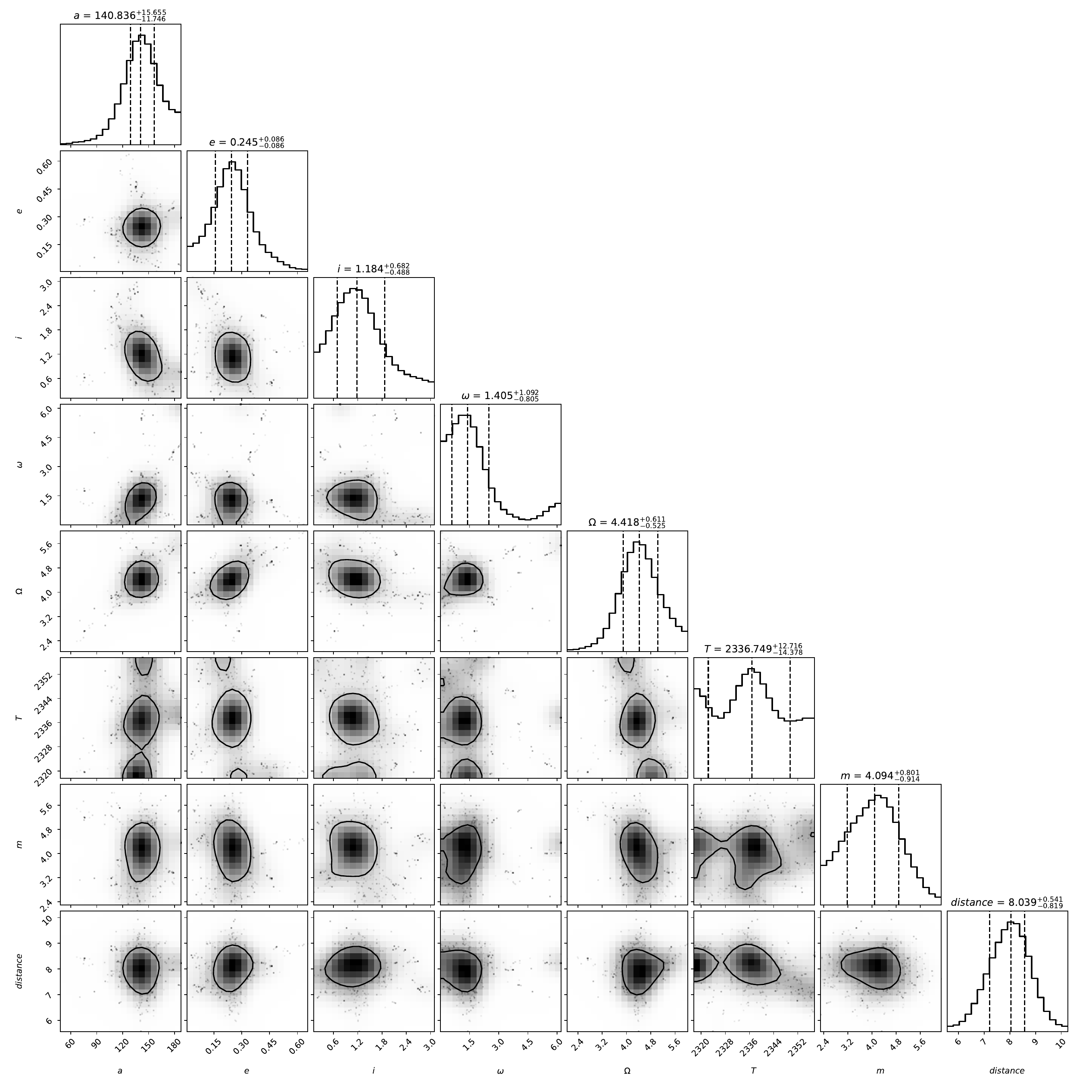}
	\caption{MCMC simulation of the Keplerian orbit of $\epsilon$ (Fig. \ref{fig:orbits_1}).}
\label{fig:mcmc_epsilon}
\end{figure*}

\begin{figure*}%
	\centering
	\includegraphics[width=1.\textwidth]{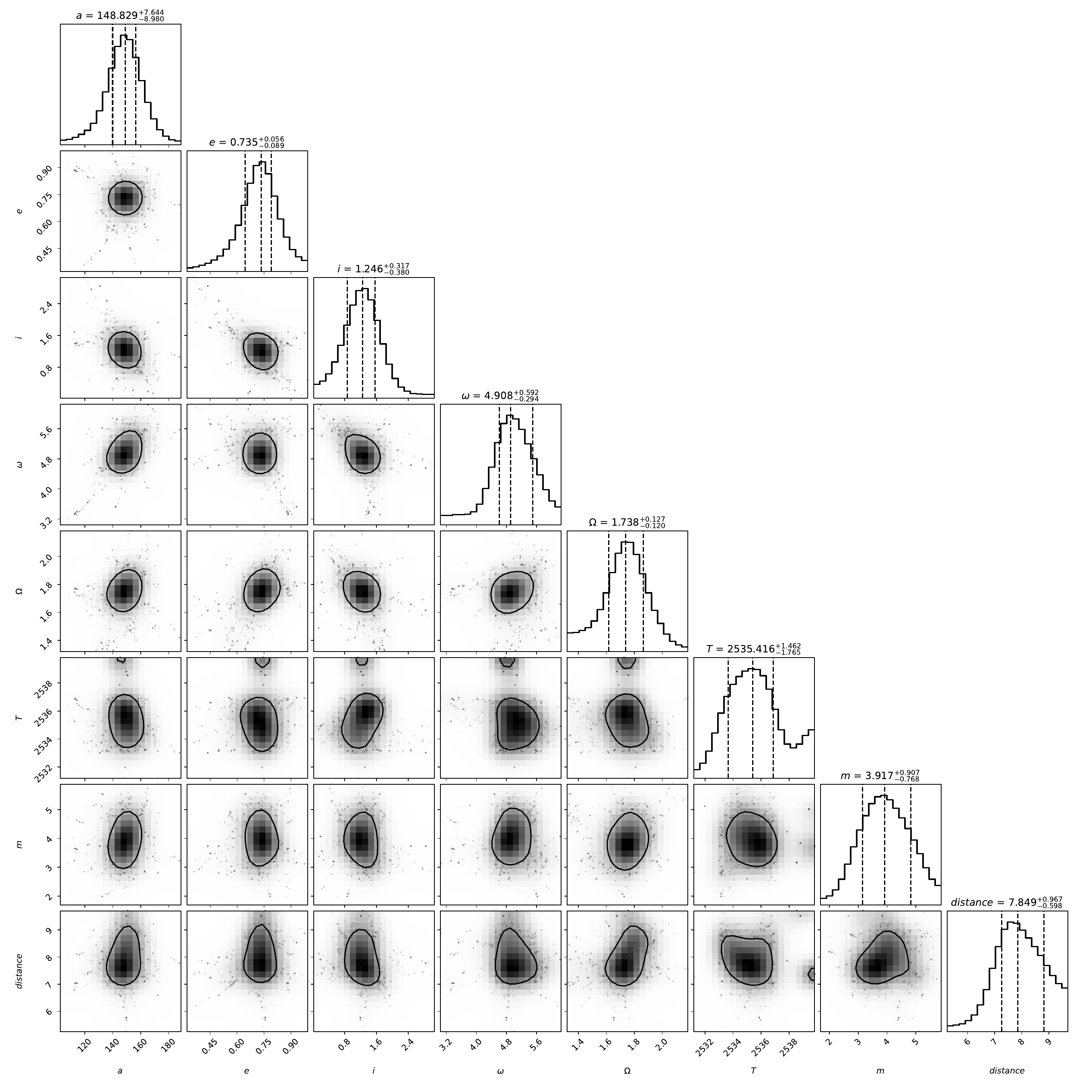}
	\caption{MCMC simulation of the Keplerian orbit of $\eta$ (Fig. \ref{fig:orbits_1}).}
\label{fig:mcmc_eta}
\end{figure*}

\begin{figure*}%
	\centering
	\includegraphics[width=1.\textwidth]{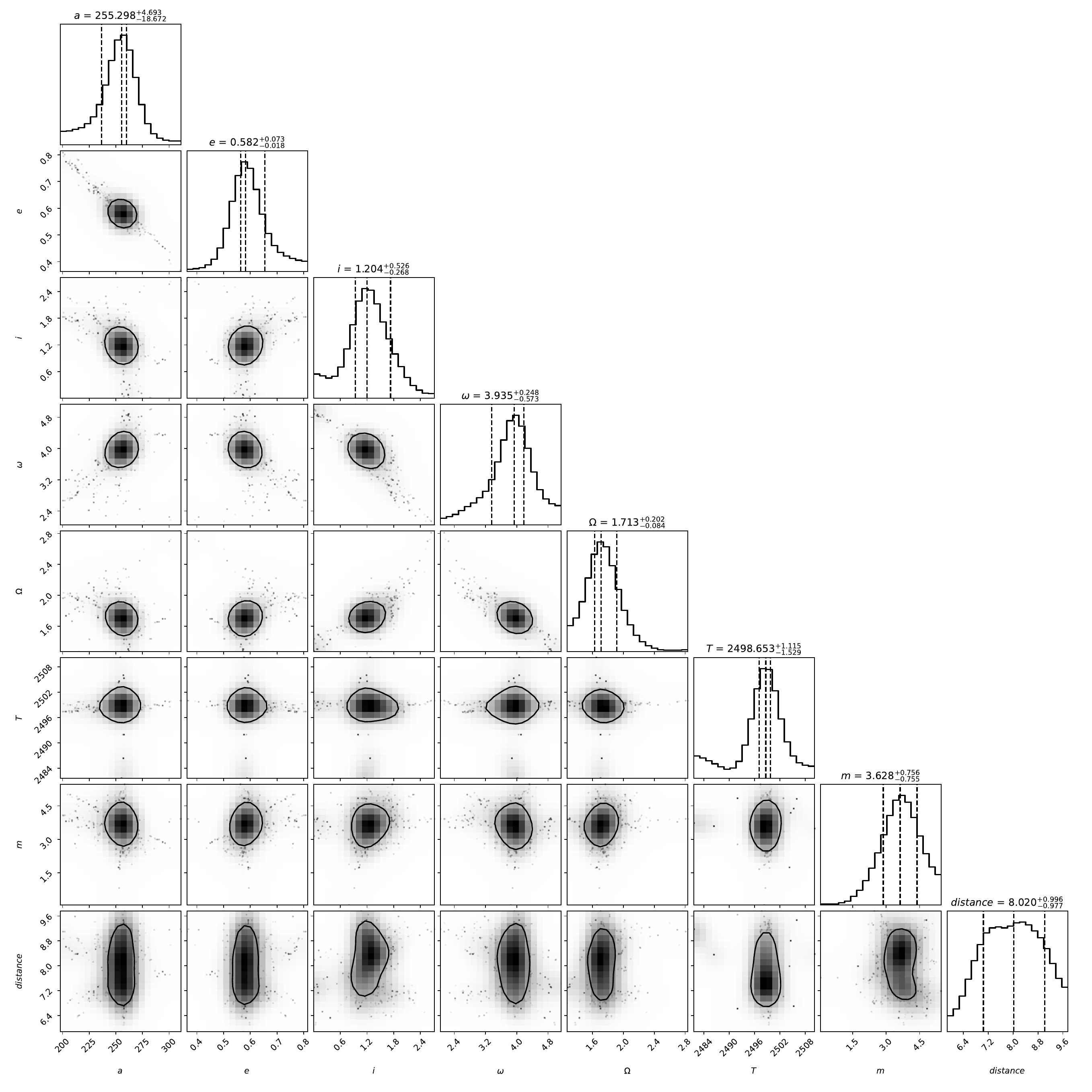}
	\caption{MCMC simulation of the Keplerian orbit of $\gamma$ (Fig. \ref{fig:orbits_1}).}
\label{fig:mcmc_gamma}
\end{figure*}

\begin{figure*}%
	\centering
	\includegraphics[width=1.\textwidth]{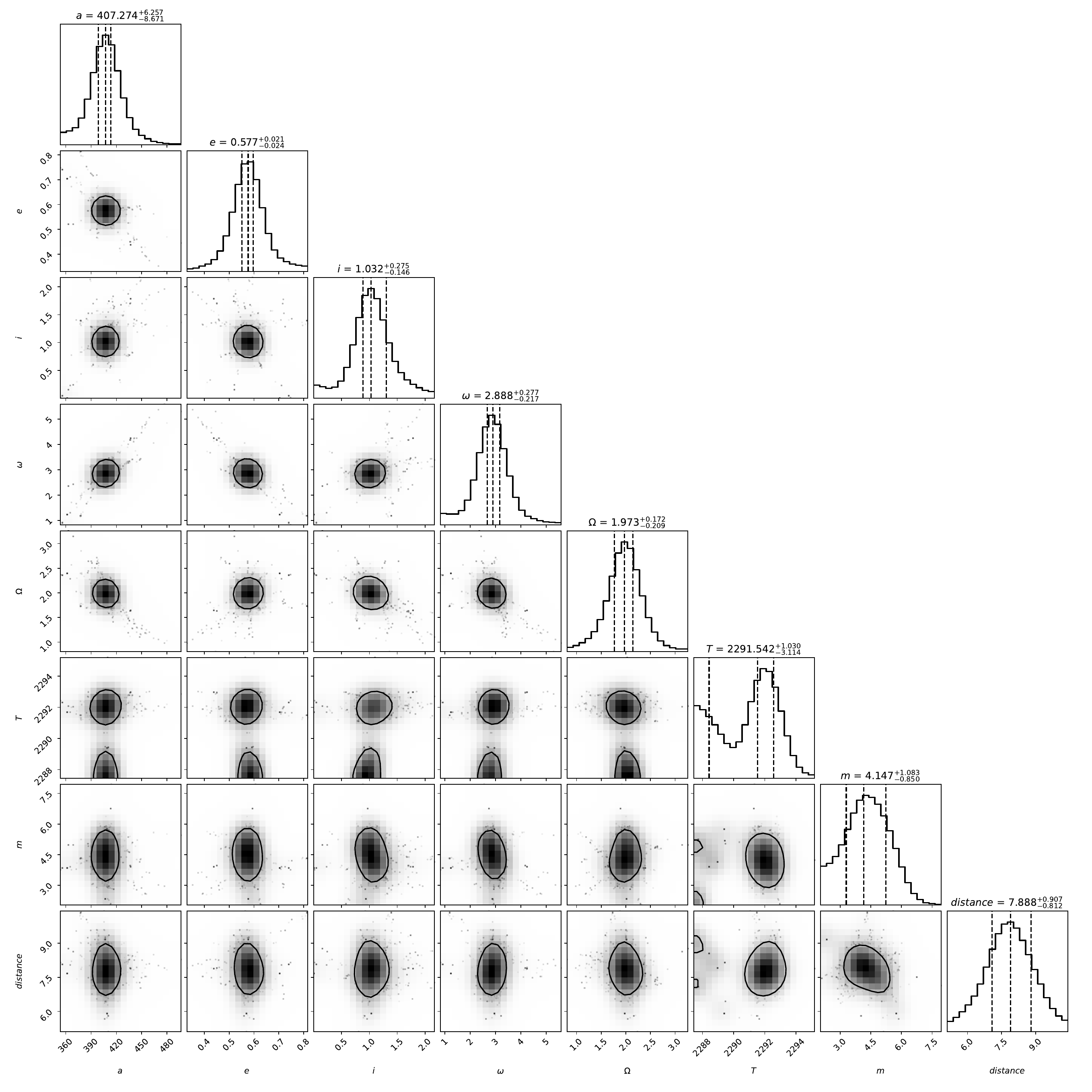}
	\caption{MCMC simulation of the Keplerian orbit of $\zeta$ (Fig. \ref{fig:orbits_1}).}
\label{fig:mcmc_zeta}
\end{figure*}

In this section, we present the MCMC simulations using the Keplerian elements listed in Table \ref{tab:final_orbit_table} as a prior. In Fig. \ref{fig:mcmc_alpha}-\ref{fig:mcmc_zeta}, we display the related results of the MCMC simulations. For every parameter, we find a compact distribution resulting in a reasonable uncertainty range. The MCMC results for the DS sources and the E stars (see Table \ref{tab:orbit_table_all_ds} and Table \ref{tab:orbit_table_estars}) will be publicly available.

\section{Additional comments of the nature of $\eta$}

\begin{figure*}%
	\centering
	\includegraphics[width=1.\textwidth]{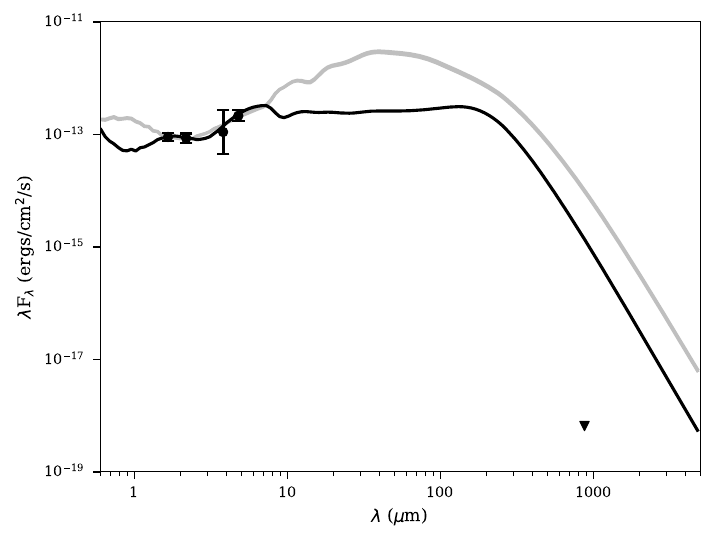}
	\caption{Spectral Energy Distribution of the candidate YSO $\eta$ using the set of synthetic models published along with \cite{Robitaille2017}. This best-fit result shows similarities with the SED presented in Paper I but including a dust envelope with a low density. The submm/radio data point marks an upper limit due to the sensitivity threshold of the ALMA observations.}
\label{fig:sed_eta}
\end{figure*}

{Due to the amount of concepts covered in Paper I, we want to add additional information on the candidate YSO $\eta$. While we do not question its nature based on the color-color analysis accompanied by the SED analysis, we want to follow-up on the missing envelope as suggested by the fit. Due to the absence of $\eta$ in the submm/radio regime, it is obvious that no dust material with a temperature $<$30 K is present in the system. If there is a {premature} envelope related to $\eta$, the material is optically thin, which would not contradict the clear detection of the system in the L- and M-band. However, we expect a low and assumably variable dust density for the case of an optically thin envelope which we could aim for with an adapted infall rate $\dot{M}$. As discussed in \cite{Robitaille2017}, this approach poses limitations and could have been compensated for with the set of synthetic SEDs based on an updated YSO model published along with the mentioned publication. The dust density of the envelope $\rho_{env}$ presented in Robitaille et al. is calculated with}
\begin{equation}
    \rho_{env}\,=\,\frac{\dot{M}}{4\pi(GM_{S}R_C)^{1/2}}
\end{equation}
where $M_{S}$ {denotes} the mass of the central stellar object and $R_{C}$\footnote{We noticed a misprint in \cite{Robitaille2017}. The centrifugal radius is R$_C$ and not R$_C^3$ \citep{Whitney1993}.} the centrifugal radius where dust material is {piled} up due to the distribution of the radiating material. Although both YSO models presented in \cite{Robitaille2011} and \cite{Robitaille2017} have to be proven highly accurate by fitting the SED of a given astronomical object, defining the dust density $\rho_{env}$ rather than the infall rate $\dot{M}$ to describe the dusty envelope seems to be a more effective way in the case of challenging objects such as $\eta$. The reason for this discrepancy is indeed the low dust density of the envelope of $\eta$ which we had to exclude in total as discussed and presented in Paper I. In Fig. \ref{fig:sed_eta}, we show the fit of the synthetic YSO models as presented in \cite{Robitaille2017}.
{As expected, the density of the optically thin dust envelope is $1\times10^{-23}\,g/cm^2$ whereas all other best-fit results show comparable scales with the analysis presented in Paper I. We list the related outcome of the fit in Table \ref{tab:sed_table_eta} in comparison with the parameters estimated in Paper I to underline the consistency of our approach.}
\begin{table*}
\setlength{\tabcolsep}{0.99pt}
\centering                                  
\begin{tabular}{|c|ccccccc|}
\toprule
ID     & \multicolumn{1}{c|}{Mass [$M_{\odot}$]} & \multicolumn{1}{c|}{Luminosity [$10^3\times L_{\odot}$]}  & \multicolumn{1}{c|}{Infall rate [$\dot{M}_{\odot}$]} & \multicolumn{1}{c|}{Radius [$R_{\odot}$]}& \multicolumn{1}{c|}{Disk mass [$M_{\odot}$]} & \multicolumn{1}{c|}{Disk size [AU]} & Envelope size [AU] \\
\hline
$\eta$, Paper I & 0.5  $\pm$ 0.2 & 2  $\pm$ 0.5 & - & 0.8 $\pm$ 0.2 & 0.005 $\pm$ 0.002 & 0.13-50 & -  \\
\hline
$\eta$, this work & 0.4  $\pm$ 0.2 & 8  $\pm$ 0.5 & $\approx 10^{-24}$ & 0.8 $\pm$ 0.2 & 0.06 $\pm$ 0.002 & 0.13-1.25 & 1.26-10  \\
\hline
\end{tabular}
\caption{Comparison of best-fit parameters describing the YSO $\eta$ using the models described in \cite{Robitaille2011} and \cite{Robitaille2017}. Using the stellar radius and luminosity, we extract the mass from the evolutionary tracks presented in \cite{peissker2023b}.}
\label{tab:sed_table_eta}
\end{table*}
{Using the stellar radius and luminosity with the evolutionary tracks shown in \cite{peissker2023b}, we find an satisfying overlap of the YSO models of \cite{Robitaille2011, Robitaille2017}. The main difference is in fact that the latter model is able to reproduce the low-density and optically thin envelope of $\eta$ with a theoretical infall rate $\dot{M}$ of $\approx 10^{-24}$. This estimate and the dust envelope density poses the limit of the model discussed in \cite{Robitaille2017} and should be considered as a motivation for further observations with, e.g., JWST.}

\section{N-body simulations of the inspiralling cluster IRS 13}
\label{sec:nbody_mock_app}

The motivation of the section is a representation of the N-body simulations that incorporate the location of the CND, the inner parsec, and the IRS 13 cluster. Due to the different dimensions of the mentioned components, smaller structures such as the CWD and CCWD are excluded. We refer the interested reader to \cite{Murchikova2019} to find a representation of the latter two components in relation to the inner parsec.
For the N-body simulations presented in this section, we use the same setup as for the results presented in Fig. \ref{fig:irs13_evo}. In addition, we include a size measure for the individual simulated stars that scales with the mass of the particles. Hence, larger dots represent a higher mass. We limit the particle size to N=50 to avoid a confusing representation of the dense cluster. The different stellar masses range from 0.1 to 100 M$_{\odot}$ following a power-law distribution with an index of -2.35 \citep{Massey1998, Kroupa2001}. We further placed the cluster inside the CND at about 5 pc.
\begin{figure*}%
	\centering
	\includegraphics[width=1.\textwidth]{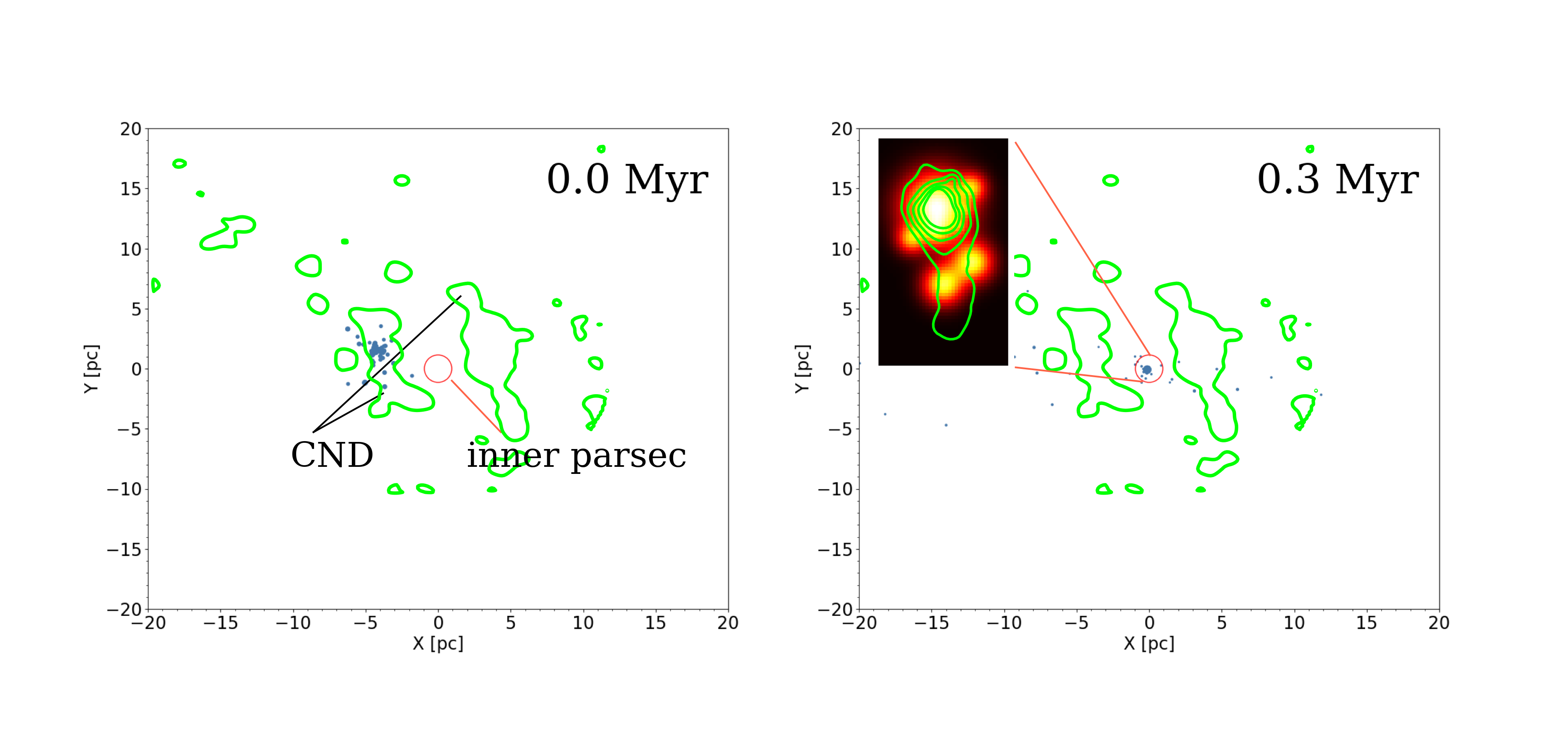}
	\caption{N-body simulations of a dense cluster carried out with AMUSE. The initial starting point of the cluster is at (0,5) pc and gets attracted by the compact and dominating mass of Sgr~A*. The circle indicates the size of the {\it inner parsec} whereas the CND (lime-colored CO contours observed with ALMA in the left plot) is located at a distance of 1-5 pc from Sgr~A* \citep{Requena-Torres2012}. After the cluster passes the CND, the denser and embedded core of the cluster rapidly sinks in towards the inner parsec consistent with IRS13. Because of soft binaries, low-mass stars are kicked out with random trajectories and velocities. Horizontal axes are conventionally oriented, that is, from a negative to a positive range. The inlet of the right plot shows the stars with the related positions of the circle located at (0,0) pc. In contrast, the orientation of the inlet is identical with the projection on the sky to ensure are correct adaptation of the [FeIII] line distribution which is represented by lime-colored contours. Here, Sgr~A* is located at (0,0).}
\label{fig:irs13_evo1}
\end{figure*}
In Fig. \ref{fig:irs13_evo1}, we show the results of the simulations consistent with the ones displayed in Fig. \ref{fig:irs13_evo}. In addition, we implemented contour lines based on CO observations carried out with ALMA. The inlet in the plot on the right of Fig. \ref{fig:irs13_evo1} shows a mock image where we used the position of the simulated particles at 0.3 Myr to create a fits file with the corresponding relative positions. This is overlaid with the [FeIII] contour lines.

\end{document}